\documentclass[11pt]{article}
\pdfoutput=1
\usepackage{jcapmod}

\usepackage{tikz}
\usepackage{tikz-cd}
\usetikzlibrary{shapes.misc,positioning,arrows,arrows.meta,bending,matrix,shapes,fit,tikzmark,calc,external,patterns,topaths,decorations.pathmorphing,decorations.markings,decorations.pathreplacing,intersections}
\usepackage{xparse}
\usepackage{xstring}

\tikzset{>=latex} 

\usepackage[export]{adjustbox}
\usepackage{shorthand}
\usepackage{mathtools}
\usepackage{booktabs}
\usepackage[english]{babel}
\usepackage{amsmath,amssymb,amsbsy,amstext, amsthm, simplewick, amsfonts}
\usepackage{graphicx}
\graphicspath{{./Figures/}}
\usepackage[small]{caption}
\usepackage{siunitx}
\usepackage{upgreek}
\usepackage{framed}
\usepackage{wrapfig}
\usepackage{multirow}
\usepackage{bbm}

\newcommand{\Cross}{$\mathbin{\tikz [x=1.4ex,y=1.4ex,line width=.2ex, black] \draw (0,0) -- (0.45,0.45) (0,0.45) -- (0.45,0);}$}%

\tikzset{fieldlines/.style={black,decoration={markings,mark=at position #1 with {\arrow[opacity=1]{latex}}},
                            postaction={decorate},line width=1},
         fieldlines/.default=0.55}
      
 \tikzset{fieldlines2/.style={black,decoration={markings,mark=at position #1 with {\arrow[opacity=1]{latex}}},
                            postaction={decorate},line width=0.5},
         fieldlines2/.default=0.55}   
 \tikzset{fieldlines3/.style 2 args={decoration={markings,mark=at position #1 with {\arrow[opacity=#2]{latex}}},
                            postaction={decorate},line width=1},
         fieldlines3/.default={0.55}{1}}  
\tikzset{cross/.style={cross out, draw=black, minimum size=2*(#1), inner sep=0pt, outer sep=0pt},
cross/.default={1pt}}

\newcounter{itemcounta}
\newcounter{itemcountb}
\newcounter{itemcountc}
\newcommand{\timeloop}[8]{
\setcounter{itemcountc}{0}
  \begin{scope}[xshift=#1,yshift=#2,rotate=180]

	\foreach[count=\i] \V in {#7} { \addtocounter{itemcountc}{\V}
	\foreach \j in {1,...,\V}{
	\filldraw ({#3*sin(360*(\i-1)/#4+180/#4+(\j-1-(\V-1)/2)*15)},{#3*cos(360*(\i-1)/#4+180/#4+(\j-1-(\V-1)/2)*15)}) circle(#5);}
	\draw  ({#3*sin(360*(\i-1)/#4+180/#4+((\V-1)/2)*15)},{#3*cos(360*(\i-1)/#4+180/#4+((\V-1)/2)*15)}) arc ({90-(360*(\i-1)/#4+180/#4+((\V-1)/2)*15)}:{90-(360*(\i-1)/#4+180/#4-((\V-1)/2)*15)}:#3);
	
    }

   	\foreach[count=\i] \LR in {#6} {
		\foreach[count=\k] \l in {#7} {
			\ifnum \k=\i
				\setcounter{itemcounta}{\l}
			\fi
			\ifnum \k=\numexpr\i-1 \relax 
				\setcounter{itemcountb}{\l}
			\fi
			\ifnum \k=\numexpr\i-1 +#4\relax 
				\setcounter{itemcountb}{\l}
			\fi	
			
		}
	\ifnum \LR =3 
			\draw ({#3*sin(360*(\i-2)/#4+180/#4)},{#3*cos(360*(\i-2)/#4+180/#4)}) arc ({90-(180/#4+360*(\i-2)/#4)}:{90-(180/#4+360*(\i-1)/#4}):#3);
	\fi
	\ifnum \LR=2
			\draw[densely dotted] ({#3*sin(360*(\i-2)/#4+180/#4)},{#3*cos(360*(\i-2)/#4+180/#4)}) arc ({90-(180/#4+360*(\i-2)/#4)}:{90-(180/#4+360*(\i-1)/#4}):#3);
	\fi
	\ifnum \LR<2

			\draw ({#3*sin(360*(\i-2)/#4+180/#4)},{#3*cos(360*(\i-2)/#4+180/#4)}) arc ({90-(180/#4+360*(\i-2)/#4)}:{90-(180/#4+360*(\i-1)/#4}):#3);
		\begin{scope}[xshift={(-1)^\LR*cos(-360*(\i-1)/#4)*7/24*(3pt+1pt *4.5)},yshift={(-1)^\LR*sin(-360*(\i-1)/#4)*7/24*(3pt+1pt *4.5)},rotate={-360*(\i-1)/#4}]
			\draw[opacity=0,fieldlines3={1}{#8}] (0,#3)--({(-1)^\LR*0.001},{#3});
		\end{scope}
	\fi

    }

  \end{scope}
}

\makeatletter

\tikzset{
  chain small blob/.style={
    draw,
    color=gray,
    line width=0.65pt,
    rounded rectangle,
    rounded rectangle arc length=180,
    inner xsep=-2pt,
    inner ysep=0.6em
  },
  chain large blob/.style={
    draw,
    color=gray,
    line width=0.65pt,
    rounded rectangle,
    rounded rectangle arc length=180,
    inner xsep=1.5pt,
    inner ysep=0.9em
  },
  chain solid edge/.style={
    line width=1.05pt
  },
  chain dashed edge/.style={
    line width=1.05pt,
    dashed
  },
  chain dashed edges/.code={\def\chain@dashededges{#1}}
}

\def\chain@dashededges{}
\newcommand{\chain@draw}[5]{%
\begin{tikzpicture}[baseline=(current bounding box.center)]
    \def\dx{0.9}
    \def\vpad{0.28}
    \def\xpad{0.28}
    \foreach \i in {1,...,#2}{
        \pgfmathsetmacro{\xx}{(\i-1)*\dx}
        \coordinate (v\i) at (\xx,0);
        \coordinate (Lv\i) at ($(v\i)+(-\vpad,0)$);
        \coordinate (Rv\i) at ($(v\i)+(\vpad,0)$);
    }
    \pgfmathtruncatemacro{\NmOne}{#2-1}
    \ifnum\NmOne>0
        \foreach \i in {1,...,\NmOne}{
            \pgfmathsetmacro{\xx}{(\i-0.5)*\dx}
            \coordinate (x\i) at (\xx,0);
            \coordinate (Lx\i) at ($(x\i)+(-\xpad,0)$);
            \coordinate (Rx\i) at ($(x\i)+(\xpad,0)$);
        }
    \fi
    \coordinate (Lall) at ($(v1)+(-0.30,0)$);
    \coordinate (Rall) at ($(v#2)+(0.30,0)$);
    \node[
        draw,
        color=white,
        line width=0.65pt,
        rounded rectangle,
        rounded rectangle arc length=180,
        fit=(Lall)(Rall),
        inner xsep=-2pt,
        inner ysep=0.6em
    ] {};
    \ifnum\NmOne>0
        \foreach \i in {1,...,\NmOne}{
            \pgfmathtruncatemacro{\ip}{\i+1}
            \def\chain@edgestyle{chain solid edge}

            \ifx\chain@dashededges\@empty\else
                \foreach \j in \chain@dashededges {
                    \ifnum\i=\j\relax
                        \xdef\chain@edgestyle{chain dashed edge}
                    \fi
                }
            \fi

            \draw[\chain@edgestyle] (v\i) -- (v\ip);
        }
    \fi
    \ifnum#1=1
        \ifnum\NmOne>0
            \foreach \i in {1,...,\NmOne}{
                \node at (x\i) {\Cross};
            }
        \fi
    \fi
    \def\largebloblist{#4}
    \ifx\largebloblist\@empty\else
        \foreach \a/\b in {#4}{
            \node[
                chain large blob,
                fit=(L\a)(R\b)
            ] {};
        }
    \fi
    \def\smallbloblist{#3}
    \ifx\smallbloblist\@empty\else
        \foreach \a/\b in {#3}{
            \node[
                chain small blob,
                fit=(L\a)(R\b)
            ] {};
        }
    \fi
    \foreach \i in {1,...,#2}{
        \draw[fill] (v\i) circle (.8mm);
    }
    \def\labellist{#5}
    \ifx\labellist\@empty\else
        \foreach \a/\b/\lab/\ddx/\ddy in {#5}{
            \node[inner sep=1pt] at ($(\a)!0.5!(\b)+(\ddx,\ddy)$) {\lab};
        }
    \fi
\end{tikzpicture}%
}
\NewDocumentCommand{\chaingraph}{m O{} O{} O{}}{%
    \chain@draw{0}{#1}{#2}{#3}{#4}%
}

\makeatother

\tikzset{
  marked chain small blob/.style={
    draw,
    color=gray,
    line width=0.65pt,
    rounded rectangle,
    rounded rectangle arc length=180,
    inner xsep=0pt,
    inner ysep=0.75em
  },
  marked chain large blob/.style={
    draw,
    color=gray,
    line width=0.65pt,
    rounded rectangle,
    rounded rectangle arc length=180,
    inner xsep=5pt,
    inner ysep=1.1em
  }
}

\NewDocumentCommand{\markedchaingraph}{m O{} O{} O{} O{1}}{%
\begin{tikzpicture}[baseline=(current bounding box.center)]
    \def\dx{1.5}
    \def\mvpad{0.22}
    \def\mxpad{0.22}
    \foreach \i in {1,...,#1}{
        \pgfmathsetmacro{\xx}{(\i-1)*\dx}
        \coordinate (v\i)   at (\xx,0);
        \coordinate (MLv\i) at ($(v\i)+(-\mvpad,0)$);
        \coordinate (MRv\i) at ($(v\i)+(\mvpad,0)$);
    }
    \pgfmathtruncatemacro{\NmOne}{#1-1}
    \ifnum\NmOne>0
        \foreach \i in {1,...,\NmOne}{
            \pgfmathsetmacro{\xx}{(\i-0.5)*\dx}
            \coordinate (x\i)   at (\xx,0);
            \coordinate (MLx\i) at ($(x\i)+(-\mxpad,0)$);
            \coordinate (MRx\i) at ($(x\i)+(\mxpad,0)$);
        }
    \fi
    \coordinate (Lall) at ($(v1)+(-0.30,0)$);
    \coordinate (Rall) at ($(v#1)+(0.30,0)$);
    \node[
        draw,
        color=white,
        line width=0.65pt,
        rounded rectangle,
        rounded rectangle arc length=180,
        fit=(Lall)(Rall),
        inner xsep=-1pt,
        inner ysep=0.6em
    ] {};
    \def\largebloblist{#3}
    \ifx\largebloblist\@empty\else
        \foreach \a/\b in {#3}{
            \node[
                marked chain large blob,
                fit=(ML\a)(MR\b)
            ] {};
        }
    \fi
    \def\smallbloblist{#2}
    \ifx\smallbloblist\@empty\else
        \foreach \a in {#2}{
            \node[
                marked chain small blob,
                fit=(ML\a)(MR\a)
            ] {};
        }
    \fi
    \ifnum\NmOne>0
        \foreach \i in {1,...,\NmOne}{
            \pgfmathtruncatemacro{\ip}{\i+1}
            \draw[line width=1.05pt] (v\i) -- (v\ip);
        }
    \fi
    \ifnum\NmOne>0
        \foreach \i in {1,...,\NmOne}{
            \node[inner sep=0pt, outer sep=0pt, scale=#5] at (x\i) {\Cross};
        }
    \fi
    \foreach \i in {1,...,#1}{
        \draw[fill] (v\i) circle (.8mm);
    }
    \def\labellist{#4}
    \ifx\labellist\@empty\else
        \foreach \a/\b/\lab/\ddx/\ddy in {#4}{
            \node[inner sep=1pt] at ($(\a)!0.5!(\b)+(\ddx,\ddy)$) {\lab};
        }
    \fi
\end{tikzpicture}%
}

\NewDocumentCommand{\twosite}{s m m}{%
\raisebox{3pt}{%
\begin{tikzpicture}[baseline=(current bounding box.center)]
  \node[
    draw,
    color=white,
    rounded rectangle,
    minimum height=0.9em,
    minimum width=3.9em,
    rounded rectangle arc length=180
  ] at (0.45,0) {};

  \IfStrEq{#2}{lp}{
    \IfBooleanTF{#1}
      {\draw[color=#3,fill=#3!20] (0,0) circle (1.75mm);}
      {\draw[color=#3] (0,0) circle (1.75mm);}
  }{}
  \IfStrEq{#2}{rp}{
    \IfBooleanTF{#1}
      {\draw[color=#3,fill=#3!20] (0.9,0) circle (1.75mm);}
      {\draw[color=#3] (0.9,0) circle (1.75mm);}
  }{}

  \IfStrEq{#2}{mid}{
    \IfBooleanTF{#1}
      {\draw[color=#3,fill=#3!20] (0.45,0) circle (1.75mm);}
      {\draw[color=#3] (0.45,0) circle (1.75mm);}
  }{}

  \IfStrEq{#2}{lt}{
    \IfBooleanTF{#1}
      {\node[draw,color=#3,fill=#3!20,rounded rectangle,
             minimum height=0.9em,minimum width=2.75em,
             rounded rectangle arc length=180] at (0.225,0) {};}
      {\node[draw,color=#3,rounded rectangle,
             minimum height=0.9em,minimum width=2.75em,
             rounded rectangle arc length=180] at (0.225,0) {};}
  }{}

  \IfStrEq{#2}{rt}{
    \IfBooleanTF{#1}
      {\node[draw,color=#3,fill=#3!20,rounded rectangle,
             minimum height=0.9em,minimum width=2.75em,
             rounded rectangle arc length=180] at (0.675,0) {};}
      {\node[draw,color=#3,rounded rectangle,
             minimum height=0.9em,minimum width=2.75em,
             rounded rectangle arc length=180] at (0.675,0) {};}
  }{}

  \IfStrEq{#2}{full}{
    \IfBooleanTF{#1}
      {\node[draw,color=#3,fill=#3!20,rounded rectangle,
             minimum height=0.9em,minimum width=3.9em,
             rounded rectangle arc length=180] at (0.45,0) {};}
      {\node[draw,color=#3,rounded rectangle,
             minimum height=0.9em,minimum width=3.9em,
             rounded rectangle arc length=180] at (0.45,0) {};}
  }{}

  \draw[fill] (0,0) circle (.5mm);
  \draw[fill] (0.9,0) circle (.5mm);
  \draw[thick] (0,0) -- (0.9,0);
  \node at (0.45,0) {\Cross};

\end{tikzpicture}}%
}

\newcommand{\loopdraw}[1]{
    \begin{tikzpicture}
        \tikzset{every picture/.style={scale=1}}
\def\n{#1}
\def\R{1}
\def\dr{0.2}
\def\gap{0}
  \draw[line width=1.05pt] (0,0) circle (\R);

  \foreach \k in {1,...,\n} {
    \pgfmathsetmacro{\angle}{90 - (\k-1)*(360/\n)}
    \fill ({\R*cos(\angle)}, {\R*sin(\angle)}) circle (.8mm);
  }
  \foreach \i in {1,...,\n} {
    \pgfmathsetmacro{\labelangle}{-90- (\i - 0.5)*(360/\n)}
    \pgfmathsetmacro{\lx}{1.4*cos(\labelangle)}
    \pgfmathsetmacro{\ly}{1.4*sin(\labelangle)}
    \node at (\lx, \ly) {$\i,\overline{\i}$};
  }
 
\end{tikzpicture}
}
\newcommand{\loopdrawtube}[3]{
    \begin{tikzpicture}
        \tikzset{every picture/.style={scale=1}}
\def\n{#3}
\def\R{1}
\def\dr{0.2}
\def\gap{0}

\newcommand{\drawtube}[2]{%
  \pgfmathsetmacro{\angleA}{90 - ((#1-1)-1)*(360/\n)}
  \pgfmathsetmacro{\angleB}{90 - ((#2-1)-1)*(360/\n)}
  \pgfmathsetmacro{\capA}{\angleA + \gap}
  \pgfmathsetmacro{\capB}{\angleB - \gap}
  \pgfmathsetmacro{\Rout}{\R + \dr}
  \pgfmathsetmacro{\Rin}{\R - \dr}
  \draw[line width=0.65pt,color=gray]
    ({\Rout*cos(\capA)}, {\Rout*sin(\capA)})
    arc[start angle=\capA,       end angle=\capB,       radius=\Rout]
    arc[start angle=\capB,       end angle=\capB - 180, radius=\dr  ]
    arc[start angle=\capB,       end angle=\capA,       radius=\Rin ]
    arc[start angle=\capA + 180, end angle=\capA,       radius=\dr  ]
    -- cycle;
}
 
\drawtube{#1-1}{#2-1}

  \draw[line width=1.05pt] (0,0) circle (\R);
 
  \foreach \k in {1,...,\n} {
    \pgfmathsetmacro{\angle}{90 - (\k-1)*(360/\n)}
    \fill ({\R*cos(\angle)}, {\R*sin(\angle)}) circle (.8mm);
  }

\end{tikzpicture}
}

\newcommand{\loopdrawmarked}[3][1]{%
\begin{tikzpicture}
  \tikzset{every picture/.style={scale=1}}

  \def\n{#2}
  \def\R{1}
  \def\crossscale{#1}

  \draw[line width=1.05pt] (0,0) circle (\R);

  \foreach \k in {1,...,\n} {
    \pgfmathsetmacro{\angle}{90 - (\k-1)*(360/\n)}
    \fill ({\R*cos(\angle)}, {\R*sin(\angle)}) circle (.8mm);
  }

  \foreach \e in {#3} {
    \pgfmathsetmacro{\edgeangle}{90 - (\e - 0.5)*(360/\n)}
    \node[
      inner sep=0pt,
      outer sep=0pt,
      scale=\crossscale
    ] at ({\R*cos(\edgeangle)}, {\R*sin(\edgeangle)}) {\Cross};
  }
\end{tikzpicture}%
}

\usepackage{selinput}

\usepackage{bm}
\usepackage{float}
\usepackage{geometry}
\usepackage{yfonts}
\usepackage{subcaption}
\usepackage{sidecap}
\usepackage{longtable}
\usepackage{anyfontsize}
\usepackage{dsfont}
\usepackage{relsize}
\usepackage{tcolorbox}

\usepackage{xcolor}
\usepackage{shuffle}
\usepackage{slashed}
\usepackage{simpler-wick}

\NewDocumentCommand{\colornucleus}{omme{_^}}{%
  \begingroup\colorlet{currcolor}{.}%
  \IfValueTF{#1}
   {\textcolor[#1]{#2}}
   {\textcolor{#2}}
    {%
     #3
     \IfValueT{#4}{_{\textcolor{currcolor}{#4}}}%
     \IfValueT{#5}{^{\textcolor{currcolor}{#5}}}%
    }%
  \endgroup
}

\definecolor{blue3}{RGB}{31, 119, 180}
\definecolor{red3}{RGB}{	214, 39, 40}
\definecolor{orange3}{RGB}{255, 127, 14}
\definecolor{green3}{RGB}{44, 160, 44}

\tikzset{cross/.style={cross out, draw=black, minimum size=2*(#1-\pgflinewidth), inner sep=0pt, outer sep=0pt},
cross/.default={3pt}}

\usepackage{array}
\newcolumntype{L}[1]{>{\raggedright\let\newline\\\arraybackslash\hspace{0pt}}m{#1}}
\newcolumntype{C}[1]{>{\centering\let\newline\\\arraybackslash\hspace{0pt}}m{#1}}
\newcolumntype{R}[1]{>{\raggedleft\let\newline\\\arraybackslash\hspace{0pt}}m{#1}}

\usepackage[framemethod=default]{mdframed}
\newmdenv[skipabove=7pt,
skipbelow=7pt,
rightline=false,
leftline=false,
topline=false,
bottomline=false,
backgroundcolor=gray!10,
linecolor=gray,
innerleftmargin=5pt,
innerrightmargin=5pt,
innertopmargin=5pt,
innerbottommargin=5pt,
leftmargin=0cm,
rightmargin=0cm,
linewidth=4pt]{eBox}
\newmdenv[skipabove=7pt,
skipbelow=7pt,
rightline=false,
leftline=false,
topline=false,
bottomline=false,
backgroundcolor=gray!10,
linecolor=gray,
innerleftmargin=5pt,
innerrightmargin=5pt,
innertopmargin=-5pt,
innerbottommargin=5pt,
leftmargin=0cm,
rightmargin=0cm,
linewidth=4pt]{eBox2}

\usepackage[framemethod=default]{mdframed}
\newmdenv[skipabove=7pt,
skipbelow=7pt,
rightline=true,
leftline=true,
topline=true,
bottomline=true,
backgroundcolor=gray!05,
linecolor=black!70,
innerleftmargin=5pt,
innerrightmargin=5pt,
innertopmargin=5pt,
innerbottommargin=5pt,
leftmargin=0cm,
rightmargin=0cm,
linewidth=0.5pt]{eBox3}

\definecolor{Red}{RGB}{214, 39, 40}
\definecolor{Blue}{RGB} {31, 119, 180}
\definecolor{Orange}{RGB}{255, 153, 51}
\definecolor{Purple}{RGB}{178, 102, 255}
\definecolor{Green}{RGB}{44, 160, 44}

\definecolor{Red4}{RGB}{234, 29, 30}

\definecolor{vio}{RGB}{19, 130, 164}
\definecolor{vioo}{RGB}{89, 2, 155}
\newcommand{\Comment}[1]{{}}
\definecolor{darkblue}{rgb}{0.15,0.35,0.55}
\definecolor{reddish}{rgb}{0.65, 0.2, 0.2}
\definecolor{darkgreen}{RGB}{50,150,0}
\definecolor{greyish}{rgb}{.90,.90,.90}
\definecolor{greyish2}{rgb}{.96,.96,.96}
\definecolor{greyish3}{rgb}{.37,.37,.37}
\definecolor{darkblue2}{rgb}{0.3,0.4,0.9}
\definecolor{Blue3}{RGB}{31, 119, 180}
\definecolor{lightgray}{cmyk}{0.1,0.2,0,0.1}
\definecolor{lightgray2}{cmyk}{0.1,0.1,0,0.1}

\definecolor{tubeRed}{HTML}{D55E00}      
\definecolor{tubeBlue}{HTML}{0072B2}     
\definecolor{tubeGreen}{HTML}{009E73}    
\definecolor{tubePurple}{HTML}{CC79A7}   
\definecolor{tubeOrange}{HTML}{E69F00}   
\definecolor{tubeGray}{HTML}{555555}

\usepackage[linktocpage=true]{hyperref}
\hypersetup{
colorlinks=true,
citecolor=darkblue,
linkcolor=reddish,
urlcolor=darkblue,
pdfauthor={},
pdftitle={},
pdfsubject={}
}

\newcommand{\bee}{\begin{equation*}}
\newcommand{\eee}{\end{equation*}}
\def\ba{\begin{equation}\begin{aligned}}
\def\ea{\end{aligned}\end{equation}}

\tikzset{cross/.style={cross out, draw=black, minimum size=2*(#1-\pgflinewidth), inner sep=0pt, outer sep=0pt},
cross/.default={3pt}}
\definecolor{purple3}{RGB}{148,0,211}

\definecolor{red4}{RGB}{	231, 86, 147}
\definecolor{blue4}{RGB}{108, 193, 212}

\usepackage{colortbl}
\definecolor{lightgreen}{cmyk}{0.2, 0, 0.2, 0.2}
\definecolor{lightgray2}{cmyk}{0.1,0.1,0,0.1}
\definecolor{Red2}{RGB}{214, 39, 40}
\definecolor{Blue2}{RGB} {31, 119, 180}
\definecolor{Orange2}{RGB}{255, 127, 14}
\definecolor{Green2}{RGB}{44, 160, 44}

\makeatletter
\newlength{\apb@width}
\newcommand{\autoparbox}[2][c]{\settowidth{\apb@width}{#2}\parbox[#1]{\apb@width}{#2}}

\makeatother


\def\beq{\begin{equation}}
\def\eeq{\end{equation}}
\def\be{\begin{equation}}
\def\ee{\end{equation}}

\def\k{\vec k}

\def\a{{\hat \alpha}}
\def\b{{\hat \beta}}

\newcommand{\rd}{{\rm d}}

\DeclareMathOperator{\E}{e}

\def\vecx{\vec x}
\def\veck{\vec k}

\allowdisplaybreaks[1]
\begin{document}

\newgeometry{top=2cm, bottom=2cm, left=2cm, right=2cm}

\begin{titlepage}
\setcounter{page}{1} \baselineskip=15.5pt 
\thispagestyle{empty}

\begin{center}
{\fontsize{21}{18} \bf Cosmology, Cluster Algebras, and $\boldsymbol{u}$}
\end{center}

\vskip 20pt
		\begin{center}
			\noindent
			{\fontsize{14}{18}\selectfont 
				Daniel Glazer and Austin Joyce}
		\end{center}

		\begin{center}

			\vskip 8pt
			\textit{
				Kavli Institute for Cosmological Physics, Department of Astronomy and Astrophysics,\\
				The University of Chicago, Chicago, IL 60637, USA}

		\end{center}

\vspace{0.4cm}
\begin{center}{\bf Abstract}
\end{center}
\noindent
We explore the cluster algebra structures present in the wavefunction and correlators of conformally coupled scalar fields in de Sitter space.
To connect cosmological kinematics to cluster variables, we utilize the so-called $u$-variables that appear on both sides.
Since $u$-variables parameterize a rigid space, it is natural to identify the ones that appear from cosmological graphs and those that arise from cluster algebras.
This provides a mapping between kinematic variables and cluster variables.
For $n$-site chain Feynman graphs, this gives a coordinate-invariant relation to $A_{2n-2}$-type cluster algebras, while in the case of $n$-site cycle graph momentum integrands, the relevant cluster algebra is $B_{2n-1}$, as found previously.
Since kinematic variables are mapped to nonlinear combinations of cluster variables, cluster compatibility imposes strong conditions on possible symbol entries.
We use this notion of compatibility to bootstrap the symbol. In the chain case, both the cosmological wavefunction and products of lower-point functions satisfy cluster compatibility, with the correlator a particular combination, so that the wavefunction and correlator share the same cluster structure. 
In the cycle case, however, the wavefunction appears to be uniquely selected by the cluster structure, and the correlator does not display the same compatibility as the wavefunction.

\end{titlepage}
\restoregeometry

\newpage
\setcounter{tocdepth}{2}
\setcounter{page}{2}

\linespread{1.2}
\tableofcontents
\linespread{1.1}

\newpage

\section{Introduction}

We learn about cosmology by measuring correlations in the distribution of structures on the largest scales.
These cosmological correlators are interesting functions of location on the sky, and their detailed form is shaped by the cosmological evolution that produces them.
We might wonder whether {\it any}  function can be produced by some cosmic history, or if some patterns of correlations are simply inconsistent. In recent years, substantial progress has been made toward answering this question~\cite{Baumann:2022jpr,Benincasa:2022gtd,Lee:2024sks,corrlectures}. Nevertheless, mapping the space of functions that can appear in cosmology remains a formidable task. To get further insight, it is useful to consider a simplified setting.
In particular, the correlation functions of conformally coupled scalar fields have enough structure that we can characterize the function space~\cite{Arkani-Hamed:2017fdk,Hillman:2019wgh,Arkani-Hamed:2023kig}.

\vskip4pt
In de Sitter space, conformally coupled correlators (for $\phi^3$ interactions) are (generalized) polylogarithmic functions~\cite{Arkani-Hamed:2017fdk,Hillman:2019wgh,Arkani-Hamed:2023kig}. Many of the properties of these functions are captured  by their symbol~\cite{Goncharov:2010jf}, which we can write schematically 
\be
{\cal S}( F_{(n)}) = R_1\otimes R_2\otimes \cdots \otimes R_n +\cdots\,,
\label{eq:symbolintro}
\ee
where the entries $R_1, R_2,\cdots$ encode the singularities of the function $F_{(n)}$. The precise sequence of these entries determines the analytic structure of the function, including its discontinuities and derivatives. As such, the challenge of determining the space of cosmological correlators in this context can be phrased as characterizing the possible sequences of symbol entries.
In the related context of particle scattering amplitudes, this perspective has proven to be very powerful. Causality places strong constraints on the possible nonzero sequences of discontinuities via Steinmann relations~\cite{Steinmann1960a,Steinmann1960b}, which in some cases essentially uniquely determine amplitudes~\cite{Caron-Huot:2016owq,Caron-Huot:2019bsq,Caron-Huot:2019vjl}. 
Surprisingly, this problem of understanding singularities is related to the mathematics of cluster algebras~\cite{FominZelevinsky2002ClusterI,FominZelevinsky2003ClusterII,BerensteinFominZelevinsky2005ClusterIII,FominZelevinsky2007ClusterIV} which control both the possible singularities that appear~\cite{Golden:2013xva,Drummond:2019cxm,Arkani-Hamed:2019rds}, and their compatibilities~\cite{Drummond:2017ssj,Drummond:2018dfd,Caron-Huot:2018dsv}. In particular, the notion of cluster adjacency has proven very powerful to constrain the form of the symbol of amplitudes for various processes~\cite{Dixon:2016nkn,Drummond:2014ffa,Dixon:2020cnr,Chicherin:2020umh,Caron-Huot:2020bkp,Dixon:2022rse}.

\vskip4pt
In this paper we study some of the cluster algebra properties of the correlation functions of 
conformally coupled scalars in de Sitter space. Related aspects of these structures have been of substantial recent interest~\cite{Mazloumi:2025pmx,Capuano:2025myy,Paranjape:2026htn,Capuano:2026pgq,Ferro:2026oph}.
Similar to amplitudes in maximally supersymmetric theories, correlation functions and wavefunction coefficients in this theory are subject to consistency conditions that restrict their symbols, including extended Steinmann relations~\cite{Benincasa:2020aoj,Benincasa:2021qcb}.
Beyond this, the singularity alphabets of these functions have simple combinatorial interpretations~\cite{Arkani-Hamed:2017fdk,Arkani-Hamed:2023kig,Hillman:2023vas}, which we utilize in order to make a connection to cluster algebras.
Concretely, we connect cosmological observables and cluster algebras via so-called $u$-variables~\cite{Koba:1969rw,Koba:1969kh,brown2006multiplezetavaluesperiods} as an intermediary.
These variables first appeared in the context of string amplitudes, and have recently found a variety of wide-ranging applications~\cite{Arkani-Hamed:2019mrd,Arkani-Hamed:2019vag,Arkani-Hamed:2023lbd,Arkani-Hamed:2023swr}.
Both cluster algebras~\cite{Arkani-Hamed:2019plo,Arkani-Hamed:2020tuz,He:2020onr,Arkani-Hamed:2025zuf} and cosmological kinematic variables~\cite{Hillman:2023vas} produce $u$-variables in a natural way.
For our purposes, the interesting feature of $u$-variables is that they satisfy nonlinear equations that encode a notion of compatibility in an invariant manner~\cite{Arkani-Hamed:2019plo}. The fact that $u$-variables coordinatize a ``binary" geometry in a rigid way provides a natural map between cosmological kinematics and cluster variables, obtained by simply equating them on the two sides. (A somewhat similar application of $u$-variables in the scattering context can be found in~\cite{He:2021esx}.)

\vskip4pt
Utilizing this mapping, we relate the kinematic variables of cosmological wavefunctions to cluster variables in two cases. For $n$-site chain Feynman graphs, kinematics are related to the cluster variables of an $A_{2n-2}$ cluster algebra (as discovered in~\cite{Mazloumi:2025pmx,Capuano:2025myy}), while $n$-site cycle graphs are related to $B_{2n-1}$ cluster variables (as found by~\cite{Paranjape:2026htn}).
These identifications have appeared previously in the literature, but the precise realization of kinematic variables in terms of cluster variables differs in detail---in particular, all kinematic singularities have nontrivial compatibilities.
The map to cluster variables imbues the kinematic variables on which the symbol~\eqref{eq:symbolintro} depends with a notion of compatibility. Kinematic variables are separately related to graph tubings~\cite{Arkani-Hamed:2023kig,Hillman:2023vas}, which also have a natural notion of compatibility. Importantly, cluster compatibility is different from, and more restrictive than, the compatibility of tubes, essentially because the mapping between tube variables and cluster variables is nonlinear.

\vskip4pt 
It is straightforward to check in explicit examples that the cosmological wavefunction respects cluster compatibility, in the sense that in a given symbol word all cluster variables that appear are mutually compatible.
It is then natural to wonder to what extent this determines the form of the wavefunction.
We investigate this question by implementing a cluster symbol bootstrap. In addition to complete cluster compatibility, we impose simple mathematical conditions like the integrability of the symbol, as well as restrictions on the physical singularities, which manifest as a first entry condition (along with a condition that energy singularities do not repeat). 
Importantly, the details of the bootstrap depend on the precise mapping to cluster variables.
We carry out this procedure up to five sites in the chain case and three sites in the cycle case.
The results have several interesting features. 
We find that cluster compatibility with respect to $B_{2n-1}$ in the form that we impose is sufficiently constraining to determine the form of the wavefunction in the cycle case, while in the $A_{2n-2}$ (chain) case, there are multiple solutions consistent with cluster compatibility (for $n\geq 4$). These solutions correspond to the cosmological wavefunction, along with products of lower-point functions of the same transcendentality. 
One particular combination is the wavefunction, while another is the corresponding cosmological correlator. Thus, we find in the chain case that both the wavefunction and cosmological correlator share the same underlying cluster structure. We find a simple characterization of the wavefunction in this language, which is that it has a distinguishing pattern of repeating cluster variables.

\vskip4pt
Cluster algebra structures in the wavefunction have been of great interest recently~\cite{Mazloumi:2025pmx,Capuano:2025myy,Paranjape:2026htn,Capuano:2026pgq,Ferro:2026oph}. It is perhaps useful to contextualize our findings relative to these interesting works. A connection between cluster algebras and cosmological wavefunctions was first noted in~\cite{Mazloumi:2025pmx,Capuano:2025myy}.
In~\cite{Mazloumi:2025pmx,Capuano:2025myy} (as well as~\cite{Paranjape:2026htn}) they similarly relate kinematic variables of $n$-site chain graphs to cluster variables of $A_{2n-2}$-type. The material difference in the map we consider is that we set variables $a_{i\,i+1}$ (corresponding to edges of a polygon) to $1$, so that all kinematic combinations are expressed in terms of variables with nontrivial compatibilities. In~\cite{Mazloumi:2025pmx,Capuano:2025myy,Paranjape:2026htn} some kinematic variables are mapped to edges, and so are  compatible with all other variables (effectively becoming frozen variables).
 In~\cite{Capuano:2026pgq}, the kinematic variables of an $n$-site chain were assigned to cluster variables of an $A_{2n-3}$ cluster algebra, which was used, along with a particular ordered single-cluster condition based on a discontinuity constraint, to reconstruct the wavefunction.
In~\cite{Paranjape:2026htn} they also relate cycle graphs to $B_{2n-1}$ and use cluster adjacency to perform a bootstrap for both chains and cycles. Since their cluster mapping differs from ours in detail, the precise conditions imposed to isolate the wavefunction differ. (In particular, they impose an additional soft condition.)
Reference~\cite{Ferro:2026oph} uses the $A_{2n-2}$ cluster structure of chain graphs and its relation to quadrangular polylogs~\cite{rudenko2022,matveiakin2022} to provide remarkably simple all-$n$ cluster-compatible formulas for the wavefunction of chains.
The cluster variables that we employ can be related to those used in~\cite{Mazloumi:2025pmx,Capuano:2025myy,Paranjape:2026htn,Ferro:2026oph} by constructing invariants. This implies that the different cluster variables that have appeared in the $A_{2n-2}$ case can be thought of as specializations of the same underlying structure.
The conceptual novelty of the present work is to provide a natural map between cluster variables and kinematics using $u$-variables which gives all the kinematic variables nontrivial compatibilities, and connects to the binary geometry defined by the $u$-equations. 
This notion of compatibility is sufficiently strong that it only requires minimal additional input to determine the symbol.
Beyond this, we provide a complementary perspective on the cluster bootstrap in the $A$ and $B$-type cases by studying the space of solutions satisfying cluster compatibility. In particular, we explore how and when the in-in correlators display the same cluster structure  as the wavefunction in the relevant cases.

\vskip4pt
It is intriguing that cluster algebras make an appearance in cosmology.
Cluster compatibility is in a sense a generalization of Steinmann relations. Since Steinmann relations themselves are consequences of causality, it is natural to suspect that the cluster structure captures fundamental features of cosmological time evolution.
 In this regard, we expect that better understanding these connections will further unravel the 
 encoding of time in the structures in the universe.

\vspace{-4pt}
\paragraph{Outline:} In Section~\ref{sec:cosmocorr} we introduce the objects that we wish to study---cosmological correlators and the wavefunction---and describe their perturbative computation. We then show how cosmological correlators can be cast as Euler-type integrals, and describe the singularity structure of these integrals and its associated combinatorics. In
Section~\ref{sec: E to A map} we make a connection between the cosmological wavefunction and cluster algebras.
To do this, we first introduce $u$-variables and describe how cosmological graphs naturally produce them. We then describe the appearance of these variables in the cluster algebra context. 
Finally, by equating the $u$-variables from these two manifestations, we obtain explicit mappings between kinematic variables and cluster variables for Feynman graphs consisting of chains and cycles.
In Section~\ref{sec:clusterbootstrap} we utilize cluster compatibility to bootstrap the symbol of the cosmological wavefunction (and cosmological correlators in the chain case). We describe how wavefunctions and correlators are determined by the cluster structure, along with additional physical inputs.
We conclude in
Section~\ref{sec:conclusion}.

\vskip4pt
Several appendices collect technical details and background information. We provide a brief overview of some of the mathematics of cluster algebras in  Appendix~\ref{app:cluster}. In Appendix~\ref{app: Symbology} we provide an introduction to the symbol map and its relation to generalized polylogarithmic functions.
In Appendix~\ref{app: Steinmann} we interpret some aspects of cluster compatibility, and compare its constraining power to other possible physical conditions. In  Appendix~\ref{app: Correlator vs WF} we explain the relation between the wavefunction and in-in correlators, and describe the relation to our cluster symbol bootstrap.

\vspace{-4pt}
\paragraph{Conventions:} Throughout we work in $D=4$ spacetime dimensions, with mostly plus metric signature. We consider the wavefunction and coefficients of a conformally coupled scalar field with a cubic $\phi^3$ interaction. Both Feynman graphs and marked versions of these graphs play an important role. We denote the number of vertices of the Feynman graph by $n$, and the (larger) number of vertices plus markings of the marked graph by $N$. Kinematic variables are denoted by $X,Y,E$, and cluster variables are denoted by $a$, while frozen variables are denoted by $z$.

\newpage
\section{Cosmological Correlations}
\label{sec:cosmocorr}

In cosmological spacetimes, the observable quantities are the statistical correlations between operators at different spacetime points. In the idealized setup of de Sitter space, it is natural to consider equal-time correlators defined on the asymptotic future boundary. In the inflationary context, these boundary correlations are the initial conditions for the evolution of the later universe.
In order to understand structural features of cosmological correlators, it is useful to investigate simplified models deeply.
Here we briefly describe the model of interest and the perturbative computation of its correlation functions.
In the remainder we explore some of the mathematical structure of these correlations.

\subsection{Wavefunctions and Correlators}
\label{sec:WFcorr}

The model that we will consider is a conformally coupled scalar field with cubic self-interactions in four-dimensional de Sitter space.\footnote{More generally, for a conformally coupled scalar field in a  $(d+1)$-dimensional power-law FLRW background $a(\eta) = (\eta/\eta_0)^{-(1+\varepsilon)}$ with a polynomial interaction $\phi^k$,  if the effective coupling $ \lambda_{k}(\eta) \sim \left[a(\eta)\right]^{2+\frac{1}{2}(d-1)(2-k)}$ scales as $\eta^{-1}$ the resulting correlators will be expressible in terms of polylogarithmic functions.} Though this is not of direct phenomenological relevance (it is, however, related to theories of phenomenological interest~\cite{Arkani-Hamed:2015bza,Baumann:2019oyu,Baumann:2020dch}), the lessons learned apply more broadly.
The theory is governed by the action 
\begin{equation}
\label{eq:scalaraction}
    S=\int \rd \eta \rd^{3} x \sqrt{-g}\left(-\frac{1}{2} (\partial \phi)^2-H^2  \phi^{2}-\frac{\lambda}{3!} \phi^{3}\right)\,,
\end{equation}
where the spacetime geometry is de Sitter space
\begin{equation}
\rd s^{2}= a(\eta)^2\left(-\rd \eta^{2}+\rd x^{2}\right)\,,
\end{equation}
which has scale factor $a(\eta)= 1/(-H\eta)$
in terms of conformal time $\eta\in (-\infty,0]$.

\vskip4pt
We can represent the quantum state of the theory at time $\eta_\star$, $\lvert \uppsi\rangle$, in terms of Heisenberg-picture field eigenstates as
\be
\Psi[\varphi,\eta_\star] = \langle \varphi,\eta_\star\rvert \uppsi\rangle\,,
\label{eq:WFdef}
\ee
where $\lvert \varphi,\eta_\star\rangle$ satisfies $\hat\phi(\vecx, \eta_\star)\lvert \varphi,\eta_\star\rangle = \varphi(\vec x)\lvert \varphi,\eta_\star\rangle$, with $\varphi(\vecx)$ a spatial configuration of the field $\phi$. The wavefunctional~\eqref{eq:WFdef} provides a probability distribution on field configurations via the Born rule, which can be used to express correlation functions as
\be
\label{eq:cosmocorrs}
\langle O(\vecx_1)\cdots O(\vecx_n)\rangle =\frac{\displaystyle\int \mathcal{D} \varphi \, O(\vecx_1)\cdots O(\vecx_n) \,\left\lvert \Psi[\varphi]\right\rvert^2}{\displaystyle\int \mathcal{D} \varphi \,\left\lvert \Psi[\varphi]\right\rvert^2}\,.
\ee
Here $O$ represents operators constructed from the fundamental fields $\phi$. In the context of cosmology, the correlation functions~\eqref{eq:cosmocorrs} are typically called in-in correlators when they are computed in a single (in) vacuum state $\lvert 0\rangle_{\rm in}$.

\vskip4pt
The field theory wavefunctional~\eqref{eq:WFdef} contains all of the information about correlation functions in the state $\lvert\uppsi\rangle$,  encoded in a somewhat nontrivial way. In perturbation theory, the relation  between the two objects is conceptually simple. In the following we will be interested mainly in the wavefunctional $\Psi$, but will also explore some interesting features of correlation functions.

\subsubsection*{Perturbation Theory}

As a practical matter, we only really have access to~\eqref{eq:WFdef} via perturbation theory. In this regime it is convenient to parameterize the wavefunctional as
\begin{equation}
\label{eq:WFdef2}
\begin{aligned}
    \Psi[\varphi,\eta_\star] =  \exp\bigg(
    &-\frac{1}{2} \int\frac{\rd^3k_1\rd^3k_2}{{(2\pi)}^{6}}\, \Psi_2(\veck_1,\veck_2)\,\varphi_{\veck_1}\varphi_{\veck_2}\\
    &+\sum_{n=3}^\infty\frac{1}{n!}\int\frac{\rd^3k_1\cdots \rd^3k_n}{(2\pi)^{3n}}\,\Psi_n(\veck_1,\cdots, \veck_n)\,\varphi_{\veck_1}\cdots\varphi_{\veck_n}
\bigg)    
    \,,
    \end{aligned}
\end{equation}
where we have expressed the field profile in momentum space, $\varphi_{\vec k}$. Due to translation invariance, the kernels $\Psi_n$ are proportional to a momentum-conserving delta function
\be
\label{eq:WFcoeff}
\Psi_n(\veck_1,\cdots, \veck_n) = (2\pi)^{3}\delta(\veck_1+\cdots+\veck_n)\,\psi_n(\veck_1,\cdots, \veck_n)\,.
\ee
It is often useful to consider the kernels $\psi_n$ which have this delta function removed. We will call the kernels $\Psi_n$ (or $\psi_n$) {\it wavefunction coefficients}. These coefficients can be simply related to correlation functions of the fields $\varphi$ using~\eqref{eq:cosmocorrs}, which we discuss in greater detail in Appendix~\ref{app: Correlator vs WF}. 

\vskip4pt
The wavefunction coefficients that parameterize the vacuum state of an interacting field theory can be computed in perturbation theory by utilizing the path integral representation
\beq
\label{eq:wavefunctionpathintegral}
\Psi[\varphi] \ =  \hspace{-0.4cm} \int\limits_{\substack{\phi(\eta_\star) \,=\,\varphi\\[1pt] \hspace{-0.4cm}\phi(-\infty_\epsilon)\,=\,0}} 
\hspace{-0.5cm} \raisebox{-.05cm}{ ${\cal D} \phi~ e^{iS[\phi]}$} \raisebox{-.125cm}{ .}
\ee
Here the boundary conditions for the path integral are chosen so that the field vanishes at (complex-valued) early times $\eta \to -\infty_\epsilon\equiv -\infty(1-i \epsilon)$ and has profile $\varphi(\vecx)$ at time $\eta_\star$.

\begin{figure}[t]
\centering
\begin{tikzpicture}[
  scale=.8,
  line cap=round,
  line join=round,
  >=Stealth
]
  \coordinate (P1) at (0,0);
  \coordinate (P2) at (1.8,2.8);
  \coordinate (P3) at (10.8,2.8);
  \coordinate (P4) at (9.1,0);

  \coordinate (O1) at (1.75,.65);
  \coordinate (O2) at (3.2,2.1);
  \coordinate (O3) at (7.4,2.15);
  \coordinate (O4) at (8.3,1.0);

  \coordinate (L) at (4.0,-1.5);
  \coordinate (R) at (7.4,-1.1);
  \coordinate (M) at ($(L)!0.5!(R)$);

  \path[name path=frontedge] (P1) -- (P4);
  \path[name path=lineO1] (O1) -- (L);
  \path[name path=lineO2] (O2) -- (L);
  \path[name path=lineO3] (O3) -- (R);
  \path[name path=lineO4] (O4) -- (R);
  \path[name intersections={of=frontedge and lineO1, by=E1}];
  \path[name intersections={of=frontedge and lineO2, by=E2}];
  \path[name intersections={of=frontedge and lineO3, by=E3}];
  \path[name intersections={of=frontedge and lineO4, by=E4}];

  \draw[gray!45, line width=1.0pt] (O1) -- (E1);
  \draw[gray!45, line width=1.0pt] (O2) -- (E2);
  \draw[gray!45, line width=1.0pt] (O3) -- (E3);
  \draw[tubeRed!45, line width=1.0pt] (O4) -- (E4);

  \draw[black!75, line width=1.1pt] (E1) -- (L);
  \draw[black!75, line width=1.1pt] (E2) -- (L);
  \draw[black!75, line width=1.1pt] (E3) -- (R);
  \draw[tubeRed!75, line width=1.1pt] (E4) -- (R);
  
  \draw[tubeBlue!80, line width=1.6pt]
    (L) .. controls (5.1,-2.45) and (6.3,-2.55) .. (R);

  \draw[black!50, thick] (P1) -- (P2) -- (P3) -- (P4) -- cycle;

  \draw[black, line width=1.2pt] (O1) -- (O2) -- (O3) -- (O4) -- (O1);

  \fill (O1) circle (2pt);
  \fill (O2) circle (2pt);
  \fill (O3) circle (2pt);
  \fill (O4) circle (2pt);

  \node[left] at (1.75,-1.8)
    {$\displaystyle\color{tubeGreen}{\int_{-\infty}^{0}{\rd\eta}\, \frac{i\lambda}{H\eta}}$};
  \draw[->, black!35, line width=.8pt]
    (1.75,-1.8) -- ($(L)+(-.2,.0)$);

  \node[right] at (9.4,-1.4)
    {$\displaystyle\color{tubeRed}{K(k\eta)}$};
  \draw[->, black!35, line width=.8pt]
    (9.4,-1.3) -- (7.9,-.3);

  \node[below] at (5.25,-3.5)
    {$\displaystyle \color{tubeBlue}{G_{\rm B}(Y;\eta_1,\eta_2)}$};
  \draw[->, black!35, line width=.8pt]
    (5.25,-3.5) -- ($(M)+(0,-1)$);
  \fill[tubeGreen] (L) circle (2.75pt);
  
\end{tikzpicture}
\caption{\small Depiction of the Feynman rules for the computation of wavefunction coefficients. External lines are anchored on the future boundary of de Sitter space at $\eta = 0$. Lines connecting to this boundary are assigned the bulk-to-boundary propagator $K$. Internal lines are assigned the bulk-to-bulk propagator $G_{\rm B}$. Each vertex is assigned a vertex factor and integrated over all times.}
\label{fig:feyn}
\end{figure}
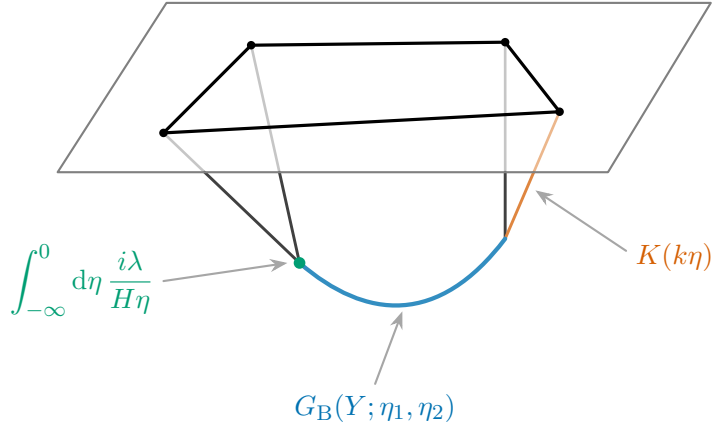

\vskip4pt
The path integral~\eqref{eq:wavefunctionpathintegral} can be approximated by a Feynman diagram expansion. In order to simplify the computation, it is convenient to rescale the fields and metric in~\eqref{eq:scalaraction} as $\phi \mapsto a(\eta)\phi$ and $g_{\mu\nu}\mapsto a(\eta)^{-2} g_{\mu\nu}$, respectively. After this Weyl transformation, the action is that of a massless scalar field in flat space but with a time-dependent coupling
\be
    \label{eq: Flat space action}
    S =\int \rd \eta \rd^{3} x\left(-\frac{1}{2}(\partial \phi)^{2}+\frac{\lambda}{3!}\frac{1}{H\eta} \phi^{3}\right)\,.
\ee
Since de Sitter space has a late-time boundary, it is
natural to take the limit $\eta_\star \to 0$, so that we are computing the wavefunction/correlation functions on this boundary.
\vskip4pt
\paragraph{Feynman Rules}~\\
The benefit of casting~\eqref{eq: Flat space action} as a flat-space field theory is that it simplifies the Feynman rules for computation of wavefunction coefficients $\psi_n$. To calculate $\psi_n$, one draws all possible Feynman diagrams with $n$ lines ending on the $\eta = \eta_\star$ surface, and assigns one of two propagators to the lines. External lines (those that connect to $\eta = \eta_\star$) correspond to bulk-to-boundary propagators
\be
K(k\eta) = e^{ik\eta}\,,
\ee
while purely internal lines correspond to bulk-to-bulk propagators
\be
G_{\mathrm{B}}\left(Y;\eta,\eta'\right)  = \frac{1}{2Y} \left(e^{iY(\eta'-\eta)} \theta(\eta-\eta')+e^{iY(\eta-\eta')} \theta(\eta'-\eta)-e^{iY(\eta+\eta')}\right)\,.
\ee
To each vertex we assign a (time-dependent) vertex factor $V_a = i\lambda/(H\eta_a)$, and then integrate over all the interaction times from $-\infty \leq \eta_a<0$. In addition, if there are loops, we integrate over undetermined loop momenta. These perturbative building blocks are summarized in Figure~\ref{fig:feyn}.

\subsection{Energy Integrals and Functions}
\label{sec: Space of Functions}

Implementing the Feynman rules, we find that wavefunction coefficients are computed by time integrals of the schematic form
\begin{equation}
    \label{eq: WF integral}
    \psi_{\mathcal{G}}(X_v,Y_e)= i^n\int_{-\infty}^0 \frac{\rd\eta_1}{\eta_1} \cdots \frac{\rd\eta_n}{\eta_n}  e^{i X_{1} \eta_{1}} \cdots  e^{i X_{n} \eta_{n}} \prod_{\rm edges} G_{\rm B}\left(Y_{e}, \eta_{a}, \eta_{b}\right)\,,
\end{equation}
where ${\cal G}$ is the underlying Feynman graph, which we have assumed to have $n$ vertices, and where the product runs over all the edges in the graph, which connect vertices $a$ and $b$.
In this expression, we have suppressed a factor of $(\lambda/H)^n$  (as well as the loop-momentum integrals over any undetermined momenta---all the loop expressions in this paper are momentum {\it integrands}). The variables $X_v$ denote the sum of $k$'s of lines connected from that vertex to the boundary at $\eta=0$, and $Y_e$ are the energies of the edges that are present. The variables $X_v$ and $Y_e$ on the left-hand side stand abstractly for the collection of $X$ and $Y$ variables appearing on the right. Note the wavefunction coefficient depends only on the external energies through the $X$ variables.

\vskip4pt
The computation of cosmological correlations in this model reduces to the challenge of doing integrals of the form~\eqref{eq: WF integral}. Fundamentally, this is difficult because the $\rd\eta/\eta$ integration measures introduce logarithmic singularities. We can get some insight into the types of functions that appear by changing integration variables~\cite{Arkani-Hamed:2017fdk,Arkani-Hamed:2023kig}. Notice that $\eta^{-1}$ has the Fourier representation
\be
\frac{1}{\eta} = -i\int_0^\infty\rd x\,e^{ix\eta}\,,
\label{eq:fouriervertex}
\ee
which expresses the time-dependent vertex factors as plane waves, at the expense of additional integrations over energy. In~\eqref{eq: WF integral}, the additional plane wave factors from~\eqref{eq:fouriervertex} serve to shift the vertex energies $X_v \mapsto X_v+x_v$. The benefit of this reorganization is that the time integral now takes precisely the form of the {\it flat-space} wavefunction in a theory with $\phi^3$ interactions, but with these shifted values of energy. We can therefore write the de Sitter wavefunction as an energy integral of the flat-space wavefunction 
\be
 \psi_{\mathcal{G}}(X_v,Y_e) = (-i)^n\int_0^\infty \rd x_1\cdots \rd x_n\, \psi^{\rm flat}_{\mathcal{G}}(X_v+x_v,Y_e)\,.
 \label{eq:dsWFint}
\ee
From this perspective, the time-dependent background is captured by integrating over all possible vertex energies involved in a process, reflecting the lack of energy conservation in cosmology.

\subsubsection*{Combinatorics of the flat-space wavefunction}

Casting the de Sitter wavefunction in the form~\eqref{eq:dsWFint} factorizes the computation into two steps. We first need to obtain the flat-space wavefunction, which we then integrate over energy. Fortunately, the flat-space wavefunction in this theory is extremely simple. The wavefunction coefficients that come from performing the time integrals are always rational functions. Further, they can be obtained by a simple combinatorial prescription~\cite{Arkani-Hamed:2017fdk}.

\vskip4pt
The wavefunction only depends on external energies via the sums entering each vertex $X_v$. It is therefore natural to suppress external lines in Feynman graphs and consider a truncated graph labeled by the kinematic information related to bulk exchanges. For example, at five points,
\be
\raisebox{-29pt}{
\begin{tikzpicture}[line width=1.pt, scale=2]

  \draw[fill=black] (0,0) -- (0.75,0);
  \draw[fill=black] (0.75,0) -- (1.5,0);

  \draw[lightgray2, line width=1.pt] (0,0) -- (-0.25,0.55);
  \draw[lightgray2, line width=1.pt] (0,0) -- (0.25,0.55);

  \draw[lightgray2, line width=1.pt] (0.75,0) -- (0.75,0.55);

  \draw[lightgray2, line width=1.pt] (1.5,0) -- (1.25,0.55);
  \draw[lightgray2, line width=1.pt] (1.5,0) -- (1.75,0.55);

  \draw[lightgray2, line width=2.5pt] (-0.45,0.55) -- (1.95,0.55);

  \draw[fill=black] (0,0) circle (.03cm);
  \draw[fill=black] (0.75,0) circle (.03cm);
  \draw[fill=black] (1.5,0) circle (.03cm);

  \node[scale=1] at (0,-.15) {\scriptsize $X_1$};
  \node[scale=1] at (0.75,-.15) {\scriptsize $X_2$};
  \node[scale=1] at (1.5,-.15) {\scriptsize $X_3$};

  \node[scale=1] at (0.375,.12) {\scriptsize $Y_{12}$};
  \node[scale=1] at (1.125,.12) {\scriptsize $Y_{23}$};
\end{tikzpicture}
}
~
\raisebox{-14pt}{
\begin{tikzpicture}
  \draw[-stealth, gray, line width=2pt] (0,0) -- (1,0);
\end{tikzpicture}
}
~~
\raisebox{-29pt}{
\begin{tikzpicture}[line width=1.pt, scale=2]

  \draw[fill=black] (0,0) -- (0.75,0);
  \draw[fill=black] (0.75,0) -- (1.5,0);

  \draw[fill=black] (0,0) circle (.03cm);
  \draw[fill=black] (0.75,0) circle (.03cm);
  \draw[fill=black] (1.5,0) circle (.03cm);

  \node[scale=1] at (0,-.15) {\scriptsize $X_1$};
  \node[scale=1] at (0.75,-.15) {\scriptsize $X_2$};
  \node[scale=1] at (1.5,-.15) {\scriptsize $X_3$};

  \node[scale=1] at (0.375,.12) {\scriptsize $Y_{12}$};
  \node[scale=1] at (1.125,.12) {\scriptsize $Y_{23}$};
\end{tikzpicture}
}
\label{eq:truncated3site}
\ee

Given a truncated Feynman graph, we can obtain the corresponding wavefunction coefficient by considering the collection of ways to enclose the vertices of the graph in a maximal number of non-overlapping circles.\footnote{Each possible circling will always include circles around each individual node and the entire graph. The additional circles present will correspond to maximal graph tubings~\cite{Arkani-Hamed:2024jbp}.}
For a single graph, there will be several inequivalent ways to do this. To each possibility, we can assign a rational function. We first assign a linear factor to each tube (circle), which is just the sum of $X$ variables corresponding to vertices enclosed by the tube plus $Y$ variables associated to lines that pierce the circle. Then, to a given maximal circling we associate the rational function obtained by dividing one by the product of all the linear factors corresponding to its constituent circles. The flat-space wavefunction is obtained by adding up the rational functions corresponding to each maximal circling~\cite{Arkani-Hamed:2017fdk}.
The procedure is most simply understood with an explicit example. The simplest nontrivial case
is provided by the three-site graph~\eqref{eq:truncated3site}, from which we obtain the flat-space wavefunction via the following diagrammatics
\begin{align}
\label{equ:3pt-tubing}
\psi_{(3)}^{\rm flat} &= 
 \raisebox{2pt}{
 \begin{tikzpicture}[baseline=(current bounding box.center)]
\draw[fill=black] (-0.75,0) -- (0.75,0);
\draw[fill=black] (0,0) circle (.5mm);
\draw[fill=black] (-0.75,0) circle (.5mm);
\draw[fill=black] (0.75,0) circle (.5mm);
\draw[color=tubeRed, line width=0.9pt] (-0.75,0) ellipse (.18cm and .18cm);
\draw[color=tubeBlue, line width=0.9pt] (0.75,0) ellipse (.18cm and .18cm);
\draw[color=tubeGreen, line width=0.9pt] (0,0) ellipse (.18cm and .18cm);
\node [
        draw, color=tubePurple, line width=0.9pt,
        rounded rectangle,
        minimum height = 1.3em,
        minimum width = 4.3em,
        rounded rectangle arc length = 180,
    ] at (-0.375,0)
    {};
    \node [
        draw, color=tubeGray, line width=0.9pt,
        rounded rectangle,
        minimum height = 1.65em,
        minimum width = 6.6em,
        rounded rectangle arc length = 180,
    ] at (-.05,0)
    {};
\end{tikzpicture}}
\,\,
+
 \raisebox{2pt}{
 \begin{tikzpicture}[baseline=(current bounding box.center)]
\draw[fill=black] (-0.75,0) -- (0.75,0);
\draw[fill=black] (0,0) circle (.5mm);
\draw[fill=black] (-0.75,0) circle (.5mm);
\draw[fill=black] (0.75,0) circle (.5mm);
\draw[color=tubeRed, line width=0.9pt] (-0.75,0) ellipse (.18cm and .18cm);
\draw[color=tubeBlue, line width=0.9pt] (0.75,0) ellipse (.18cm and .18cm);
\draw[color=tubeGreen, line width=0.9pt] (0,0) ellipse (.18cm and .18cm);
\node [
        draw, color=tubeOrange, line width=0.9pt,
        rounded rectangle,
        minimum height = 1.3em,
        minimum width = 4.3em,
        rounded rectangle arc length = 180,
    ] at (0.37,0)
    {};
    \node [
        draw, color=tubeGray, line width=0.9pt,
        rounded rectangle,
        minimum height = 1.65em,
        minimum width = 6.6em,
        rounded rectangle arc length = 180,
    ] at (0.05,0)
    {};
\end{tikzpicture}}\\[2pt]\nonumber
&= \frac{1}{{\color{tubeGray}(X_1+X_2+X_3)}{\color{tubeRed}(X_1+Y_{12})}{\color{tubeGreen}(X_2+Y_{12}+Y_{23})}{\color{tubeBlue}(X_3+Y_{23})} }\bigg(\frac{1}{\color{tubePurple}X_1+X_2+Y_{23}}+\frac{1}{\color{tubeOrange}X_2+X_3+Y_{12}} \bigg) .
\end{align}
Note that there are two possible maximal circlings in this case, which have all but one of their circles in common. We have color-coded the rational factors in the wavefunction to correspond to the circles that appear in these collections.

\vskip4pt
Flat-space wavefunction coefficients for arbitrary graphs can be obtained this way. The expressions will always be rational functions built from linear factors in the
$X$ and $Y$ variables.

\subsubsection*{Functions and Symbols}

The fact that the flat-space wavefunction is rational
implies that the de Sitter wavefunction in these models~\eqref{eq:dsWFint} can be cast as an Euler-type integral of a rational function. Both the integrand and the integration region are related to projective polytopes, and so these integrals evaluate to (generalized) Aomoto polylogarithms~\cite{Arkani-Hamed:2017fdk,Arkani-Hamed:2017ahv}. The fact that the maximal residues of the 
flat-space wavefunction are $\pm \prod_e 1/(2Y_e)$ makes it natural to rescale $\psi$ and define~\cite{Arkani-Hamed:2017fdk,Hillman:2019wgh,Arkani-Hamed:2023kig}
\be
F_{\cal G}(X_v,Y_e)  \equiv i^{n_v} \bigg(\prod_e 2Y_e\bigg) \, \psi_{\mathcal{G}}(X_v,Y_e)\,,
\label{eq:Fdef}
\ee
where $n_v$ is the number of vertices,
so that the integrand of $F$ has unit residues.
With this definition we can note that~\eqref{eq:dsWFint} implies that $F_{\cal G}$ can be written as
\be
\label{eq:integralrepF}
F_{\mathcal{G}}(X_v,Y_e) = \bigg(\prod_e 2Y_e\bigg)\int_0^\infty \rd x_1\cdots \rd x_n\, \psi^{\rm flat}_{\mathcal{G}}(X_v+x_v,Y_e)\,.
\ee
This makes manifest the fact that $F$ is a {\it real} function for real kinematic arguments~\cite{Chowdhury:2025ohm,Chowdhury:2026dwm}.

\vskip4pt
The polylogarithmic functions~\eqref{eq:Fdef} can be expressed as iterated integrals involving rational factors $R_a$, of the form
\be
\label{eq:iteratedint}
  f_{(n)} 
   = \int \rd \log R_{1} \circ \cdots \circ \rd \log R_{n} .
\ee
The number of integrations $n$ is called the {\it transcendentality}. The functions~\eqref{eq:iteratedint} can be rather complex and so it is useful to have a way to summarize some of the physical information that we might want to know about correlations.
As an example, the discontinuities and differentials of the functions $f_{(n)}$ are determined by the rational factors $R_a$, and so it is convenient to keep track of these factors (and their ordering) via what is called the {\it symbol}~\cite{goncharov2001multiple,goncharov2009simple,Goncharov:2010jf}
\be
{\cal S}( f_{(n)}) = R_1\otimes R_2\otimes \cdots \otimes R_n \,.
\label{eq:symboldef}
\ee
Here the product $\otimes$ behaves in the same way as an ordinary product if we think of the $R_a$ factors as implicitly $\log R_a$.
We summarize some of the mathematical features of the symbol in Appendix~\ref{app: Symbology}. The symbol captures some of the ``invariant" information about the actual function, including its sequences of discontinuities and differentials. One benefit is that it manifests identities satisfied by polylogarithmic functions---typically a given function has many different representations that all share the same symbol. The price paid for this simplicity is that the symbol is not fully the function itself; pieces of lower transcendentality with coefficients that are transcendental numbers (like $\pi$, $\log 2$, etc.) are projected out. However, it is often possible to ``integrate" the symbol by supplying boundary conditions that the function should obey. In the context of the wavefunction, a natural such boundary condition is that the function vanishes when any of the $Y_e$ are set to zero, which is manifest from~\eqref{eq:Fdef}, along with the fact that $\psi_{\cal G}$ is regular at these points.  

\vskip4pt
There are efficient algorithms to compute the symbol of~\eqref{eq:Fdef}, either from its integral form~\cite{Arkani-Hamed:2017fdk,Arkani-Hamed:2017ahv}, or using recursion relations that follow from the singularity structure of the energy integrand~\cite{Hillman:2019wgh}. As such, our focus is not on obtaining the symbol {\it per se}, but rather on elucidating the combinatorial and geometric underpinnings that control its structure.

\subsection{Combinatorics of Singularities}
\label{sec:singularities}

The correlators (or wavefunctions) we are considering are interesting functions, whose symbols we want to characterize.
The first step toward understanding the structure of the symbol is to understand what its entries (the $R_a$ in~\eqref{eq:symboldef}) can be.  These entries are often referred to as {\it letters}, and the collection of letters is called the symbol {\it alphabet}.

\vskip4pt
In general it is difficult to catalog the possible singularities of the transcendental functions that arise in physics. However, it is a remarkable feature of cosmological correlators that we are able to completely enumerate their possible singularities extremely simply (at least at tree level). This is possible essentially because of the integral representation~\eqref{eq:dsWFint}. The singularities of the integrand (the flat-space wavefunction) are easy to characterize. The wavefunction can be singular when the energy flowing into a subgraph happens to vanish~\cite{Arkani-Hamed:2017fdk}, corresponding to the vanishing of some sum of $X$ and $Y$ variables. These singularities can of course also be singularities of the integrated de Sitter wavefunction, but the act of integration can also introduce new singularities.
There is a nice combinatorial structure underlying the possible singularities~\cite{Arkani-Hamed:2023kig,Hillman:2023vas}. It is already clear that the singularities of the integrand are related to circlings of a graph, and the new singularities of the integrated object can be given a similar interpretation.

\vskip4pt
To catalog the singularities, we introduce the concept of a {\it marked graph}. Given a truncated Feynman graph, we mark all internal lines with a cross. We then assign label $X_v+\sum Y_e$ to each vertex, which is the $X$ variable associated to that vertex plus the sum of all edge energies attached to that vertex. To the crosses we also assign the label $-2Y_e$, where $Y_e$ is the energy of the edge marked by the cross. As an example,
the truncated graph corresponding to~\eqref{eq:truncated3site} turns into the marked graph
\be
\label{eq:3sitemarked}
\raisebox{-16pt}{
\begin{tikzpicture}[line width=1.pt, scale=2]

  \draw[fill=black] (0,0) -- (0.75,0);
  \draw[fill=black] (0.75,0) -- (1.5,0);

  \draw[fill=black] (0,0) circle (.03cm);
  \draw[fill=black] (0.75,0) circle (.03cm);
  \draw[fill=black] (1.5,0) circle (.03cm);

  \node[scale=1] at (0,-.15) {\scriptsize $X_1$};
  \node[scale=1] at (0.75,-.15) {\scriptsize$X_2$};
  \node[scale=1] at (1.5,-.15) {\scriptsize$X_3$};

  \node[scale=1] at (0.375,.12) {\scriptsize$Y_{12}$};
  \node[scale=1] at (1.125,.12) {\scriptsize$Y_{23}$};
\end{tikzpicture}
}
~~
\raisebox{-1pt}{
\begin{tikzpicture}
  \draw[-stealth, gray, line width=2pt] (0,0) -- (1,0);
\end{tikzpicture}
}
~~
\raisebox{-18pt}{
\begin{tikzpicture}[line width=1.pt, scale=2]

  \draw[fill=black] (0,0) -- (0.75,0);
  \draw[fill=black] (0.75,0) -- (1.5,0);

  \draw[tubeRed,fill=tubeRed] (0,0) circle (.03cm);
  \draw[tubeGreen,fill=tubeGreen] (0.75,0) circle (.03cm);
  \draw[tubeBlue,fill=tubeBlue] (1.5,0) circle (.03cm);
  
    \node[scale=1.55] at (.375,0) {\Cross};
    
      \node[scale=1.55] at (1.125,0) {\Cross};
      
        \node[scale=1] at (-.16,-.2) {\scriptsize ${\color{tubeRed} X_1+Y_{12}}$};
  \node[scale=1] at (0.79,-.2) {\scriptsize ${\color{tubeGreen} X_2+Y_{12}+Y_{23}}$};
  \node[scale=1] at (1.75,-.2) {\scriptsize ${\color{tubeBlue} X_3+Y_{23}}$};

  \node[scale=1] at (0.375,.17) {\scriptsize$-2Y_{12}$};
  \node[scale=1] at (1.125,.17) {\scriptsize$-2Y_{23}$};

\end{tikzpicture}
}
\ee
where we have colored the vertex labels for clarity. Given an $n$-site initial Feynman graph, we denote the number of vertices (original vertices plus crosses) of the marked graph by $N$.

\vskip4pt
The possible singularities of the wavefunction/correlator are given by the tubes of the marked graph. Recall that a {\it tube} is a connected (proper) subgraph, which can simply be represented by encircling the included vertices, where we treat the marking as a vertex. We assign energy variables to tubes, given by the sum of enclosed vertex energies. Note that tubes that have vertices of the original graph as their outermost vertices only involve sums of energy variables, while those that have crosses as their outermost vertices will involve expressions of the form $X-Y$. These latter singularities correspond to degenerations of the kinematics where some of the momenta become collinear, and for this reason are called {\it folded} singularities. Importantly, these singularities are absent on the physical sheet in the Bunch--Davies vacuum~\cite{Chen:2006nt,Holman:2007na,Flauger:2013hra,Arkani-Hamed:2018kmz,Green:2020whw}.

\vskip4pt
Graph tubings form a partially ordered set, where compatibility means that tubes do not intersect or enclose adjacent sites without being proper subsets~\cite{carr2005coxetercomplexesgraphassociahedra,devadoss2006realizationgraphassociahedra}. See Appendix~\ref{app: Structures of An and Bn} for more details.  The compatibility relations between graph tubings are captured by a polytope called the {\it graph associahedron}, and the graph associahedron therefore captures the singularities of wavefunction coefficients~\cite{Hillman:2023vas}.

\vskip4pt
Let us illustrate the possible singularities and their combinatorics using the simplest example, the two-site chain. The corresponding marked graph has three vertices:
\be
\raisebox{-16pt}{
\begin{tikzpicture}[line width=1.pt, scale=2]

  \draw[fill=black] (0,0) -- (0.65,0);
  \draw[fill=black] (0.65,0) -- (1.3,0);

  \draw[fill=black] (0,0) circle (.035cm);
  \draw[fill=black] (1.3,0) circle (.035cm);

  \node[scale=1.55] at (.65,0) {\Cross};

  \node[scale=1] at (-.05,-.2) {$X_1+Y$};
  \node[scale=1] at (1.25,-.2) {$X_2+Y$};

  \node[scale=1] at (0.65,.2) {$-2Y$};
\end{tikzpicture}
}
\label{eq:2sitemarkedgraph}
\ee
The possible singularities of the de Sitter wavefunction are in one-to-one correspondence with single tubes of this marked graph, so the alphabet is given by
\begin{align}
\twosite*{lp}{tubeBlue} ~&= X_1+Y  & \twosite*{rp}{tubeBlue} ~&=X_2+Y\\ 
\twosite*{lt}{tubeRed} ~&= X_1-Y & \twosite*{rt}{tubeRed} ~&=X_2-Y\\ 
\twosite*{mid}{tubeGreen} ~&=-2Y& \twosite*{full}{tubeGray} ~&= X_1+X_2\,.
\end{align}
A small subtlety is that we have included the total energy of the graph, corresponding to encircling the full graph. Strictly speaking, this is not a proper subset, but it does enter as a singularity of the final answer. (We will see that this subtlety is part of the reason that compatibility between symbol letters is slightly more complicated than simply compatibility between graph tubings.) The singularities $X_1+Y$, $X_2+Y$, and $X_1+X_2$ are all singularities of the flat-space wavefunction $\psi_{(2)}^{\rm flat}$, while the singularities $X_1-Y$ and $X_2-Y$ appear in the de Sitter wavefunction after integration. The singularity $Y=0$ does not actually appear in cosmological observables, but is nevertheless important combinatorially.\footnote{In the conformally coupled case it is always possible to rescale the wavefunction to remove any singularities in $Y$ alone, but not for generic masses~\cite{Baumann:2026atn}.}
 The various tubings that determine the possible singularities of the cosmological wavefunction can be used to label the graph associahedron of a three-site marked chain, which is the pentagon~\cite{Hillman:2023vas}
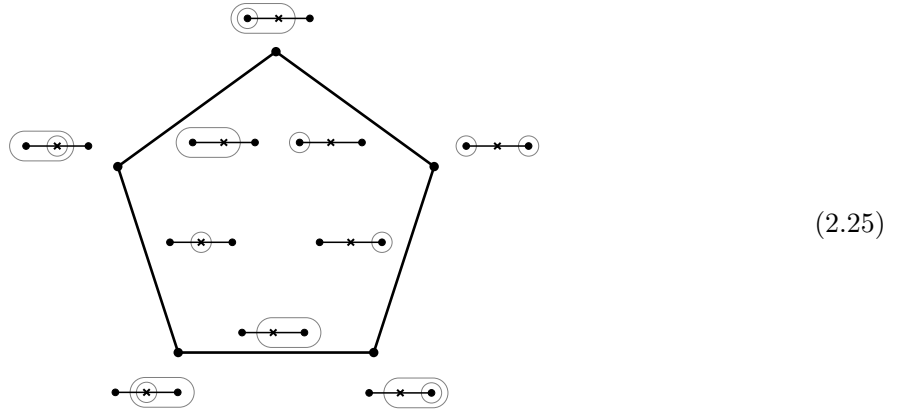
\begin{equation}
\raisebox{-70pt}{
\begin{tikzpicture}[
  scale=2.2,
  every node/.style={font=\footnotesize},
  vtx/.style={circle,fill=black,inner sep=1.25pt},
  pent/.style={line width=1.pt},
  qarrow/.style={
    draw=darkblue!60,
    -{Stealth[length=4pt,width=4pt]},
    line width=0.85pt
  }
]
\def\r{1.0}
\def\roff{0.4}
\def\foff{0.16}

\coordinate (A) at (90:\r);
\coordinate (B) at (162:\r);
\coordinate (C) at (234:\r);
\coordinate (D) at (306:\r);
\coordinate (E) at (18:\r);

\draw[pent] (A) -- (B) -- (C) -- (D) -- (E) -- cycle;

\node[vtx] at (A) {};
\node[vtx] at (B) {};
\node[vtx] at (C) {};
\node[vtx] at (D) {};
\node[vtx] at (E) {};

\newcommand{\seedright}[3]{%
  \begin{scope}[shift={#1}]
    \node[anchor=east,inner sep=1pt] (L) at (-0.14,0) {$#2$};
    \node[anchor=west,inner sep=1pt] (R) at ( 0.14,0) {$#3$};
    \draw[qarrow] (L.east) -- (R.west);
  \end{scope}
}
\newcommand{\seedleft}[3]{%
  \begin{scope}[shift={#1}]
    \node[anchor=east,inner sep=1pt] (L) at (-0.14,0) {$#2$};
    \node[anchor=west,inner sep=1pt] (R) at ( 0.14,0) {$#3$};
    \draw[qarrow] (R.west) -- (L.east);
  \end{scope}
}

\node at ($(E)+(18:\roff)$)
  {\scalebox{.55}{{\markedchaingraph{2}[v1,v2][][][1.8]}}};

\node at ($(A)+(90:.5*\roff)$)
  {\scalebox{.55}{{\markedchaingraph{2}[v1][v1/x1][][1.8]}}};

\node at ($(B)+(162:\roff)$)
  {\scalebox{.55}{{\markedchaingraph{2}[x1][v1/x1][][1.8]}}};

\node at ($(C)+(234:.74*\roff)$)
  {\scalebox{.55}{{\markedchaingraph{2}[x1][x1/v2][][1.8]}}};

\node at ($(D)+(306:.75*\roff)$)
  {\scalebox{.55}{{\markedchaingraph{2}[v2][x1/v2][][1.8]}}};

\node at ($($(E)!0.5!(A)$)+(234:1.55*\foff)$)
  {\scalebox{.55}{{\markedchaingraph{2}[v1][][][1.8]}}};

\node at ($($(A)!0.5!(B)$)+(306:1.55*\foff)$)
  {\scalebox{.55}{{\markedchaingraph{2}[][v1/x1][][1.8]}}};

\node at ($($(B)!0.5!(C)$)+(18:2.1*\foff)$)
  {\scalebox{.55}{{\markedchaingraph{2}[x1][][][1.8]}}};

\node at ($($(C)!0.5!(D)$)+(90:.75*\foff)$)
  {\scalebox{.55}{{\markedchaingraph{2}[][x1/v2][][1.8]}}};

\node at ($($(D)!0.5!(E)$)+(162:2.1*\foff)$)
  {\scalebox{.55}{{\markedchaingraph{2}[v2][][][1.8]}}};

\end{tikzpicture}
}
\label{eq:graphassocex}
\end{equation}
Here we see that the facets are in correspondence with the possible singularities (with the exception of the total energy singularity). It is convenient to give these energy singularities names. We will denote them by $E_I$, where $I$ labels the possible tubes.

\newpage
\section{Cosmological Cluster Algebras}
\label{sec: E to A map}

Our goal is to understand the structures underlying the symbol of the wavefunction coefficients $\psi$ of the model~\eqref{eq:scalaraction}. Given a Feynman graph, ${\cal G}$, we can parameterize the symbol abstractly as
\begin{equation}
    {\cal S}(F_{\cal G})=\sum_{I_1,\cdots,I_n} C_{I_1\cdots I_n}\, E_{I_1}\otimes E_{I_2}\otimes \cdots \otimes E_{I_n}\,,
\end{equation}
where the sum runs over all the $E_I$, which correspond to the possible tubes of the marked version of the Feynman graph, as described in Section~\ref{sec:singularities}. In this parameterization, the coefficients $C_{I_1\cdots I_n}$ determine the symbol by dictating what letters appear together in a given term.

\vskip4pt
The challenge of constructing the symbol can therefore be phrased as determining what letters are ``compatible", in the sense that they appear together in the symbol.
It is consequently natural that the mathematical apparatus of cluster algebras~\cite{FominZelevinsky2002ClusterI} can be brought to bear on this problem.
A cluster algebra is a commutative algebra whose generators (cluster variables) can be arranged into groups called clusters, which can be transformed via a process called mutation. Cluster variables that appear together in some cluster (or seed) are called compatible, while those that never do are called incompatible. 
(See Appendix~\ref{app:cluster} for a brief introduction to the ideas from cluster algebras that are useful in the present context.)
Roughly, cluster algebras provide a mechanism by which singularities can be compatible or incompatible with each other, which determines the structure of the symbol in cases admitting such a cluster structure. 
In order to realize this, we need to find a way to express the energy variables that appear inside the symbol in terms of cluster variables of some cluster algebra.

\vskip4pt
Recall that in Section~\ref{sec:singularities}, we enumerated the possible singularities of a wavefunction coefficient by relating them to tubings of a marked graph. Since these tubings have a natural notion of compatibility (depicted, for example, in~\eqref{eq:graphassocex}), we might imagine that it is this notion of compatibility that will be reflected in the symbol. However, this is not the case. We can see this already in the simplest example. The four-point wavefunction/correlator coming from a single exchange has the symbol~\cite{Arkani-Hamed:2015bza,Arkani-Hamed:2017fdk,Arkani-Hamed:2018kmz,Hillman:2019wgh}
\be
\label{eq:2sitesymbol}
{\cal S}(F_{(2)}) = \frac{X_1+Y}{X_1+X_2}\otimes \frac{X_2-Y}{X_2+Y}+\frac{X_2+Y}{X_1+X_2}\otimes \frac{X_1-Y}{X_1+Y}\,.
\ee
We see that this expression has, for example, $X_1+Y$ and $X_2-Y$ appearing as adjacent entries, despite being incompatible in the sense of~\eqref{eq:graphassocex} because they correspond to two tubes that are adjacent. This suggests that we cannot directly map the energies associated to tubings to cluster variables. Another hint that this is the case is that cluster variables are inherently dimensionless, while the energy tube variables have mass dimension one. As such, we should expect that cluster variables will be related to dimensionless ratios of singularities.

\vskip4pt
We make a connection between energy variables and cluster variables using the fact that they both have polytopal origins.
Each of these polytopes admits a ``binary" realization in terms of $u$-variables~\cite{Koba:1969rw,Koba:1969kh,brown2006multiplezetavaluesperiods,Arkani-Hamed:2019mrd,Arkani-Hamed:2019plo,Arkani-Hamed:2020tuz,He:2020onr}.
This realization is in a sense coordinate invariant. This means that in cases where the cluster polytope and graph associahedron are the same---which happens for $n$-site chain graphs and the cluster algebra $A_{2n-2}$  and $n$-site cycle graphs and $B_{2n-1}$ cluster algebras---there is a natural way to relate the two objects. By setting the $u$-variables of the cluster polytope and the marked graph associahedron equal to each other we derive a map between energy variables and cluster variables in these cases.

\vskip4pt
Here we first briefly review $u$-variables and their appearance in cosmological graphs and cluster algebras  before using them to connect the two objects.

\subsection{$u$ Variables}

We have seen several polytopes arise which encode compatibility between different objects. For example, both the tubings of a chain graph and the clusters of $A$-type cluster algebras are related to the associahedron. However, since these polytopes are defined in different spaces, it is not clear if or how they are related. We therefore want to find a more invariant parameterization of their content.

\vskip4pt
An invariant realization of compatibility is provided by the ``binary geometries" of~\cite{Arkani-Hamed:2020tuz,Arkani-Hamed:2019mrd,Arkani-Hamed:2019plo,He:2020onr}, which are expressed in terms of {\it $u$-variables}, introduced by~\cite{Koba:1969rw,Koba:1969kh} and~\cite{brown2006multiplezetavaluesperiods}. The defining feature of $u$-variables is that they satisfy nonlinear equations of the form
\begin{equation}
    \label{eq: u equations}
    u_a +\prod_{b} u_b^{(b||a)}=1\,,
\end{equation} 
where $a, b$ index the collection of variables, and $(b||a)$ is the compatibility degree between $b$ and $a$. An interesting feature of these equations is that if we restrict to the (real) positive region $u_a \geq 0$ for all $u_a$, then $0\leq u_a\leq 1$, so if we approach the boundary $u_a \to 0$, then from~\eqref{eq: u equations}, all of the $u_b$ that are incompatible with $u_a$ must go to $1$. In this sense, the space of $u$-variables captures the compatibility in a rigid way: $u$-variables that are compatible can simultaneously be set to zero, while the incompatible variables get set to $1$ in this limit.

\vskip4pt
Given that the equations~\eqref{eq: u equations} are over-constrained, it is not obvious that they should have an interesting solution space. In fact they do, and two mechanisms for generating solutions to these equations will appear as central elements in our construction.

\subsubsection*{Marked graph tubings}

The singularities of the cosmological wavefunction have a natural set of $u$-variables related to them~\cite{Hillman:2023vas}. Given a marked graph labeled as in Section~\ref{sec:singularities}, we can assign $u$-variables as follows~\cite{Hillman:2023vas,He:2020onr}. We consider the collection of tubes of the graph $T_{\cal G}$. The sum of vertex energies enclosed by the tube we denote $E_T$. Given an initial tube $T$, we consider all possible minimal extensions of the tube (corresponding to all possible ways of enlarging the tube to enclose its immediately adjacent crosses or vertices). We then take the product over all such minimal extensions, with power $+1$ if the number of additional vertices is even, or $-1$ if the number is odd~\cite{He:2020onr}. Concretely, we have the formula
\be
u_T = \frac{E_T\, E_{T+2}\cdots}{E_{T+1}\,E_{T+3}\cdots}\,.
\label{eq:uvardef1}
\ee
This formula has an interesting inverse, which expresses the ratio of a tube energy to the total energy of the graph in terms of
the product of $u$-variables associated to tubes that contain $T$:
\be
    \frac{E_T}{E_{\cal G}}=\prod_{ T' \supseteq T} u_{T'}\,.
    \label{eq:inverseEurelation}
\ee

It is illuminating to consider an explicit example. Given the two-site marked graph~\eqref{eq:2sitemarkedgraph}, the $u$-variables are~\cite{Hillman:2023vas}
\begin{align}
\twosite{lp}{black} ~&\leadsto~ u_{13} = \frac{X_1+Y}{X_1-Y}  & \twosite{rp}{black} ~&\leadsto~ u_{35}=\frac{X_2+Y}{X_2-Y}\\ 
\twosite{lt}{black} ~&\leadsto~ u_{14}= \frac{X_1-Y}{X_1+X_2} & \twosite{rt}{black} ~&\leadsto~ u_{25}=\frac{X_2-Y}{X_1+X_2}\\ 
\twosite{mid}{black} ~&\leadsto~ u_{24}=-\frac{2Y(X_1+X_2)}{(X_1-Y)(X_2-Y)}\,.
\end{align}
Here we have labeled the $u$-variables such that their compatibility relations are those of the corresponding chords of a pentagon. One can check directly that they satisfy the $u$-equations
\be
u_{13} + u_{24}\,u_{25} = 1\,,
\label{eq:graphuequations}
\ee
along with its cyclic permutations. Together these five equations parameterize a two-dimensional space (which we can for example take to be $X_1/Y$ and $X_2/Y$). We can also verify that~\eqref{eq:inverseEurelation} is satisfied, for example $(X_1+Y)/(X_1+X_2) = u_{13}\,u_{14}$. Notice that the $u$-variables express the same compatibility relations as the tubes that they derive from.

\subsubsection*{Cluster algebras}

Finite-type cluster algebras also  provide a natural mechanism to produce $u$-variables~\cite{Arkani-Hamed:2019plo,Arkani-Hamed:2020tuz,He:2020onr}.
Recall that cluster variables can be categorized as either mutable or frozen. Mutable cluster variables, $a_k$,
can be {\it mutated} to produce new variables
\be
a_k \mapsto a_k' = \mu_k(a_k)\,,
\label{eq:exchangerelation}
\ee
while frozen variables $z_k$ cannot. Relations of the type~\eqref{eq:exchangerelation} are called {\it exchange relations} (because they exchange cluster variables between clusters). Amongst all the exchange relations, a subset are related to $u$-variables. These {\it primitive} exchange relations are of the schematic form
\begin{equation}
    \mu_k(a_k)\, a_k= M_k[z]\, S_k[a] + M'_k[z]\,,
\label{eq:primitiveexchangerelation}
\end{equation}
where $M_k$ and $M_k'$ are monomials in the frozen variables, while $S_k$ is a monomial in mutable cluster variables. The remarkable feature of primitive exchange relations is that one of the terms in the exchange binomial contains only frozen variables~\cite{Arkani-Hamed:2020tuz}. (If we eliminate frozen variables by setting $z_k=1$, primitive exchange relations are then of the form $\mu_k(a_k)\, a_k=S_k+1$.)

\vskip4pt
From a primitive exchange relation of the form~\eqref{eq:primitiveexchangerelation}, we can produce a corresponding $u$-variable as~\cite{Arkani-Hamed:2020tuz}
\begin{equation}
    \label{eq: u variables from primitive exchange}
    u_k=\frac{M_k[z]\, S_k[a]}{\mu_k(a_k) \, a_k}\,.
\end{equation}
Given a cluster algebra, in principle one can identify primitive exchange relations by explicitly enumerating all exchanges. This is often impractical, but we can take advantage of the fact that all finite-type cluster algebras have an automorphism $\tau$ on the set of cluster variables for which the exchange relation between $a_k$ and $\tau(a_k)$ is primitive~\cite{Arkani-Hamed:2020tuz}.\footnote{In the notation of Appendix~\ref{app: Structures of An and Bn}, the automorphism $\tau$ for 
 $A_N$ or $B_N$ cluster algebras can be understood as a cyclic rotation of the indices of the polygon used to label the cluster variables, $\tau(a_{i\,j})=a_{i-1\,j-1}$. (See~\cite{Arkani-Hamed:2020tuz} for explicit forms of the automorphisms of other cluster algebras.)}
These $u$-variables obey~\eqref{eq: u equations} for any assignment of frozen variables (including $z_k = 1$). See Appendix~\ref{app:clusterdefs} for more details.

\vskip4pt
It is again useful to illustrate this construction with an example. We consider the $A_2$ cluster algebra---in this case everything is so simple that it can be enumerated explicitly. We describe this example in detail in Appendix~\ref{app:cluster}, but here we summarize the main features.
The cluster algebra is naturally associated to a quiver
\be
\raisebox{-5pt}{
\begin{tikzpicture}[
  quiver/.style={
    draw=tubeBlue!60,
    -{Stealth[length=5pt,width=5pt]},
    line width=0.8pt
  },
]
  \node (1) at (0,2) {$a_1$};
  \node (2) at (2,2) {$a_2$};
  \draw[quiver] (1) -- (2);
\end{tikzpicture} 
}\,,
\ee
which corresponds to the initial cluster containing two variables $\{a_1,a_2\}$. (Here we are not including any frozen variables for simplicity.) By mutating, we produce a total of $5$ cluster variables ($a_1, \cdots,  a_5$) 
appearing in $5$ distinct seeds.
The mutation operation in this case is very simple. Acting on either the left or right node, it reverses the arrow and produces a new variable according to the rule
\be
\mu_k(a_k)\,a_k = \prod_{i\to k} a_i+\prod_{k\to j} a_j\,,
\ee
where the first term is a product of variables at nodes pointing into $k$, while the second term is the product of variables for nodes that $k$ points to. In this case, only one of these two terms will ever be nontrivial (the other is just $1$). We can enumerate all the exchange relations obtained by sequences of mutations:
\be
a_{i+3}\,a_i = a_{i+4}+1\,,
\label{eq:A2exc}
\ee
where $i \in \{1,\cdots, 5\}$ and the index labels are cyclic, and so should be interpreted mod $5$. From this, we see that in this special case all of the exchange relations are actually primitive.\footnote{Relatedly, this implies that the cluster variables themselves---in this particular case---satisfy $u$-equations. However, thinking of the cluster variables directly as $u$-variables does not generalize beyond this special case.}
We can therefore use~\eqref{eq: u variables from primitive exchange} to infer
\be
u_i \equiv \frac{a_{i+4}}{a_{i+3} \,a_i}\,.
\ee
We describe how to derive the same variables from an algebra automorphism in Appendix~\ref{app:cluster}.
It is then straightforward to check that these $u$-variables satisfy the $u$-equations
\be
1-u_i = u_{i+2}\,u_{i+3}\,,
\ee
as a consequence of the exchange relations~\eqref{eq:A2exc}. In order to make contact with the graph tubing case, we can map the indexing as
\be
\begin{aligned}
    u_{1}&\equiv  u_{13} \,,   & \hspace{1.7cm} u_{2}& \equiv u_{14} \,,&\hspace{1.7cm} u_{3}&\equiv u_{24}\,,\\[1pt]
    u_{4}&\equiv u_{25}\,, & u_{5}&\equiv u_{35}\,,&&
\end{aligned}
\ee
after which the variables satisfy the same $u$-equation~\eqref{eq:graphuequations} and its cyclic permutations. The benefit of this parameterization is that it makes the connection between the cluster algebra and graph tubings obvious.

\subsection{From $a$ to $X$ and $Y$ via $u$}
\label{sec:aXYu}

We have now seen $u$-variables and equations appear from the enumeration of singularities of cosmological correlators, and separately in finite-type cluster algebras. The benefit of the $u$-variable perspective is that the geometry they parameterize is rigid, and so the two appearances are in a sense the same. This provides a natural way to make a connection between the alphabet of cosmological singularities and cluster variables. We simply set the $u$-variables obtained from the two constructions to be equal.
 This will allow us to imbue the singularities with a (rather restrictive and nontrivial) notion of compatibility. Importantly, this notion is {\it different} from the requirement that the tubings related to the letters are compatible in the mathematical sense.

\vskip4pt
We describe the mapping between singularities and cluster variables in two cases: the first relates $n$-site chains and $A_{2n-2}\equiv A_{N-1}$ cluster variables~\cite{Mazloumi:2025pmx,Capuano:2025myy}, while the second relates $n$-site one-loop graphs and $B_{2n-1}\equiv B_{N-1}$~\cite{Paranjape:2026htn}.\footnote{A relation between the letters of the two-site cycle and two $A_3$ algebras was also noted in~\cite{Mazloumi:2025pmx}.} 
In order to write these relations in a simple way we label the cluster variables and graph tubes by chords and faces of a polygon. We review the labeling scheme here; more details can be found in Appendix~\ref{app: Structures of An and Bn}.
For more general cosmological graphs, it is unclear whether there is a (generalized) cluster algebra sharing the same $u$-equations.

\subsubsection*{Chains and $\boldsymbol{A_N}$}
We first describe the mapping between chain graphs and $A$-type cluster algebras. To do this, we explain how to construct $u$-variables in each case and then equate them.

\vskip4pt
Let us describe the labeling scheme for graph tubings and cluster variables. It is convenient to introduce labels ``between" the sites of the graph:
\begin{equation*}
{\scalebox{.9}{%
\markedchaingraph{3}
  [][]
  [v1/x1/{$1$}/{-21pt}/{-8pt},
   v1/x1/{$2$}/{0pt}/{-8pt},
   x1/v2/{$3$}/{0pt}/{-8pt},
   v2/x2/{$4$}/{0pt}/{-8pt},
   x2/v3/{$5$}/{0pt}/{-8pt}
   ][1.8]%
}}
\;\cdots\;
{\scalebox{.9}{%
\markedchaingraph{2}
  [][]
  [
   x1/v2/{$\cdots$ ~~$N$}/{-13pt}/{-8pt},
   v2/x2/{$N+1$}/{8pt}/{-8pt}
   ][1.8]%
}}
\end{equation*}
so that the tubes (and energies) can be labeled as $T_{ij}$, with $1\leq i <j \leq N+1$. Here $N$ is the total number of vertices and crosses of the marked graph ($N= 2n-1$, where $n$ is the number of vertices in the original graph). In this labeling scheme, the tube $T_{ij}$ encircles all the sites between the numbers $i$ and $j$. We assign each tube the energy $E_{i\,j}=\sum x_{v}$ where $x_v$ are the energies of the sites inside the tube (including the crosses). 
An $N$-site marked chain graph can be associated to the cluster algebra $A_{N-1}$. The cluster variables can be labeled by the chords of an $(N+2)$-gon as $a_{ij}$ with $1\leq i<j-1<N+2$ (excluding $(i,j) = (1,N+2)$, which is not a diagonal).

\vspace{4pt}
With these labeling schemes, we can construct $u$-variables using~\eqref{eq:uvardef1} and~\eqref{eq: u variables from primitive exchange}, and then set
\be
u(a_{ij}) = u(T_{i\,j-1})\,.
\label{eq:uequaluequation}
\ee
Explicitly, with this indexing convention the $u$-variables are given by
\be
   u_{i j}=\frac{a_{i\,j-1}a_{i-1\,j}}{a_{i-1\,j-1}a_{i\,j}}z_{i j}=\frac{E_{i\,j-1}E_{i-1\,j}}{E_{i-1\,j-1}E_{i\,j}}\,,
   \label{eq:uaEequation}
\ee
where $z_{ij}$ is the (frozen) universal coefficient associated with the cluster variable $a_{ij}$. In this formula some index assignments should be understood to be $1$, concretely $E_{0i} =E_{i\, N+2}=1$ and $a_{i\,i+1}=1$. In addition, the indices should be understood to be cyclic, so for example $a_{i0} = a_{i\,N+2}$.

\vskip4pt
The labeling convention has been chosen so that the $u_{ij}$ are labeled the same as the cluster variables of $A_{N-1}$. 
For odd $N$ (which is the physical case because $N = 2n-1$) the equations have a closed form solution for either $a_{ij}$ or $E_{ij}$. 
To see this, we split the indices into two cyclic sets
\begin{equation}
    A=\{i,i+1,\cdots,j-1\}\,, \quad{\rm and}\quad
    B=\{j,j+1,\cdots,N+2,1,\cdots,i-1\}\,,
\end{equation}
and define 
\begin{equation}
  Z_{i\, j} \equiv \prod_{\substack{1\le k\le i\\[1pt] j+1\le m\le N+2}} z_{km}\,.
\end{equation}
The solution for $a_{i j}$ is then 
\begin{equation}\label{eq: General a and E relation}
    a_{ij}=\frac{E_{ij}}{Z_{i\, j}} \times \begin{cases}
      \displaystyle{\prod_{k \in A}} \left(\frac{E_{k\,k+1}}{Z_{k\, k+1}}\right)^{(k-i\mod 2)}\, \left(\frac{E_{k\,k+1}}{Z_{k\, k+1}}\right)^{-(k-i+1\mod 2)},\qquad\, \text{if $j-i$ is odd}\\
      \displaystyle{ \prod_{k \in B}} \left(\frac{E_{k\,k+1}}{Z_{k\, k+1}}\right)^{(k-j\mod 2)}\, \left(\frac{E_{k\,k+1}}{Z_{k\, k+1}}\right)^{-(k-j+1\mod 2)},\qquad  \text{if $j-i$ is even}
    \end{cases}\,,
\end{equation}
where $k-j$ denotes the cyclic distance from $j$ to $k$  (rather than just the difference of the labels).
Equation~\eqref{eq:uaEequation} can also be solved for $E_{ij}$ as 
\begin{equation}
    \label{eq: E in a An}
    \frac{E_{i j}}{E_{1\,N+1}}=\frac{ a_{ij}\,  Z_{i\, j}}{ a_{i\,N+2} \, a_{j\,N+2}} \frac{ a_{1\,N+2}\, a_{N+1\,N+2}}{a_{1\,N+1}\, Z_{1\, N+1}}\,,
\end{equation}
which expresses the ratio $E_{ij}/E_{1\,N+1}$ in terms of cluster variables. Though this expression determines dimensionless ratios of energy variables, it is conceptually helpful to imagine solving for $E_{ij}$ (this is ambiguous, but the ambiguous overall rescaling of the $E$ variables always drops out). When we set the frozen variables to $1$, we obtain a relation between $E_{ij}$ and $a_{ij}$
\be
 E_{i j}  \propto \frac{ a_{ij}}{ a_{i\,N+2} \, a_{j\,N+2}}\,.
 \label{eq:AnEamap}
\ee
The additional factors of $a_{i\, N+2}$ are the source of much of the nontriviality of this mapping.

\vskip4pt
It is worth noting that the expression~\eqref{eq:uaEequation} bears an un-coincidental resemblance to a cross-ratio, which was also utilized by~\cite{Capuano:2025myy,Ferro:2026oph}. We explain this feature in the following inset.

\begin{eBox3}
\noindent
{\small
{\bf Grassmannian Representation of $A_N$:}
Here we describe the derivation of~\eqref{eq:uaEequation} and~\eqref{eq:uequaluequation} from cross-ratios, reflecting the fact that both $A_N$ and the energy alphabet are naturally related to the Grassmannian ${\rm Gr}(2,k)$, the configuration space of $k$ points on $\mathbb{P}^1$. (See also~\cite{Capuano:2025myy,Ferro:2026oph}.)

\vskip4pt
The cluster variables of $A_{N'}$ can be viewed as Pl\"ucker coordinates---or $2\times2$ minors---of the matrix representation of ${\rm Gr}(2,N'+3)$ given by 
\begin{equation}
    G'_{(2,N'+3)}= 
    \begin{pmatrix}
        1 & a_{2\,N'+3} & a_{3\,N'+3} & \cdots&a_{N'+1\,N'+3}&1& 0 \\
        0 & 1 & a_{13} & a_{14} &\cdots&a_{1\,N'+2} &1 \\
    \end{pmatrix}\,.
\end{equation}
This indexing labels the cluster variable $a_{ij}$ both as the minor of columns $i$ and $j$, and as the chord of an $(N'+3)$-gon (described
in Appendix~\ref{app: Structures of An and Bn}).
Simultaneously, we can  describe the energy singularity alphabet of an $N$-site marked chain as Pl\"ucker coordinates of the Grassmannian
\begin{equation}
    G_{(2,N+2)}= 
    \begin{pmatrix}
        1 & 1 & 1 & \cdots&1&1& 0 \\
        0 & E_{12} & E_{13} & E_{14} &\cdots&E_{1\,N+1} &1 \\
    \end{pmatrix}\,,
\end{equation}
so that the $E_{ij}$ variables appearing in~\eqref{eq:uaEequation} are the $ij^{\rm th}$ minors of this matrix. Matrix representations of the Grassmannian are only defined up to a left action of ${\rm GL}(2,\mathbb{R})$. The invariant coordinates are cross ratios of the column vectors.
If we set $N' = N-1$, call the cross ratios of each matrix $u_{ij}$, and then demand that $G$ and $G'$ represent the same Grassmannian, 
we recover~\eqref{eq:uaEequation} (with $z=1$).
}
\end{eBox3}

We can also understand the relation between our identification of cluster variables and that of~\cite{Ferro:2026oph} using this Grassmannian representation. In~\cite{Ferro:2026oph} they define the cross ratios
\be
q_{ijkl} \equiv  \frac{\Delta_{ij}\Delta_{kl}}{\Delta_{jk}\Delta_{li}} =  \frac{E_{ij}E_{kl}}{E_{jk}E_{li}}\,.
\ee
From this, we see that the $u$-variables are related to a particular assignment of indices
\be
q_{i\,j-1\,i-1\,j} \ =u_{ij}\,.
\label{eq:qtou}
\ee
While the $\Delta_{ij}$ and $a_{ij}$ variables are not directly equal, their cross ratios are. Conceptually, it is satisfying that the natural ``invariant" structures---$u$-variables and cross-ratios---are related in a simple way. It would be interesting to determine if the additional structure carried by $u$-variables manifests in the representation employed in~\cite{Ferro:2026oph} in terms of quadrangular polylogs~\cite{rudenko2022,matveiakin2022}.

\paragraph{Example: $\boldsymbol{A_2}$}~\\
Here we make the mapping explicit in the simple example of the $A_2$ cluster algebra (corresponding to a two-site Feynman graph, or three-site marked chain).
The singularity alphabet is
\be
\label{eq:2sitesingularities}
\begin{aligned}
    \mathcal{E}_{\cal G}&=\left\{E_{12},E_{23},E_{34},E_{13},E_{24},E_{14}\right\}\\
    &=\{X_1+Y,\, -2Y,\, X_2+Y,\, X_1-Y,\, X_2-Y,\, X_1+X_2\} \,,
\end{aligned}
\ee
and the cluster algebra consists of the mutable variables  $ \mathcal{A}_{A_2}=\{a_{13},a_{14},a_{24},a_{25},a_{35}\}$,
along with a frozen variable $z_{ij}$ for each mutable variable.
The $u$-variables~\eqref{eq:uaEequation} can be written
\be
\begin{aligned}
    u_{13}&=\frac{a_{35} \, z_{13}}{a_{13}\,  a_{25}}=\frac{E_{12}}{E_{13}} & \hspace{1cm}
    u_{14}&=\frac{a_{13}\,  z_{14}}{a_{14}\, a_{35}}=\frac{E_{13}}{E_{14}}
    & \hspace{1cm} 
    u_{35}&=\frac{a_{25}\,  z_{35}}{a_{24}\, a_{35}}=\frac{E_{34}}{E_{24}}
    \\
    u_{24}&=\frac{a_{14}\,  z_{24}}{a_{13} \, a_{24}}=\frac{E_{14} E_{23}}{E_{13} E_{24}} & u_{25}&=\frac{a_{24}\,  z_{25}}{a_{14}\,  a_{25}}=\frac{E_{24}}{E_{14}} \,,
\end{aligned}
\ee
which relate energy variables and cluster variables. The equations can be solved explicitly for the cluster variables in terms of energies:
\be
\begin{aligned}
    a_{13}&=\frac{E_{13} z_{25} z_{35}}{E_{34} z_{14}} &\hspace{1cm}
    a_{14}&=\frac{E_{14} E_{23} z_{13} z_{35}}{E_{12} E_{34} z_{24}} &\hspace{1cm}
    a_{35}&=\frac{E_{12} z_{24} z_{25}}{E_{23} z_{13}}\\
    a_{24}&=\frac{E_{24} z_{13} z_{14}}{E_{12} z_{25}} &a_{25}&=\frac{E_{34} z_{14} z_{24}}{E_{23} z_{35}}\,,
\end{aligned}
\label{eq:atoEexplicit}
\ee
and also for the energy variables in terms of cluster variables\footnote{From this example we can see the difference from~\cite{Paranjape:2026htn}, where for example $E_{12}/E_{14}$ is mapped to the chord $\Delta_{12}$, which is compatible with everything.}
\be
\begin{aligned}
    \frac{E_{12}}{E_{14}}&=\frac{ z_{13} z_{14}}{a_{14} a_{25}}
    &\hspace{1cm}
    \frac{E_{13}}{E_{14}}&=\frac{a_{13}z_{14}}{a_{14} a_{35}}
    &\hspace{1cm}
     \frac{E_{34}}{E_{14}}&=\frac{ z_{25}z_{35}}{a_{14} a_{35}}
    \\
    \frac{E_{23}}{E_{14}}&=\frac{z_{14} z_{24} z_{25}}{a_{14} a_{25} a_{35}}
   &\frac{E_{24}}{E_{14}}&=\frac{a_{24} z_{25}}{a_{14} a_{25}}\,.
\end{aligned}
   \label{eq:EtoamapA2}
\ee
In this case it is consistent to set all the frozen variables to one ($z_{ij}=1$). Notice that the notions of compatibility are very different for the energy variables and cluster variables because they are nonlinearly related. In Section~\ref{sec:clusterbootstrap} we will see how cluster compatibility in terms of the $a_{ij}$ can be used to fix the symbol of the cosmological wavefunction and correlators.

\subsubsection*{One Loop and $\boldsymbol{B_N}$}

The energy variables of one-loop cycle graphs can be imbued with a cluster structure in a similar way. In this case the labeling scheme is a bit more complicated because merely specifying two labels does not tell us whether they are enclosed by a tube going clockwise or counterclockwise, and we need to include both possibilities.
A convenient way to resolve this ambiguity is to assign {\it two} labels to each edge of the graph $(i,\bar{i})$, organized clockwise around the graph. For an $N$-site marked graph, we have $\bar{i}=i+N$. As an example, consider a two-site Feynman graph
\be
\raisebox{-35pt}{
\begin{tikzpicture}[
  every node/.style={font=\footnotesize}
]
  \node at (0,0) {\rotatebox{90}{$\loopdrawmarked[2]{2}{1,2}$}};
  \node at (-1.1, .6) {$2$};
    \node at (-.6, 1.1) {$\overline{2}$};
      \node at (1.1, .6) {$\overline 3$};
    \node at (.6, 1.1) {$3$};
      \node at (-1.1, -.6) {$\overline 1$};
    \node at (-.6, -1.1) {$1$};
      \node at (1.1, -.6) {$4$};
    \node at (.6, -1.1) {$\overline 4$};
\end{tikzpicture}
}
\label{eq:marked2siteloop}
\ee
With this indexing we can label the tubes of a given graph by specifying two edge labels as $T_{[i,j]}$ with $1\leq i <j\leq  N$, where $i,j$ can be either barred or unbarred.
We use the convention that if both labels $i,j$ are unbarred or if both are barred $\bar i,\bar j$, this specifies the tube enclosing the vertices between these labels clockwise. On the other hand, if one index is barred and the other unbarred, for example $T_{[i, \bar j]}$, this corresponds to the tube enclosing the vertices between $i$ and $ \bar j$ counterclockwise. The tube $T_{[i,\bar{i}]}$ is the tube that encloses the total graph for any $i$. This labeling scheme is redundant because $T_{[i,j]} = T_{[\bar i,\bar j]}$ for any $i,j$, barred or unbarred.
We label energy variables as $E_{[i,j]}=\sum x_{v}$, which is the sum of site energies $x_v$ for sites enclosed in the tube $T_{[i,j]}$.
We index the cluster variables of the cluster algebra $B_{N-1}$ in the same way:
\be
 \mathcal{A}_{B_{N-1}}=
   ~  \{a_{[i,j]}, a_{[i,\bar j]} \}~~{\rm with}~~ 1\leq i \leq N~~{\rm and}~~i\leq j\leq N\,,
\ee
where again $\bar i = i+N$, and the variables $a_{[i,i]}$, $a_{[i,i+1]}$, and $a_{[1,2N]}$ are excluded. This is equivalent to the labeling of cluster variables by pairs of diagonals of a $2N$-gon; see Appendix~\ref{app: Structures of An and Bn} for details.

\vskip4pt
As in the $A_N$ case, we can obtain a mapping between energy variables and cluster variables by equating their $u$-variables as $ u(a_{[i,j]})=u(E_{[i,j-1]})$.
These equations take the explicit form 
\be
\begin{aligned}
\label{eq: u in E and A for Bn}
    u_{[i,j]}&=\frac{a_{[i,j-1]}^\epsilon a_{[i-1,j]}^\epsilon}{a_{[i-1,j-1]}a_{[i,j]}}z_{[i,j]}=\frac{E_{[i, j-1]}E_{[i-1, j]}}{E_{[i-1, j-1]}E_{[i,j]}} \\
    u_{[i,\bar i]}&=\frac{a_{[i,\bar i -1]}}{a_{[i-1,\bar i-1]}a_{[i,\bar i]}}z_{[i,\bar i]}=\frac{E_{[i, \bar i -1]}}{E_{[i,\bar i]}} \,,
\end{aligned}
\ee
where we should interpret $E_{[0,i]}=E_{[i,2N]}$ and $ a_{[0,i]}= a_{[i,2N]}$, for any $i$, along with $a_{[i,i+1]}=a_{[1,2N]}=1$. The variable $a_{[i,j]}^\epsilon$ is defined by
\begin{equation}
    a_{[k,l]}^\epsilon=\begin{cases}
        a_{[k,l]}^2& ~{\rm if}~l = \bar k \\
        a_{[k,l]} & ~\text{otherwise}
    \end{cases}\,.
\end{equation}
The equations~\eqref{eq: u in E and A for Bn} have a closed-form solution for $E$ variables in terms of cluster variables:
\begin{equation}
    \label{eq: E in a Bn}
    \frac{E_{[i, j]}}{E_{[i, \bar i]}}=\frac{a_{[i,j]}^\epsilon}{a_{[i,\bar{i}]} a_{[j,\bar{j}]}} G_z[i,j]\,,
\end{equation}
where $G_z$ is a monomial in the frozen variables given by
\begin{align}
    G_z[i,j]= 
\prod_{\substack{1\le k\le i\\[1pt] j+1\le \ell\le \bar k}}
z_{[k,\ell]}
\,
\prod_{\substack{j+1\le k\le N\\[1pt] k+N\le \ell\le \bar N}}
z_{[k,\ell]} \quad{\rm and}\quad
    G_z[i,\bar{j}]=\prod_{\substack{i+1\le k\le j\\[1pt] \bar k\le \ell\le \bar j}}
z_{[k,\ell]}\,,
\end{align}
with $G_z[i,\bar{i}]\equiv 1$.
It is also possible to invert this relation (or equivalently solve~\eqref{eq: u in E and A for Bn} for the cluster variables) for generic frozen variables $z_{[i,j]}$. If, however, we specialize to $z_{[i,j]}= 1$, the solution~\eqref{eq: E in a Bn} is unaffected, but its inverse is no longer well-defined. This is one of the reasons that the introduction of frozen cluster variables is necessary.

\paragraph{Example: $\boldsymbol{B_3}$}~\\
It is illuminating to spell out all the details for the simplest loop example, a two-site Feynman diagram, which corresponds to a four-site marked graph~\eqref{eq:marked2siteloop}. This graph can be related to the $B_3$ cluster algebra.
The energy singularities can be labeled as
\begin{equation}
\label{eq:B3-energy-singularities}
\begin{alignedat}{3}
E_{[1,2]} &= X_1+Y_{12}+Y_{21},
&\qquad\quad
E_{[2,3]} &= -2Y_{12},
&\qquad\quad
E_{[3,4]} &= X_2+Y_{12}+Y_{21},
\\
E_{[1,\bar 4]} &= -2Y_{21},
&
E_{[1,3]} &= X_1-Y_{12}+Y_{21},
&
E_{[2,\bar 4]} &= X_1+Y_{12}-Y_{21},
\\
E_{[2,4]} &= X_2-Y_{12}+Y_{21},
&
E_{[1,\bar 3]} &= X_2+Y_{12}-Y_{21},
&
E_{[1,4]} &= X_1+X_2+2Y_{21},
\\
E_{[3,\bar 4]} &= X_1-Y_{12}-Y_{21},
&
E_{[2,\bar 3]} &= X_1+X_2+2Y_{12},
&
E_{[1,\bar 2]} &= X_2-Y_{12}-Y_{21} \,,
\end{alignedat}
\end{equation}
while the
 total energy can be written in many equivalent ways, for example
 $E_T \equiv X_1+X_2
= E_{[1,\bar 1]}
= E_{[2,\bar 2]}
= E_{[3,\bar 3]}
= E_{[4,\bar 4]} $.
We want to relate these letters to $B_3$ cluster variables, which are
\begin{equation}
    \mathcal{A}_{B_3}=\left\{a_{[1,3]},a_{[1,4]},a_{[1,\bar{1}]},a_{[1,\bar{2}]},a_{[1,\bar{3}]},a_{[2,4]},a_{[2,\bar{2}]},a_{[2,\bar{3}]},a_{[2,\bar{4}]},a_{[3,\bar{4}]},a_{[3,\bar{3}]},a_{[4,\bar{4}]}\right\}\,.
\end{equation}
For each mutable variable, we also introduce a frozen variable $z_{[i,j]}$.

\vskip4pt
With these variables, 
the equations~\eqref{eq: u in E and A for Bn} take the explicit form
\begin{equation}
\label{eq:B3-u-equations}
\begin{alignedat}{2}
u_{[1,3]}
&=
\frac{z_{[1,3]}a_{[3,\bar 4]}}
     {a_{[1,3]}a_{[2,\bar 4]}}
=
\frac{E_{[1,2]}E_{[3,\bar 4]}}
     {E_{[1,3]}E_{[2,\bar 4]}}
&\qquad
u_{[1,4]}
&=
\frac{a_{[1,3]}z_{[1,4]}a_{[4,\bar 4]}^2}
     {a_{[1,4]}a_{[3,\bar 4]}}
=
\frac{E_{[1,3]}E_T}
     {E_{[1,4]}E_{[3,\bar 4]}}
\\[2pt]
u_{[1,\bar 1]}
&=
\frac{a_{[1,4]}z_{[1,\bar 1]}}
     {a_{[1,\bar 1]}a_{[4,\bar 4]}}
=
\frac{E_{[1,4]}}{E_T}
&\qquad
u_{[1,\bar 2]}
&=
\frac{a_{[2,4]}a_{[1,\bar 1]}^2z_{[1,\bar 2]}}
     {a_{[1,4]}a_{[1,\bar 2]}}
=
\frac{E_{[2,4]}E_T}
     {E_{[1,4]}E_{[1,\bar 2]}}
\\[2pt]
u_{[1,\bar 3]}
&=
\frac{a_{[1,\bar 2]}z_{[1,\bar 3]}}
     {a_{[2,4]}a_{[1,\bar 3]}}
=
\frac{E_{[3,4]}E_{[1,\bar 2]}}
     {E_{[2,4]}E_{[1,\bar 3]}}
&\qquad
u_{[2,4]}
&=
\frac{a_{[1,4]}z_{[2,4]}}
     {a_{[1,3]}a_{[2,4]}}
=
\frac{E_{[1,4]}E_{[2,3]}}
     {E_{[1,3]}E_{[2,4]}}
\\[2pt]
u_{[2,\bar 2]}
&=
\frac{a_{[1,\bar 2]}z_{[2,\bar 2]}}
     {a_{[1,\bar 1]}a_{[2,\bar 2]}}
=
\frac{E_{[1,\bar 2]}}{E_T}
&\qquad
u_{[2,\bar 3]}
&=
\frac{a_{[1,\bar 3]}a_{[2,\bar 2]}^2z_{[2,\bar 3]}}
     {a_{[1,\bar 2]}a_{[2,\bar 3]}}
=
\frac{E_T E_{[1,\bar 3]}}
     {E_{[1,\bar 2]}E_{[2,\bar 3]}}
\\[2pt]
u_{[2,\bar 4]}
&=
\frac{a_{[2,\bar 3]}z_{[2,\bar 4]}}
     {a_{[1,\bar 3]}a_{[2,\bar 4]}}
=
\frac{E_{[1,\bar 4]}E_{[2,\bar 3]}}
     {E_{[1,\bar 3]}E_{[2,\bar 4]}}
&\qquad
u_{[3,\bar 3]}
&=
\frac{a_{[2,\bar 3]}z_{[3,\bar 3]}}
     {a_{[2,\bar 2]}a_{[3,\bar 3]}}
=
\frac{E_{[2,\bar 3]}}{E_T}
\\[2pt]
u_{[3,\bar 4]}
&=
\frac{a_{[2,\bar 4]}a_{[3,\bar 3]}^2z_{[3,\bar 4]}}
     {a_{[2,\bar 3]}a_{[3,\bar 4]}}
=
\frac{E_T E_{[2,\bar 4]}}
     {E_{[2,\bar 3]}E_{[3,\bar 4]}}
&\qquad
u_{[4,\bar 4]}
&=
\frac{a_{[3,\bar 4]}z_{[4,\bar 4]}}
     {a_{[3,\bar 3]}a_{[4,\bar 4]}}
=
\frac{E_{[3,\bar 4]}}{E_T}.
\end{alignedat}
\end{equation}
These equations can be solved either for the energy variables in terms of cluster variables, or vice versa.  For example, we can solve for the ratio of the energy singularities to the total energy as
\begin{equation}
\label{eq:B3-inverse-map}
\begin{alignedat}{3}
\frac{E_{[1,2]}}{E_T}
&=
\frac{G_z[1,2]}
     {a_{[1,\bar1]}a_{[2,\bar2]}}
&\quad\qquad
\frac{E_{[2,3]}}{E_T}
&=
\frac{G_z[2,3]}
     {a_{[2,\bar2]}a_{[3,\bar3]}}
&\quad\qquad
\frac{E_{[3,4]}}{E_T}
&=
\frac{G_z[3,4]}
     {a_{[3,\bar3]}a_{[4,\bar4]}}
\\[2pt]
\frac{E_{[1,\bar4]}}{E_T}
&=
\frac{G_z[1,\bar4]}
     {a_{[1,\bar1]}a_{[4,\bar4]}}
&
\frac{E_{[1,3]}}{E_T}
&=
\frac{a_{[1,3]}G_z[1,3]}
     {a_{[1,\bar1]}a_{[3,\bar3]}}
&
\frac{E_{[2,\bar4]}}{E_T}
&=
\frac{a_{[2,\bar4]}G_z[2,\bar4]}
     {a_{[2,\bar2]}a_{[4,\bar4]}}
\\[2pt]
\frac{E_{[2,4]}}{E_T}
&=
\frac{a_{[2,4]}G_z[2,4]}
     {a_{[2,\bar2]}a_{[4,\bar4]}}
&
\frac{E_{[1,\bar3]}}{E_T}
&=
\frac{a_{[1,\bar3]}G_z[1,\bar3]}
     {a_{[1,\bar1]}a_{[3,\bar3]}}
&
\frac{E_{[1,4]}}{E_T}
&=
\frac{a_{[1,4]}G_z[1,4]}
     {a_{[1,\bar1]}a_{[4,\bar4]}}
\\[2pt]
\frac{E_{[3,\bar4]}}{E_T}
&=
\frac{a_{[3,\bar4]}G_z[3,\bar4]}
     {a_{[3,\bar3]}a_{[4,\bar4]}}
&
\frac{E_{[2,\bar3]}}{E_T}
&=
\frac{a_{[2,\bar3]}G_z[2,\bar3]}
     {a_{[2,\bar2]}a_{[3,\bar3]}}
&
\frac{E_{[1,\bar2]}}{E_T}
&=
\frac{a_{[1,\bar2]}G_z[1,\bar2]}
     {a_{[1,\bar1]}a_{[2,\bar2]}} \,.
\end{alignedat}
\end{equation}
Alternatively, the cluster variables can be written in terms of energies as
\begin{equation}
\begin{alignedat}{2}
    a_{[1,\bar{2}]}&=\frac{E_{[3,4]} z_{[1,4]} z_{[2,4]} E_{[1,\bar{2}]} z_{[2,\bar{4}]} z_{[3,\bar{4}]}
   z_{[4,\bar{4}]}^2}{E_{[2,3]} E_{[1,\bar{4}]} z_{[1,\bar{3}]}} 
    & ~ a_{[1,4]}&=\frac{E_{[1,4]} z_{[2,\bar{2}]} z_{[2,\bar{3}]} z_{[2,\bar{4}]} z_{[3,\bar{3}]} z_{[3,\bar{4}]}
   z_{[4,\bar{4}]}}{E_{[1,\bar{4}]} z_{[1,\bar{1}]}} \\
    a_{[1,\bar{1}]}&=\frac{E_{[3,4]} a_{[3,\bar{3}]} z_{[2,\bar{4}]} z_{[3,\bar{4}]} z_{[4,\bar{4}]}}{E_{[1,\bar{4}]} z_{[1,\bar{1}]} z_{[1,\bar{2}]} z_{[1,\bar{3}]}} & ~~\,  a_{[1,\bar{3}]}&=\frac{E_{[3,4]} a_{[3,\bar{3}]}^2 E_{[1,\bar{3}]} z_{[2,\bar{4}]} z_{[3,\bar{4}]} z_{[4,\bar{4}]}}{E_T E_{[1,\bar{4}]} z_{[1,\bar{1}]} z_{[1,\bar{2}]} z_{[1,\bar{3}]} z_{[2,\bar{2}]} z_{[2,\bar{3}]} z_{[3,\bar{3}]}} \\   
       a_{[1,3]}&=\frac{E_{[1,3]} E_{[3,4]} a_{[3,\bar{3}]}^2 z_{[2,\bar{4}]} z_{[3,\bar{4}]}}{z_{[1,4]} E_T E_{[1,\bar{4}]} z_{[1,\bar{1}]}^2 z_{[1,\bar{2}]} z_{[1,\bar{3}]}} 
   & 
    ~ a_{[2,4]}&=\frac{E_{[2,4]} z_{[1,4]} z_{[2,4]}
   E_T z_{[1,\bar{1}]} z_{[1,\bar{2}]} z_{[1,\bar{3}]} z_{[2,\bar{2}]} z_{[2,\bar{3}]} z_{[3,\bar{3}]} z_{[4,\bar{4}]}}{E_{[2,3]} E_{[3,4]} a_{[3,\bar{3}]}^2} \\
    a_{[2,\bar{2}]}&=\frac{z_{[1,4]} z_{[2,4]} E_T z_{[1,\bar{1}]} z_{[1,\bar{2}]} z_{[2,\bar{2}]} z_{[4,\bar{4}]}}{E_{[2,3]} a_{[3,\bar{3}]}} & 
    ~ a_{[2,\bar{3}]}&=\frac{z_{[1,4]} z_{[2,4]} E_{[2,\bar{3}]} z_{[1,\bar{1}]} z_{[1,\bar{2}]} z_{[2,\bar{2}]} z_{[4,\bar{4}]}}{E_{[2,3]}
   z_{[3,\bar{3}]}} \\
 a_{[3,\bar{4}]}&=\frac{E_{[3,\bar{4}]} z_{[1,\bar{1}]} z_{[1,\bar{2}]}
   z_{[1,\bar{3}]} z_{[2,\bar{2}]} z_{[2,\bar{3}]} z_{[3,\bar{3}]}}{E_{[3,4]} z_{[4,\bar{4}]}} &     a_{[2,\bar{4}]}&=\frac{z_{[1,4]} z_{[2,4]} E_T E_{[2,\bar{4}]} z_{[1,\bar{1}]}^2 z_{[1,\bar{2}]}^2 z_{[1,\bar{3}]} z_{[2,\bar{2}]}^2 z_{[2,\bar{3}]}}{E_{[2,3]} E_{[3,4]} a_{[3,\bar{3}]}^2 z_{[3,\bar{4}]}} \\
    a_{[4,\bar{4}]}&=\frac{E_T z_{[1,\bar{1}]} z_{[1,\bar{2}]} z_{[1,\bar{3}]} z_{[2,\bar{2}]} z_{[2,\bar{3}]} z_{[3,\bar{3}]}}{E_{[3,4]} a_{[3,\bar{3}]}} & ~ z_{[1,3]}&=\frac{E_{[1,2]} E_{[3,4]} z_{[2,4]} z_{[2,\bar{2}]} z_{[2,\bar{4}]} z_{[4,\bar{4}]}}{E_{[2,3]} E_{[1,\bar{4}]}
   z_{[1,\bar{1}]} z_{[1,\bar{3}]} z_{[3,\bar{3}]}} 
\end{alignedat}
\end{equation}
Here we see explicitly that this inverse map involves one of the frozen variables $z_{[1,3]}$, and $a_{[3,\bar 3]}$ is undetermined (and so can be fixed to any value). We see that in order to be able to solve for the cluster variables, we cannot set all of the frozen variables to $1$.

\newpage
\section{A Cluster Symbol Bootstrap}
\label{sec:clusterbootstrap}

The relations between energy singularities and cluster variables~\eqref{eq:AnEamap} and~\eqref{eq: E in a Bn} directly produce a notion of compatibility between symbol entries. 
The obvious question to investigate is whether this cluster compatibility is actually respected by cosmological wavefunctions/correlators. As a first check, we can use the map~\eqref{eq:EtoamapA2} to write the two-site wavefunction symbol~\eqref{eq:2sitesymbol} in $A_2$ cluster variables
\be
{\cal S}(F_{(2)}) =a_{14}\otimes \frac{1}{a_{13} a_{24}}+a_{25}\otimes
   \frac{a_{25}}{a_{24} a_{35}}+a_{35}\otimes
   \frac{a_{35}}{a_{13} a_{25}}\,.
\ee
We see that this symbol expression is manifestly cluster compatible. (Each $a_{ij}$ appearing in the right slot shares at least one index with the $a_{ij}$ appearing in the left slot.) This is particularly notable because the symbol in terms of $X$ and $Y$ energy variables~\eqref{eq:2sitesymbol} had some adjacent entries that corresponded to adjacent tubings, which are not compatible in the graph associahedron sense. 
To underscore this point, this happens because the notion of compatibility inherited from cluster compatibility is 
{\it different} from that of the tubings related to energy singularities.

\vskip4pt
It is natural to wonder whether cluster compatibility is strong enough to uniquely specify the answer. 
Indeed, this same question was studied in~\cite{Capuano:2026pgq,Paranjape:2026htn,Ferro:2026oph}.\footnote{Our analysis is complementary to these investigations.  In detail, the mapping between kinematics and cluster variables is different from~\cite{Capuano:2026pgq,Paranjape:2026htn} (in particular no energy variables are mapped to frozen variables). As a result cluster compatibility is more constraining in our case, so we do not need to impose either a discontinuity condition~\cite{Capuano:2026pgq} or vanishing as $Y\to 0$~\cite{Paranjape:2026htn}. The cluster structure utilized in~\cite{Ferro:2026oph} can be related to the one here via~\eqref{eq:qtou}. Their focus was on directly constructing the wavefunction in a manifestly cluster-compatible form, here we explore the uniqueness of the wavefunction and in-in correlators as cluster-compatible objects.} 
Here we also investigate this, finding that---combined with minimal physical conditions---cluster compatibility restricts the symbol to be one with a cosmological interpretation in the cases considered. An interesting feature in the chain case is that both the correlator and wavefunction are cluster compatible.

\subsection{Bootstrap Constraints}
\label{sec:bstrapconstraints}

Our goal is to reconstruct the symbol of cosmological observables from some set of requirements. We  can parameterize the most general symbol in the form 
\begin{equation}
    \label{eq: generic symbol in energies Sec 4}
    {\cal S}( F_{\cal G})=\sum_{E_{I_1},\cdots, E_{I_n} \in\, \mathcal{E}_{\cal G}} C_{I_1 \cdots I_n}\, E_{I_1}\otimes E_{I_2}\otimes \cdots \otimes E_{I_n}\,,
\end{equation}
where the sum runs over the energy singularities $E_I$ associated to the marked graph ${\cal G}$ (described in Section~\ref{sec:singularities}), and $I$ is a label that indexes them.
The coefficients $C_{I_1\cdots I_n}$ then specify the symbol, and we can rephrase the goal as being to determine these coefficients. Remarkably this can be done with relatively few requirements.
Concretely, we impose the following conditions:
\begin{itemize}

\item
{\bf Mathematics:} A basic constraint is that the symbol is integrable to an actual function:
\begin{enumerate}
    \item The symbol must be integrable.
\end{enumerate}
Since the symbol encodes the differentials of the function, the requirement of integrability is essentially that mixed partial derivatives commute.
This translates to constraints on the various symbol entries, as we describe more fully in Appendix~\ref{app: Symbology}. Therefore, the first constraint we impose is the mathematical one of integrability.

\item
{\bf Physics:} Aside from purely mathematical considerations, there are some physical properties that we require the symbol to satisfy.

The first is a version of the requirement that the symbol be expressible in terms of dimensionless ratios of the energy singularities. In particular, by deforming the flat-space wavefunction integrand into the complex plane and integrating around a small loop that contains no singularities, we find that the sum of the discontinuities of the wavefunction coefficient must vanish~\cite{Arkani-Hamed:2017fdk,Benincasa:2018ssx,Hillman:2019wgh}. We therefore impose:
\begin{enumerate}
    \setcounter{enumi}{1}
    \item The sum of the leading discontinuities of the wavefunction coefficient must vanish. 
\end{enumerate}
As a practical matter, we utilize the fact that the discontinuity across the total energy singularity is minus the sum of all other discontinuities, so that the letters in the first entry can be written as dimensionless ratios with respect to the total energy.

We also impose a first-entry condition by forbidding folded singularities on the physical sheet, since these singularities are absent for adiabatic vacuum initial conditions:
\begin{enumerate}
    \setcounter{enumi}{2}
    \item The first letter in a symbol word can only be a physical singularity.
\end{enumerate}
This means that folded singularities and  internal energies $Y$ do not appear by themselves in the first slot.

Finally we impose a constraint on the pattern of energy singularities that can be thought of as philosophically capturing part of Steinmann relations~\cite{Benincasa:2020aoj,Benincasa:2021qcb} or the kinematic flow satisfied by the wavefunction/correlators~\cite{Arkani-Hamed:2023kig}. The simplest version that we can impose is 
\begin{enumerate}
    \setcounter{enumi}{3}
    \item The energy singularities in a symbol word do not repeat.
\end{enumerate}
This means, for example, if $E_1 \,\otimes\, E_2 \,\otimes \cdots \otimes\, E_n$ is a term in the symbol, then no pairs of the $E_a$ are the same.
This is in a sense the minimal constraint that we could impose on compatibilities of energies. See Appendix~\ref{app: Steinmann} for more details.

\item
{\bf Cluster Algebra Constraint:} Interestingly, the only cluster-algebraic constraint that we need to impose is the obvious one:
\begin{enumerate}
    \setcounter{enumi}{4}
    \item The cluster variables in a symbol word must all be mutually compatible.
\end{enumerate}
This requires that the entries appearing within a symbol word must appear in at least one cluster with each of the other entries.\footnote{A similar condition is imposed by~\cite{Capuano:2026pgq}, albeit with a smaller $A_{2n-3}$ algebra, and by~\cite{Ferro:2026oph}, though with a slightly different cluster mapping.}

\end{itemize}

In the following we will impose each of these constraints on a generic ansatz for the symbol. Remarkably we will find that they are sufficient to completely fix the symbol to be that of a physical object, in the cases where a cluster structure is present. In the chain case, for sufficiently many sites these constraints admit multiple solutions, which have the interpretation as being the symbol of either the wavefunction or of products of lower-point wavefunctions, with a particular linear combination corresponding to the correlator.

\subsection{Implementation}
\label{sec:symbolconstraints}

Here we describe the systematics of implementing the bootstrap constraints, and then describe some explicit examples to illustrate the procedure.

\paragraph{Integrability}~\\
For a symbol to be integrable to a parent function, it must obey~\eqref{eq: Symbol integrability condition}. Expressed in terms of  energies, this condition is 
\begin{equation}
    \label{eq: Integrability condition on energy symbol}
    \sum_{E_{I_1},\cdots,E_{I_n} \in \mathcal{E}_{\cal G}}  C_{I_1\ldots I_n} \, E_{I_1}\otimes \cdots \otimes E_{I_{p-1}} \otimes E_{I_{p+2}}\otimes\cdots \otimes E_{I_n} \, \rd \log E_{I_p} \wedge  \rd \log E_{I_{p+1}}=0\,,
\end{equation}
for all $1\leq p\leq n-1$.
One can directly calculate~\eqref{eq: Integrability condition on energy symbol} and infer the conditions $C_{I_1\cdots I_n}$ must obey to make the expression vanish. To simplify this, we can note that integrability only really constrains adjacent entries in the symbol, so ultimately we just need to solve
\begin{equation}
    \label{eq: short energy int condition}
    \sum_{E_{I_p},\, E_{I_{p+1}} \in \,\mathcal{E}_{\cal G}}  \hat C_{I_p\,I_{p+1}} \, \rd \log E_{I_p} \wedge  \rd \log E_{I_{p+1}}=0\,,
\end{equation}
which we can then apply to adjacent entries in the large tensor $C_{I_1\cdots I_n}$.
For the graphs of interest, this equation implies the following conditions. Let energies $E_J$ and $E_K$ be associated to the tubes $T_J$ and $T_K$.
\begin{itemize}

\item If $T_J$ and $T_K$ intersect or are disjoint and not adjacent, integrability imposes
\be
\hat C_{JK} = \hat C_{KJ}\,.
\ee

\item Consider $T_K\subset T_J$ (meaning one tube is a subset of the other). The integrability condition depends on the complement of $T_K$ in $T_J$, which we denote as $\bar T_J$. If the complement is connected, i.e., $\bar T_J \in \mathcal{T}_{\cal G}$, then integrability imposes
\begin{equation}\label{eq: subset integrability condition}
  \hat C_{JK}- \hat C_{KJ}=\hat C_{K \bar{J}}-\hat C_{\bar{J} K}\,,
\end{equation}
where $\bar J$ is the index of the tube enclosing the complement $\bar{T}_J=T_{\bar J}$. If the complement is not connected, then integrability imposes
\begin{equation}
  \hat C_{JK}=\hat C_{KJ}\,.
\end{equation}

\item Tubes that are disjoint and adjacent do not generate an integrability condition, but their coefficients appear on the right-hand side of~\eqref{eq: subset integrability condition}. In the case that $T_J=T_K$, the above relations become trivial. 

\end{itemize}
Coefficients which satisfy all of these conditions satisfy~\eqref{eq: short energy int condition}. To extend these conditions to the coefficients found in~\eqref{eq: Integrability condition on energy symbol}, we simply send 
\begin{equation}
    \hat C_{JK} \longrightarrow\, C_{I_1\cdots I_{p-1}\, JK\, I_{p+2}\cdots I_n}\quad{\rm for~all}~~   1\leq p\leq n-1\,,
\end{equation}
in the above equations.

\paragraph{Physical Constraints}~\\
The physical constraints are relatively straightforward to enforce in terms of the coefficients in~\eqref{eq: generic symbol in energies Sec 4}. First,   we forbid folded singularities in the first slot. More formally, if we denote the set of folded singularities as $\{E_F\}$, then we set
\be
C_{J\,I_2\cdots I_n} = 0\qquad{\rm for}~~E_J\in\{E_F\}\,.
\ee
The discontinuity condition is enforced by requiring the total energy discontinuity to be equal to minus the sum over all other legal discontinuities\footnote{This can be satisfied by setting any of the legal discontinuities to minus the sum of the rest. The total energy is simply a convenient choice.}
\begin{equation}
    \sum_{I_1\neq I_T} C_{I_1\cdots I_n}=-C_{I_TI_2\cdots I_n}\,.
\end{equation}
The no repeated energy condition can be written as
\begin{equation}
    C_{I_1 \cdots J\cdots J\cdots I_n}=0,\quad{\rm if~any~indices~match} \,.
\end{equation}
We now turn to the implementation of cluster compatibility.

\paragraph{Cluster Algebra Constraints}~\\
The most nontrivial constraints come from requiring that the symbol can be written in terms of cluster variables in such a way that
term-by-term the letters appearing are mutually compatible.
Using~\eqref{eq: E in a An} and~\eqref{eq: E in a Bn} we can map the symbol in energy variables~\eqref{eq: generic symbol in energies Sec 4} to an analogous expression in cluster variables:
\begin{equation}
    \label{eq: generic symbol in cluster variables}
    {\cal S}( F_{\cal G})=\sum_{a_{I_1},\cdots, a_{I_n} \in\, \mathcal{A}_{\cal G}} f(C)_{I_1\cdots I_n}\, a_{I_1}\otimes a_{I_2}\otimes \cdots \otimes a_{I_n}\,,
\end{equation}
where the coefficients $f(C)_{I_1\cdots I_n}$ appearing here are linear combinations of the coefficients in~\eqref{eq: generic symbol in energies Sec 4}. To calculate these sums explicitly, one can use the relations in Section~\ref{sec: E to A map} to write the energy singularities in terms of cluster variables. 
In the cases of interest, we can write this as
\begin{equation}
  E_{ij}=\frac{a_{ij}}{\tilde{a}_i \, \tilde{a}_j}\,,
\end{equation}
where $\tilde{a}_i\equiv a_{i\, N+3}$ for $A_N$ and $\tilde{a}_i\equiv a_{[i,\bar{i}]}$ for $B_N$. This is equivalent to~\eqref{eq: E in a An} and~\eqref{eq: E in a Bn} upon multiplying both sides by the total energy and setting frozen variables to 1. 
After this, we can use the properties of the symbol~\eqref{eq: symbol sum rule} to expand, until each symbol letter is a single cluster variable.

\vskip4pt
It is useful to notice that the un-tilded cluster variables $a_{ij}$ appear only if $E_{ij}$ does in the word of interest. On the other hand, $\tilde{a}_i$ appears in the expression for $E_{ik}$, for any value $k$. We can therefore abstractly write
\begin{equation}
  f(C)_{I_1\cdots I_n} =\sum_{E_{J_1},\cdots, E_{J_n} \in \,\mathcal{E}_{\cal G}} C_{J_1\cdots J_n} \prod_{k=1}^n \sigma_{I_k \, J_k}\,,
\end{equation}
where we have defined 
\begin{equation}
  \sigma_{I_k\, J_k}=
  \begin{cases}
    1 & {\rm if}~I_k= J_k \\
    -1 & {\rm if}~ I_k \text{ and } J_k \text{ share one index (with other index $N+3$ [$A_N$] or $i+N+1$ [$B_N$])}\\
    0 & \text{otherwise}
  \end{cases}\,.
\end{equation}
Recall that $I,J$ are multi-indices, so that e.g., $I = (i,j)$.
In terms of these coefficients, cluster compatibility is simply the condition that $ f(C)_{I_1\cdots I_n}=0$ if $a_{I_j}$ and $a_{I_k}$ are not compatible.

\subsubsection{Chains}

The constraints are simplest to implement on chain graphs. An (unmarked) chain with $n$ sites corresponds to a transcendentality-$n$ function, with a symbol alphabet that can be enumerated by tubing its marked $N= 2n-1$ site counterpart. We can parameterize the most general possible symbol and then impose the conditions described in Section~\ref{sec:symbolconstraints}. It is easiest to illustrate the procedure via explicit examples.

\paragraph{Two-Site Chain}~\\
The simplest example is provided by the two-site chain, which corresponds to a single exchange, leading to a four-point function. The corresponding wavefunction/correlator has transcendentality two, so the most general symbol is of the form
\begin{equation}
    \label{eq: generic symbol two site chain}
    {\cal S}( F_{(2)})=\sum_{E_{I_1},\, E_{I_2} \in\, \mathcal{E}_{\cal G}} C_{I_1I_2}\, E_{I_1}\otimes E_{I_2}\,,
\end{equation}
where the possible singularity alphabet can be obtained from the marked graph, as in~\eqref{eq:2sitesingularities}. It is convenient to label the energy singularities as
\be
\left\{E_{12},E_{23},E_{34},E_{13},E_{24},E_{14}\right\} \equiv \{E_1,E_2,E_3,E_4,E_5,E_6\} \,.
\ee
This choice has no intrinsic significance; it is just to simplify notation.

\vskip4pt
The ansatz~\eqref{eq: generic symbol two site chain} has $36$ free coefficients that we would like to determine.
As a first step, we can impose integrability~\eqref{eq: Integrability condition on energy symbol}, which implies
\be
\begin{aligned}
C_{12}-C_{21}
&=C_{24}-C_{42}=C_{41}-C_{14}
&
C_{13}-C_{31}&=0
\\
C_{23}-C_{32}
&=C_{35}-C_{53}=C_{52}-C_{25}
&\hspace{2cm}
C_{45}-C_{54}&=0
\\
C_{34}-C_{43}
&=C_{46}-C_{64}=C_{63}-C_{36}
&
C_{26}-C_{62}&=0
\\
C_{15}-C_{51}
&=C_{56}-C_{65}=C_{61}-C_{16}\,. &
\end{aligned}
\ee
In addition, imposing the first entry condition that there are no folded singularities, demanding that the sum of all first entry discontinuities vanishes, and forbidding the appearance of repeated energy entries imply the following conditions
\be
\begin{aligned}
    C_{2J}&=C_{4J}=C_{5J}=0, &\hspace{1cm} &{\rm for~all}~J\\
    C_{6J}&=-C_{1J}-C_{3J}, & &{\rm for~all}~J\\
    C_{II}&=0,& &{\rm for~all}~I\,.
\end{aligned}
\ee
\vskip4pt
Solving this system of equations, we find that the most general symbol consistent with integrability and the other physical requirements we want to impose is the two-parameter family
\be
\begin{aligned}
 {\cal S}( F_{(2)}) =~&  c_1 \left(\frac{X_1+Y}{X_1+X_2}\otimes \frac{X_1+X_2}{X_2-Y}+\frac{X_2+Y}{X_1+X_2}\otimes \frac{X_1-Y}{X_1+X_2}\right)\\
    &+c_2 \left(\frac{X_1+Y}{X_1+X_2}\otimes \frac{X_2-Y}{X_2+Y}+\frac{X_2+Y}{X_1+X_2}\otimes \frac{X_1-Y}{X_1+Y}\right) \,,
\end{aligned}
\label{eq:symbol2siteansatzXY}
\ee
where the precise relation between $c_1,c_2$ and the $C_{IJ}$ coefficients depends on how we choose to eliminate parameters to solve the equations.

\vskip4pt
We now want to impose a cluster algebra structure on the symbol. To do this, we use the map~\eqref{eq: E in a An} for the $A_2$ case (which takes the explicit form~\eqref{eq:EtoamapA2}), so that we have
\begin{align}
  {\cal S}( F_{(2)}) &=c_1\left(a_{25}\, a_{14}\otimes \frac{a_{24}}{a_{14} a_{25}}+a_{35}\, a_{14}\otimes \frac{a_{14} a_{35}}{a_{13}}\right)\\
  &+c_2 \left(a_{25}\, a_{14}\otimes \frac{a_{25}}{a_{24} a_{35}}+a_{35}\, a_{14}\otimes
   \frac{a_{35}}{a_{13} a_{25}}\right) \,. \nonumber
\end{align}

Notice that the expression proportional to $c_1$---when fully expanded out---contains terms like $a_{25}\otimes a_{14}$ and $a_{35}\otimes a_{14}$, which are not cluster compatible. Therefore, cluster compatibility of the letters in the symbol forces us to set $c_1 = 0$. There is therefore a unique (up to overall rescaling) consistent symbol. 
In terms of cluster variables it can be written in the manifestly cluster compatible form
\be
{\cal S}(F_{{(2)}}) =a_{14}\otimes \frac{1}{a_{13} a_{24}}+a_{25}\otimes
   \frac{a_{25}}{a_{24} a_{35}}+a_{35}\otimes
   \frac{a_{35}}{a_{13} a_{25}}\,.
   \label{eq:twosite}
\ee
(In energy variables, it is given by the term proportional to $c_2$ in~\eqref{eq:symbol2siteansatzXY}.) If desired, we can fix the normalization by requiring any of the discontinuities to be normalized correctly~\cite{Arkani-Hamed:2017fdk,Arkani-Hamed:2018kmz,Hillman:2019wgh,Baumann:2020dch}.

\vskip4pt
It is interesting that there is a {\it unique} possibility for the symbol that is compatible with all the conditions that we have imposed. Note that cluster compatibility played an important role. Without imposing cluster compatibility there are two possibilities.\footnote{Here the cluster algebra constraints only remove one parameter's worth of freedom, but this is an artifact of the smallness of the two-site case. In more complicated examples they are substantially more powerful.} A further notable feature is that there are {\it two} physical objects that satisfy all the criteria that we have imposed: the two-site wavefunction and the two-site correlator.\footnote{The functions themselves differ by some overall factors of $k_a$ that come from power spectra that appear in the definition of the correlator, but which we have  absorbed into the definition of $F$.} As can be verified by direct computation, these objects share the same symbol, and only differ by pieces of lower transcendentality (in this case a multiple of the zeta value $\pi^2/6$, see Appendix~\ref{app: Correlator vs WF} for explicit formulas). We can select one or the other by imposing appropriate boundary conditions  to integrate the symbol. A convenient boundary condition in the wavefunction case is the vanishing of the function as $Y\to 0$~\cite{Hillman:2019wgh,Paranjape:2026htn}.

\paragraph{Three-Site Chain}~\\
As another example, we describe the bootstrap of the three-site chain. In this case the symbol has transcendentality three. 
Its alphabet  can be enumerated by considering the tubes of the marked Feynman graph, which
has five sites~\eqref{eq:3sitemarked}.
 We can then make an ansatz for the most general possible symbol and then impose the conditions from Section~\ref{sec:bstrapconstraints} on the energy singularities.
 This produces a $63$-parameter family of possible consistent symbols.
 After this, we express the symbol in cluster variables using~\eqref{eq: E in a An}. In this case the relevant cluster algebra is $A_4$, which has $14$ cluster variables arranged into $42$ clusters of size $4$. Demanding that the symbol respects cluster compatibility between all entries in a given word, we again find the symbol is unique:
\be
\begin{aligned}
    \mathcal{S}(F_{(3)})=~&\frac{1}{a_{16} a_{27}} \otimes \left(\frac{a_{26}}{a_{24} a_{36}}\otimes \frac{a_{47}}{a_{46} a_{57}}-\frac{a_{27} a_{36}}{a_{26} a_{37}}\otimes \frac{a_{24} a_{57}}{a_{25} a_{47}}+\frac{a_{37}}{a_{36} a_{57}}\otimes
   \frac{a_{25}}{a_{24} a_{35}}\right)\\
   &+\frac{1}{a_{16} a_{37} a_{47}}\otimes \left(\frac{a_{37}}{a_{13} a_{27}}\otimes \frac{a_{47}}{a_{46} a_{57}}+\frac{a_{47}}{a_{46} a_{57}}\otimes \frac{a_{37}}{a_{13} a_{27}}\right) \\
   &+\frac{1}{a_{16} a_{57}}\otimes \left(\frac{a_{15}}{a_{14} a_{35}}\otimes \frac{a_{37}}{a_{13} a_{27}}+\frac{a_{47}}{a_{14} a_{27}}\otimes \frac{a_{25}}{a_{24} a_{35}}-\frac{a_{14} a_{57}}{a_{15} a_{47}}\otimes \frac{a_{27}
   a_{35}}{a_{25} a_{37}}\right) \\
   &-\frac{a_{14}}{a_{16} a_{47}}\otimes \left(\frac{1}{a_{13} a_{24}}\otimes \frac{a_{47}}{a_{46} a_{57}}+\frac{a_{47}}{a_{46} a_{57}}\otimes \frac{1}{a_{13} a_{24}}\right)\\
   &-\frac{a_{36}}{a_{16} a_{37}} \otimes \left(\frac{a_{37}}{a_{13} a_{27}}\otimes \frac{1}{a_{35} a_{46}}+\frac{1}{a_{35} a_{46}}\otimes \frac{a_{37}}{a_{13} a_{27}}\right)\,.
\end{aligned}
\label{eq:3sitesymbolcluster}
\ee
In this case the cluster compatibility conditions greatly reduce the number of free parameters. An interesting feature of~\eqref{eq:3sitesymbolcluster} is that when it is written back in terms of $X$ and $Y$ variables, it corresponds to a real function when its kinematic arguments are real. Recall from~\eqref{eq:Fdef} that the actual wavefunction is given by (up to factors of $Y$) $\psi_3 \sim iF_3$. Since the most transcendental part of the three-site correlator is proportional to ${\rm Re}[\psi_3]$ (see Appendix~\ref{app: Correlator vs WF}), the correlator has transcendentality two in the physical region~\cite{Chowdhury:2023arc,Glew:2025arc,Arkani-Hamed:2025mce,Chowdhury:2026dwm}.\footnote{Note, however, that for complexified kinematics the correlator will have the same transcendentality as the wavefunction, and its leading symbol will coincide with that of the wavefunction.}
This feature is shared by all graphs with an odd number of vertices~\cite{Chowdhury:2026dwm}.
It would be interesting to understand in more detail whether the $A_4$ cluster structure of the three-site wavefunction is represented in this lower-transcendentality object, or whether just the cluster structure of its lower-point constituents plays a role.

\paragraph{Four Sites and Beyond}~\\
The generalization to Feynman graphs with more sites has the same general features. There is, however, an interesting new phenomenon that arises starting at four sites; there are {\it multiple} cluster-compatible solutions with leading transcendentality.

\vskip4pt
An $n$-site Feynman graph generates a wavefunction of transcendentality $n$. As before we can make an ansatz for the most general symbol of length $n$ and construct its alphabet by considering the corresponding $N$-site marked graph. After imposing integrability and physical conditions, we finally require that the symbol be compatible with an $A_{2n-2}$ cluster algebra.

\vskip4pt
The presence of multiple solutions to these constraints is a reflection of the fact that
wavefunction coefficients and in-in correlators have {\it different} symbols starting at four sites, and the freedom is precisely to add disconnected products of lower-point functions. In order for the disconnected product to have the same transcendentality, it is necessary for the graph to be splittable into two pieces that are themselves at least two-site chains. (The one-site chain contributes $\pi$ to the correlator and so is projected out of the leading symbol.) It is first at four sites that we can disconnect an internal line and have it be a product of two pieces whose transcendentalities sum up to that of the original graph. See Appendix~\ref{app: Correlator vs WF} for more details.

\vskip4pt
It is interesting that the wavefunction and correlator respect the {\it same} cluster algebra, not least because the singularity alphabet of correlators is strictly smaller than that of wavefunctions~\cite{Chowdhury:2023arc,Benincasa:2024ptf,Glew:2025arc,Arkani-Hamed:2025mce,Pimentel:2026kqc,Chowdhury:2026dwm}.
In the full space of cluster-compatible functions, the wavefunction and in-in correlator are of particular interest, so it would be convenient to have a prescription to isolate them.
There are several ways that one can isolate the wavefunction. One possibility is to require that the various singularities are normalized correctly~\cite{Arkani-Hamed:2023kig}. However, it is often more convenient to require that the wavefunction vanishes when we take any of the internal energies to vanish $Y \to 0$~\cite{Paranjape:2026htn}. Interestingly, there is a purely cluster-algebraic way to isolate the wavefunction. It turns out that wavefunction coefficients have the (semi-mysterious) property that when written in cluster variables, only variables appearing in the first slot ever appear elsewhere, and they only ever reappear once. In contrast, any letters appearing for the first time in the second entry onwards never repeat. The wavefunction is the unique object with this property. Mechanically, correlators do not satisfy this because they contain products of lower-point functions, which lead to a different pattern of repeating variables. It would be nice to have a similar diagnostic for the in-in correlator.

\vskip4pt
Following this procedure one can in principle continue to arbitrarily long chains. 
We have checked directly up to five sites that the full space of cluster compatible solutions corresponds to the wavefunction, in-in correlator (at general kinematics), and products of disconnected lower-point functions.\footnote{As noted earlier, for real kinematics, the correlator is lower transcendentality for odd $n$.} Additionally, one can prove directly, using the recursion relation derived in~\cite{Hillman:2019wgh}, that wavefunction coefficients are cluster compatible for arbitrarily many sites (see also~\cite{Ferro:2026oph}).\footnote{As we explain in Appendix~\ref{app: Correlator vs WF}, the addition of disconnected pieces will preserve compatibility, but we have not characterized the full space of compatible objects for arbitrary length chains.}

\subsubsection{Loops}

We can also consider cosmological observables arising from loop diagrams. We will particularly be interested in one-loop graphs with an arbitrary number of interaction vertices. In this case, an $n$-site Feynman graph leads to a symbol governed by a $B_{2n-1}$ cluster algebra. We follow the same procedure to bootstrap this symbol from general principles.

\paragraph{Two-Site Loop}~\\
The simplest example we can consider is provided by the two-site loop. The corresponding marked graph is of the form~\eqref{eq:marked2siteloop}. The possible tubes of this graph correspond to the singularity alphabet
\begin{align}
    \mathcal{E}_{\cal G}&=\left\{E_{[1,2]},E_{[2,3]},E_{[3,4]},E_{[1,\bar{4}]},E_{[1,3]},E_{[2,\bar{4}]},E_{[2,4]},E_{[1,\bar{3}]},E_{[1,4]},E_{[3,\bar{4}]},E_{[2,\bar{3}]},E_{[1,\bar{2}]},E_T\right\}\\
    &=\Big\{X_1+Y_{12}+Y_{21},-2 Y_{12},X_2+Y_{12}+Y_{21},-2 Y_{21}, X_1-Y_{12}+Y_{21},X_1+Y_{12}-Y_{21},X_2-Y_{12}+Y_{21}, \nonumber\\
    &\quad \quad X_2+Y_{12}-Y_{21},X_1+X_2+2 Y_{21},X_1-Y_{12}-Y_{21},X_1+X_2+2 Y_{12},X_2-Y_{12}-Y_{21},X_1+X_2 \Big\} \nonumber \,,
\end{align}
where we have labeled the energy singularities as in Section~\ref{sec:aXYu}.

\vskip4pt
As in the chain case, we can parameterize the most general symbol---which has transcendentality two---in the form~\eqref{eq: generic symbol two site chain} and impose all of the conditions, including compatibility with $B_3$ cluster structure, using~\eqref{eq: E in a Bn}. We find a unique symbol, which can be written in energy variables
\be
{\cal S}(F_{\rm loop}^{(2)}) = \frac{E_{[1,2]}}{E_T}\otimes \frac{E_{[3,4]} E_{[1,\bar{2}]}}{E_{[2,4]} E_{[1,\bar{3}]}}+\frac{E_{[2,\bar{3}]}}{E_T}\otimes \frac{E_{[1,\bar{3}]} E_{[2,\bar{4}]}}{E_{[1,2]} E_{[3,4]}}+\frac{E_{[3,4]}}{E_T}\otimes \frac{E_{[1,2]} E_{[3,\bar{4}]}}{E_{[1,3]} E_{[2,\bar{4}]}}+\frac{E_{[1,4]}}{E_T}\otimes \frac{E_{[1,3]} 
   E_{[2,4]}}{E_{[1,2]} E_{[3,4]}}\,.
\ee
We can check that this object is cluster compatible by expressing it in terms of $B_3$ cluster variables
\begin{align}
{\cal S}(F_{\rm loop}^{(2)}) =&\,
\frac{a_{[1,4]} z_{[1,\bar{1}]}}{a_{[1,\bar{1}]} a_{[4,\bar{4}]}}\otimes \frac{a_{[1,3]} a_{[2,4]}}{z_{[1,3]} z_{[1,\bar{3}]} z_{[2,\bar{3}]} z_{[3,\bar{3}]}^2 z_{[3,\bar{4}]}}+\frac{a_{[2,\bar{3}]} z_{[3,\bar{3}]}}{a_{[2,\bar{2}]} a_{[3,\bar{3}]}}\otimes \frac{a_{[1,\bar{3}]} a_{[2,\bar{4}]}}{z_{[1,3]} z_{[1,4]} z_{[1,\bar{1}]}^2 z_{[1,\bar{2}]} z_{[1,\bar{3}]}} \\[2pt]
  &+\frac{z_{[1,\bar{1}]} z_{[1,\bar{2}]} z_{[1,\bar{3}]} z_{[2,\bar{2}]} z_{[2,\bar{3}]} z_{[3,\bar{3}]}}{a_{[3,\bar{3}]} a_{[4,\bar{4}]}}\otimes \frac{z_{[1,3]} a_{[3,\bar{4}]}}{a_{[1,3]} a_{[2,\bar{4}]}}+\frac{z_{[1,3]} z_{[1,4]} z_{[1,\bar{1}]} z_{[3,\bar{3}]} z_{[3,\bar{4}]} z_{[4,\bar{4}]}}{a_{[1,\bar{1}]} a_{[2,\bar{2}]}}\otimes \frac{a_{[1,\bar{2}]} z_{[1,\bar{3}]}}{a_{[2,4]} a_{[1,\bar{3}]}}\nonumber \,.
\end{align}
In contrast to the two-site chain, the wavefunction and the correlator have different symbols in this case, and it is the symbol of the {\it wavefunction} that is cluster compatible.
In detail, this happens because the additional disconnected pieces included in the correlator are not compatible in the sense of $B_3$ (see Appendix~\ref{app: Correlator vs WF}). The main difference with the chain case is that for chains the various disconnected pieces are each compatible with respect to some $A_{N'}$ subalgebra of the $A_N$ corresponding to the full chain, but this does not happen for the loop case.

\paragraph{Loops with More Sites}~\\
We can also consider cycle graphs with additional sites and repeat the procedure. Conceptually, nothing changes, an $n$-site Feynman graph wavefunction coefficient has a symbol that is fixed by a $B_{2n-1}$ cluster algebra structure, which one can again prove directly by recursion. In all the cases we checked, there is a unique such cluster compatible object (the wavefunction), consistent with the fact that subgraphs do not have a subalgebra cluster structure (see Appendix~\ref{app: Correlator vs WF}).

\paragraph{Banana/Melon Graphs}~\\
There is an interesting class of loop diagrams with a simple cluster-algebraic structure at the level of the correlator: two-site banana/melonic graphs. These are $\ell$-loop diagrams of the form\footnote{Aside from the two-site one-loop case, these diagrams do not arise in the computation of correlators for $\phi^3$ theory in $D=4$ de Sitter space, but by considering other power-law cosmologies, dimensions, and interaction vertices so that the effective coupling is $\eta^{-1}$, they can appear. }
\be
\raisebox{-30pt}{
\begin{tikzpicture}[
    vertex/.style={circle, fill=black, inner sep=1.8pt},
    prop/.style={line width=0.95pt},scale=.75
]
    \coordinate (L) at (-1.5,0);
    \coordinate (R) at ( 1.5,0);
    \draw[prop] (L) arc[start angle=180, end angle=0, radius=1.5];
    \draw[prop] (L) arc[start angle=180, end angle=360, radius=1.5];
    \draw[prop] (L) to[out=55,  in=125] (R);
    \draw[prop] (L) to[out=-55, in=-125] (R);
    \node at (0,.1) {$\boldsymbol\vdots$};
    \node[vertex] at (L) {};
    \node[vertex] at (R) {};
\end{tikzpicture}
}
\ee
The novelty of these graphs is that the time integrand for the correlator takes the form of a two-site {\it chain} graph, albeit with shifted energy variables~\cite{Donath:2024utn}. Concretely we can replace the energies of the internal lines $Y_1, Y_2,\cdots, Y_{\ell+1}$ with $Y_T \equiv Y_1+Y_2+\cdots +Y_{\ell+1}$ so that
\be
\langle F_{\rm melon}^{(\ell)}(X_1,X_2,Y_1,\cdots)\rangle =  \frac{1}{2^{\ell+1} Y_1 Y_2\cdots Y_{\ell+1}}\langle F_2(X_1,X_2,Y_T)\rangle \,,
\ee
where we have written $\langle F\rangle$ to emphasize that this is the correlator. Since these correlators can be written in terms of a two-site chain with an effective internal energy $Y_T$, they therefore have an $A_2$ cluster algebra structure. This is just~\eqref{eq:twosite}, where the cluster variables correspond to $X$ and $Y$ variables as in~\eqref{eq:atoEexplicit}, where the $E$ variables are expressed in terms of $X_1,X_2$ and $Y_T$.

\newpage
\section{Conclusion}
\label{sec:conclusion}

Correlation functions in de Sitter space display a remarkable amount of rich and interesting mathematical structure. 
The nontrivial features of these functions are reflections of the physics that generated cosmological correlations.
The challenge is to understand the encoding of fundamental principles in the properties of the correlators. In this regard, it is essential to better understand the mathematical features that shape observables in simplified settings.
In this paper, we have explored the cluster-algebraic properties of one such simplified setting (that of chain and cycle graphs of conformal scalars in de Sitter space).
In this example, the function space is sufficiently simple (generalized polylogs) that many of its features are captured by the symbol.
The locations of singularities and their sequences of compatibilities encode aspects of locality and causality (e.g., via Steinmann relations), and so the fact that the symbol's structure is controlled by a cluster algebra in this setting is remarkable.

\vskip4pt
In order to make a connection between the kinematic variables that the cosmological wavefunction depends on and cluster algebras, we utilized the fact that each of these produces a realization of $u$-variables. Since $u$-variables parameterize a somewhat rigid geometry, it is natural to identify them, which provides a bridge. In the cases of chain and cycle graphs, the binary geometry coincides with that of either an $A_N$ or $B_N$ type cluster algebra, respectively.

\vskip4pt
The mapping between kinematic variables and cluster variables imbues the singularities of the cosmological wavefunction with a nontrivial notion of compatibility. The obvious question is whether physical objects are uniquely determined by this compatibility. In Section~\ref{sec:clusterbootstrap} we investigated this question by bootstrapping the symbol of the wavefunction and correlators from simple physics principles supplemented with cluster compatibility. The physical conditions that we imposed were fairly minimal: in addition to requiring integrability, we restricted the first entry to only contain physical (non-folded) singularities, and forbade repeating singularities in the energy variables.\footnote{This can be thought of as an abstracted weak version of Steinmann relations or kinematic flow, see Appendix~\ref{app: Steinmann}.}
After this, 
we found that imposing complete cluster compatibility was sufficiently strong to reduce the solution space to objects with a physical interpretation. In the chain case, the space of consistent symbols corresponds to the wavefunction plus products of lower-point disconnected wavefunctions, so that both the wavefunction and correlator appear. In the cycle case, the wavefunction appears to be uniquely selected by the cluster structure.

\vskip4pt
There are a number of natural and interesting questions raised by our investigation. It is useful to elaborate on some of them.

\vskip4pt
Perhaps the most obvious direction is to understand how these constructions can be generalized. In the setting of conformally coupled scalars, it would be nice to apply cluster algebra machinery to more complicated graphs. We expect that this should be possible because cosmological $u$-variables exist for arbitrary graphs~\cite{Hillman:2023vas}. The challenge is that the graph associahedron that captures their compatibilities does not appear to be the cluster polytope of any finite-type cluster algebra in general.\footnote{The simplest challenging case is the four-site star, where one can check that there is no finite-type cluster algebra that produces the $u$-equations that arise from the corresponding marked graph.}
However, instead of a challenge, this perhaps can be viewed as inspirational; it strongly suggests that there exists some generalization of cluster algebras that is relevant for generic graphs. It would be very interesting to elucidate this and understand its implications for cosmology.\footnote{Possibly relevant are the Laurent phenomenon algebras of~\cite{lam2016laurentphenomenonalgebras}. For the four-site star the corresponding Laurent phenomenon algebra does not produce the correct $u$-equations, but perhaps there is some further generalization.}
Beyond the generalization to more complicated graphs, it would be interesting to understand how cluster compatibility controls the subleading structure of the symbol in power-law FLRW spaces, and for wavefunctions involving massive fields. Each of these situations represents a controlled deformation away from the limit we have considered in this paper, and can be expanded in polylogarithmic functions order-by-order in the deformation parameter.
 In the FLRW case, we expect that the cluster structure will survive and control the subleading terms in the expansion. Intuitively, the existence of kinematic flow for general power laws~\cite{Arkani-Hamed:2023kig}, and the coaction enjoyed by the FLRW wavefunction~\cite{McLeod:2026jpz,McLeod:2026kpo} suggest this will be the case~\cite{Paranjape:2026htn}. On the other hand, the first subleading-in-mass correction to the symbol away from the conformally coupled value already violates cluster compatibility in the most naive generalization.\footnote{At first subleading order in the deviation of the mass from the conformal value $\xi = \nu -\tfrac{1}{2}$, the  symbol for the first correction for a massive exchange is~\cite{Gasparotto:2024bku,Baumann:2026atn}
\be
\begin{aligned}
{\cal S}(F_2^{(\xi)})={}&
a_{14}\otimes a_{13}\otimes
\frac{1}{a_{24}\,a_{35}^{2}}
+
a_{14}\otimes a_{24}\otimes
\frac{1}{a_{13}\,a_{25}^{2}}
+
a_{25}\otimes a_{24}\otimes
\frac{1}{a_{25}\,a_{35}}
+
a_{25}\otimes a_{25}\otimes
\left(a_{24}\,a_{35}^{2}\right)
\\[4pt]
&+
a_{25}\otimes a_{35}\otimes
\frac{1}{a_{24}\,a_{35}^{2}}
+
a_{35}\otimes a_{13}\otimes
\frac{1}{a_{25}\,a_{35}}
+
a_{35}\otimes a_{25}\otimes
\frac{1}{a_{13}\,a_{25}^{2}}
+
a_{35}\otimes a_{35}\otimes
\left(a_{13}\,a_{25}^{2}\right),
\end{aligned}
\ee
where $\nu\equiv \sqrt{\frac{9}{4}-\tfrac{m^2}{H^2}}$.
We can see that this is not cluster compatible; for example $a_{13}$ and $a_{24}$ appear together.
}
One intuition for this is that the massive wavefunction satisfies a modified kinematic flow in which tubes can grow and overlap~\cite{Baumann:2026atn}, which violates the simplest notions of compatibility. It would be interesting to understand if there is a refinement that captures this massive case.

\vskip4pt
The cluster variable mapping considered here is closely related to the parameterization of~\cite{Ferro:2026oph}. There, they were able to find simple manifestly cluster-compatible expressions for the wavefunction in terms of special classes of polylogarithmic functions~\cite{rudenko2022,matveiakin2022}. The arguments of these functions are nearly the $u$-variables considered here. It would be interesting to understand if the additional structure of $u$-variables (e.g., the fact that they satisfy nonlinear equations) manifests at the level of these functions. It would also be interesting to see if these natural variables can be used to generalize the insights of~\cite{Ferro:2026oph} to more general graphs.

\vskip4pt
We have seen some glimpses of cluster structure in cosmological correlators themselves (rather than the wavefunction), but it is important and interesting to understand this in greater depth. In the chain case, we found in explicit examples (and give a general argument in Appendix~\ref{app: Correlator vs WF})
 that both the cosmological wavefunction and associated correlators have the same cluster-algebraic structure. The wavefunction has an interesting property: only variables that appear in the first slot ever reappear in the symbol. It would be interesting to determine whether the correlator has a similar distinguishing feature in cluster variables that selects it uniquely. More generally, we found that correlators associated to cycle graphs do not have a $B_N$ cluster structure. This suggests that the general cluster structure underlying correlators (if one exists) will be different from that of the wavefunction. It is natural to suspect that better understanding the smaller singularity alphabet of correlators will play an important role in disentangling these features.

\vskip4pt
In the chain and cycle cases, cluster algebraic considerations control the structure of the wavefunction symbol. The wavefunction also satisfies a system of first-order differential equations~\cite{Arkani-Hamed:2017fdk,Hillman:2019wgh,Arkani-Hamed:2023kig,Arkani-Hamed:2023bsv,Baumann:2025qjx,De:2023xue,De:2024zic,Fan:2024iek,He:2024olr,McLeod:2026jpz,McLeod:2026kpo,Glew:2025ypb,Capuano:2025ehm,Fu:2026dqb,Gasparotto:2024bku}, which directly determine the symbol. In a sense the connection matrices that appear in these differential equations are like $n$-th roots of the symbol. Given this,
it is natural to expect that these matrices can be determined by cluster-algebraic consistency as well.
It would be very interesting to elucidate these features. 
A natural goal would be to develop a cluster algebraic understanding of the graphical kinematic flow that underlies these equations from this perspective.

\vskip4pt
Our deepest aspirations are to uncover new principles that govern the physics of the very early universe. In this regard, the connections between cosmological correlators, $u$-variables, and cluster algebras are inspiring. Since the cluster structure appears related to the locality and causality of quantum field theory, it would be deeply illuminating to see the cosmological wavefunction arise from some mathematical principles where these physical properties appear as consequences of some other considerations.
One possible avenue towards this goal is to understand the spacetime interpretation of cluster mutation.
While cluster compatibility played an essential role in our construction, the appearance of multiple clusters was somewhat formal, and does not have a clear physical interpretation.
Understanding how different singularities or scattering processes mutate into each other may imbue the spacetime structure with further combinatorial life.

\paragraph{Acknowledgements:}
Thanks to Santiago Ag\"u\'i Salcedo, Nima Arkani-Hamed, Paolo Benincasa, John Joseph Carrasco, Saba Etezad-Razavi, Ross Glew, Shruti Paranjape, Houri Tarazi, Anastasia Volovich, He-Chen Weng, and most especially Aaron Hillman  for helpful discussions.
AJ and DG are supported by DOE award DE-SC0025323 and by the Kavli Institute for Cosmological Physics at the University of Chicago.

\appendix

\newpage
\section{Cluster Crash Course}
\label{app:cluster}

In the main text, one of our goals was to understand the features of functions that arise in cosmological correlation functions. An important aspect of this is the sequence of possible singularities of these functions. 
The possible locations of such singularities are polynomials in kinematic variables, and so it is natural that mathematical structures related to the compatibility of polynomials arise.
Similar questions have been of great interest recently in the study of cluster algebras~\cite{FominZelevinsky2002ClusterI,FominZelevinsky2003ClusterII,BerensteinFominZelevinsky2005ClusterIII,FominZelevinsky2007ClusterIV}.

\vskip4pt
Here we provide a brief introduction to the ideas from cluster algebras that are needed in the main text. Our goal is to describe the central concepts for a cosmology-minded audience. The subject of cluster algebras is an active area of mathematical research and more detailed discussions can be found in~\cite{williams2014cluster,Daniel_2015,Arkani-Hamed:2020tuz,le2017approachclusterstructuresmoduli,2016arXiv160805735F,FominZelevinsky2007ClusterIV,Papathanasiou:2022lan}.

\subsection{A Constructive Definition}
\label{app:clusterdefs}

At a high level, a cluster algebra is a way to imbue a commutative ring with a notion of compatibility between generators. 
This is done by grouping the generators of the ring of Laurent polynomials into overlapping subsets called {\it clusters}. The number of elements in these subsets, $N$, is called the rank of the algebra. The elements of a cluster can be transformed by an operation called {\it mutation} to produce a new cluster. Iterating these mutations produces a set of clusters, with overlapping members. 
Variables that appear together inside some cluster are called {\it compatible}, while those that never do are called {\it incompatible}.
In the cosmological context, the cluster variables will be expressible in terms of $X$ and $Y$ energies of cosmological observables, and the mapping to cluster variables will encode the compatibility between different singularities in the $X$ and $Y$ variables.

\vskip4pt
As a practical matter, it is most useful to give an operational definition of a cluster algebra by describing how to construct one.

\paragraph{Quivers, Clusters, and Seeds}
~\\
As input data, we first require a {\it quiver}, which is a directed graph on $N$ vertices, $\{1,\ldots ,N\}$, where we forbid directed cycles of length one or two (meaning we disallow arrows that begin and terminate on the same vertex $i\rightarrow i$, and we disallow closed loops between two vertices $i\rightleftarrows j$). An example of a quiver is
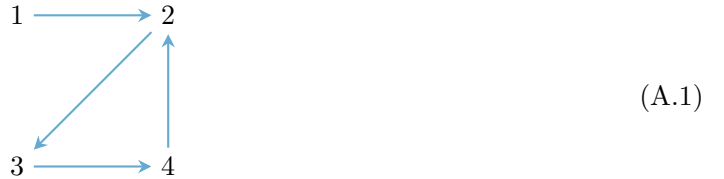
\begin{equation}
\raisebox{-30pt}{
\begin{tikzpicture}[
  quiver/.style={
    draw=tubeBlue!60,
    -{Stealth[length=5pt,width=5pt]},
    line width=0.8pt
  },
]
  \node (1) at (0,2) {$1$};
  \node (2) at (2,2) {$2$};
  \node (3) at (0,0) {$3$};
  \node (4) at (2,0) {$4$};

  \draw[quiver] (1) -- (2);
  \draw[quiver] (4) -- (2);
  \draw[quiver] (3) -- (4);
  \draw[quiver] (2) -- (3);
\end{tikzpicture}
}
\label{eq:quiver1}
\end{equation}
where we have labeled the vertices of the graph.
A quiver can be represented by an antisymmetric matrix $B(Q)$ with integer entries, called the exchange matrix, where $B_{ij}$ is the number of arrows from vertex $i$ to vertex $j$ (the convention is that positive numbers point towards $j$).
For example, the quiver~\eqref{eq:quiver1} corresponds to the matrix
\begin{equation}
    B_{ij}(Q^{(1)})=\left(
\begin{array}{cccc}
 0 & 1 & 0 & 0 \\
 -1 & 0 & 1 & -1 \\
 0 & -1 & 0 & 1 \\
 0 & 1 & -1 & 0 \\
\end{array}
\right)\,.
\end{equation}

\vskip4pt
Given a quiver, we can obtain a new one by an operation called {\it mutation}, which is an automorphism on the space of quivers. The mutation of a quiver $Q$ at vertex $k$, which we denote by $\mu_k(Q)$,  corresponds to the following sequence of operations:
\begin{enumerate}
        \item For any path that connects two vertices through vertex $k$, of the form $i\rightarrow k \rightarrow j$, add an arrow directly connecting $i\rightarrow j$.
        \item Reverse the direction of all arrows pointing in or out of vertex $k$.
        \item Delete any two-cycles, $i \rightleftarrows j$, generated by the previous two steps.
\end{enumerate}
The output of this operation is a new quiver (in general distinct from the original one). We can express the mutation operation directly on the components of the exchange matrix as\footnote{This mutation operation can also be applied to more general
skew-symmetrizable matrices~\cite{2016arXiv160805735F}. A matrix $B_{ij}$ is skew-symmetrizable if $d_i \, B_{ij} = -d_j \, B_{ji}$ for some positive integers $d_1,\cdots,d_N$. That is, by rescaling each of its rows we can obtain an antisymmetric matrix. In this case, we can keep track of these rescaling factors and assign them to each of the vertices of a quiver, so that this more general case can also be associated to a decorated quiver. The corresponding $B_{ij}$ can then be mutated using the same recipe~\eqref{eq: quiver mutation rule}.}
\begin{equation}
    \label{eq: quiver mutation rule}
    B_{ij}\big(\mu_k (Q)\big)=
        \begin{cases}
            -B_{ij}(Q) & \text{if }~~ i~{\rm or}~j=k \\
            B_{ij}(Q) & \text{if } ~~B_{ik}B_{kj}\leq 0 \\
            B_{ij} (Q)+ \lvert B_{ik}(Q)\rvert B_{kj}(Q) & \text{if }~~ B_{ik}B_{kj} > 0 ~.
        \end{cases}
\end{equation}
As an example, we can consider mutating the quiver~\eqref{eq:quiver1} on vertex $2$, which yields the new quiver
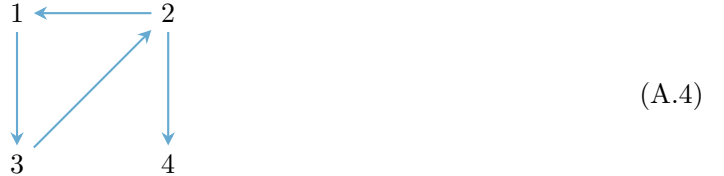
\begin{equation}
\raisebox{-30pt}{
\begin{tikzpicture}[
  quiver/.style={
    draw=tubeBlue!60,
    -{Stealth[length=5pt,width=5pt]},
    line width=0.8pt
  },
]
  \node (1) at (0,2) {$1$};
  \node (2) at (2,2) {$2$};
  \node (3) at (0,0) {$3$};
  \node (4) at (2,0) {$4$};

  \draw[quiver] (2) -- (1);
  \draw[quiver] (1) -- (3);
  \draw[quiver] (3) -- (2);
  \draw[quiver] (2) -- (4);
\end{tikzpicture}
}
\label{eq:quiver2}
\end{equation}
which can be represented as the matrix 
\begin{equation}
    B_{ij}(Q^{(2)})= B_{ij}\big(\mu_2(Q^{(1)})\big)=\left(
\begin{array}{cccc}
 0 & -1 & 1 & 0 \\
 1 & 0 & -1 & 1 \\
 -1 & 1 & 0 & 0 \\
 0 & -1 & 0 & 0 \\
\end{array}
\right)\,.
\end{equation}
Note that choosing to mutate at the same vertex again reproduces the original quiver $\mu_k^2(Q)=Q$.

\vskip4pt
The second ingredient we need to construct a cluster algebra is the 
eponymous cluster. 
A \textit{cluster} is a set of $N$ variables, $A=\{a_1,\ldots,a_N\}$, assigned to the $N$ vertices of the quiver $Q$. 
Together, a cluster, $A$, and an exchange matrix, $B$, define a {\it seed}, denoted $S=(A,B)$.
Similar to the underlying quiver, cluster variables transform under mutation. If we mutate vertex $k'$, then the cluster variables transform as
\begin{equation}
        \label{eq: cluster var mutation rule}
        \mu_{k'}(a_k)= \begin{cases}
        \frac{1}{a_k} \left(\prod_{B_{ik}>0} \, a_i^{B_{ik}}+\prod_{B_{ik}<0} \,a_i^{-B_{ik}} \right) & {\rm if}~~ k=k' \\
        a_k & {\rm if}~~k \neq k'
        \end{cases}\,.
\end{equation} 
These equations are called exchange relations. They have a simple graphical interpretation in terms of the underlying quiver.
The variable associated to a vertex $k$ is mutated to a new variable $a_k' = \tfrac{1}{a_k}\left(\prod_{i\to k} a_i+\prod_{k\to j} a_j\right)$, which is simply the sum of the product of cluster variables associated to vertices $i$ with arrows pointing to $k$, plus the product of variables associated to vertices $j$ that $k$ points to. As an example, consider the quiver~\eqref{eq:quiver1} with variables $\{a_1,\cdots, a_4\}$. Mutating the variable $a_2$ produces a new variable according to the rule
\begin{equation}
\raisebox{-30pt}{
\begin{tikzpicture}[
  quiver/.style={
    draw=tubeBlue!60,
    -{Stealth[length=5pt,width=5pt]},
    line width=0.8pt
  },
]
  \node (1) at (0,2) {$a_1$};
  \node (2) at (2,2) {$a_2$};
  \node (3) at (0,0) {$a_3$};
  \node (4) at (2,0) {$a_4$};

  \draw[quiver] (1) -- (2);
  \draw[quiver] (4) -- (2);
  \draw[quiver] (3) -- (4);
  \draw[quiver] (2) -- (3);
\end{tikzpicture}
}
~
\raisebox{-0pt}{
\begin{tikzpicture}[
  quiver/.style={
    draw=black!25,
    -{Stealth[length=8pt,width=8pt]},
    line width=2pt
  },
]
  \node (1) at (0,0) {};
  \node (2) at (2,0) {};
  \node (3) at (1,.4) {$\mu_2$};

  \draw[quiver] (1) -- (2);
\end{tikzpicture}
}
~
\raisebox{-30pt}{
\begin{tikzpicture}[
  quiver/.style={
    draw=tubeBlue!60,
    -{Stealth[length=5pt,width=5pt]},
    line width=0.8pt
  },
]
  \node (1) at (0,2) {$a_1$};
  \node (2) at (2,2) {\phantom{$a_2$}};
  \node[anchor=west] at (1.75,2) {$a_2'=\tfrac{1}{a_2}\left(a_1a_4+a_3\right)$};
  \node (3) at (0,0) {$a_3$};
  \node (4) at (2,0) {$a_4$};

  \draw[quiver] (1) -- (2);
  \draw[quiver] (4) -- (2);
  \draw[quiver] (3) -- (4);
  \draw[quiver] (2) -- (3);
\end{tikzpicture}
}
\label{eq:quiver12}
\end{equation}
Note that only one variable in the cluster is changed. Here we have held the underlying quiver fixed and mutated the cluster variable, but more generally we will want to mutate the entire seed, transforming both the variable and the quiver. This operation acts in the obvious manner $\mu_k(S)=\left(\mu_k(A),\mu_k(B)\right)$.

\vskip4pt
We now have enough technology to describe what a cluster algebra is. Given an initial seed $(A,B)$, we perform all possible sequences of mutations on its vertices.
We then write the union of all cluster variables that appear associated to vertices. A \textit{\textbf{cluster algebra}}  is the algebra generated by all cluster variables. Variables that appear together in a given seed are said to form a cluster. The number of vertices of the underlying graph $N$ is called the {\it rank} of the cluster algebra (which is also the number of cluster variables in a cluster).

\vskip4pt
\noindent
\textit{An Example: $A_2$ Cluster Algebra}~\\
In order to make the definitions more concrete, it is useful to explicitly construct a simple cluster algebra. Here we describe the 
$A_2$ cluster algebra. This cluster algebra can be constructed from the initial seed
\be
\raisebox{-5pt}{
\begin{tikzpicture}[
  quiver/.style={
    draw=tubeBlue!60,
    -{Stealth[length=5pt,width=5pt]},
    line width=0.8pt
  },
]
  \node (1) at (0,2) {$a_1$};
  \node (2) at (2,2) {$a_2$};
  \draw[quiver] (1) -- (2);
\end{tikzpicture} 
}\,,
\label{eq:A2seed1}
\ee
which has initial cluster variables $A=\{a_1,a_2\}$ assigned to a two-site quiver. This corresponds to the exchange matrix
\begin{equation}
    B =\left(
    \begin{array}{cc}
    0 & 1 \\
    -1 & 0 \\
    \end{array}
    \right) \,.
\end{equation}
We now consider all possible sequences of mutations. First mutate at vertex $1$, this generates a new seed
\be
\raisebox{-5pt}{
\begin{tikzpicture}[
  quiver/.style={
    draw=tubeBlue!60,
    -{Stealth[length=5pt,width=5pt]},
    line width=0.8pt
  },
]
  \node (1) at (-.75,2) {$a_3=\frac{1+a_2}{a_1}$};
  \node (2) at (2,2) {$a_2$};
  \draw[quiver] (2) -- (1);
\end{tikzpicture}
}\,,
\ee
where we have introduced a new cluster variable associated to vertex $1$ as  $a_3 \equiv \mu_1(a_1)=\frac{1+a_2}{a_1}$.
If we were to
mutate again on vertex $1$ we would return to the initial seed, so we now mutate on vertex $2$. This generates
\be
\raisebox{-5pt}{
\begin{tikzpicture}[
  quiver/.style={
    draw=tubeBlue!60,
    -{Stealth[length=5pt,width=5pt]},
    line width=0.8pt
  },
]
  \node (1) at (-.75,2) {$a_3=\frac{1+a_2}{a_1}$};
  \node (2) at (2.75,2) {$a_4=\frac{1+a_1+a_2}{a_1 a_2}$};
  \draw[quiver] (1) -- (2);
\end{tikzpicture}
}\,,
\ee
where we have defined another new cluster variable  $a_4 \equiv \mu_2(a_2)=\frac{1+a_3}{a_2}=\frac{1+a_1+a_2}{a_1 a_2}$.
The pattern is now somewhat clear, we can alternately mutate vertices $1$ and $2$, which flips the orientation of the quiver and generates a new variable. The next three iterations of this procedure generate the new variables
\begin{align}
    a_5&=\mu_1(a_3)=\frac{1+a_4}{a_3}=\frac{1+a_1}{ a_2}\,,\\
    a_6&=\mu_2(a_4)=\frac{1+a_5}{a_4}=a_1\,,\\
    a_7&=\mu_1(a_5)=\frac{1+a_1}{a_5}=a_2\,.
\end{align}
Notice that after generating $5$ distinct cluster variables, further mutations produce variables we have already seen. In particular, the $5^{\rm th}$ mutation produces the seed
\be
\raisebox{-5pt}{
\begin{tikzpicture}[
  quiver/.style={
    draw=tubeBlue!60,
    -{Stealth[length=5pt,width=5pt]},
    line width=0.8pt
  },
]
  \node (2) at (0,2) {$a_2$};
  \node (1) at (2,2) {$a_1$};
  \draw[quiver] (1) -- (2);
\end{tikzpicture} 
}\,,
\ee
which is equivalent to the original seed~\eqref{eq:A2seed1}.  We see that there are $5$ distinct clusters. They can be used to label the vertices of a pentagon:
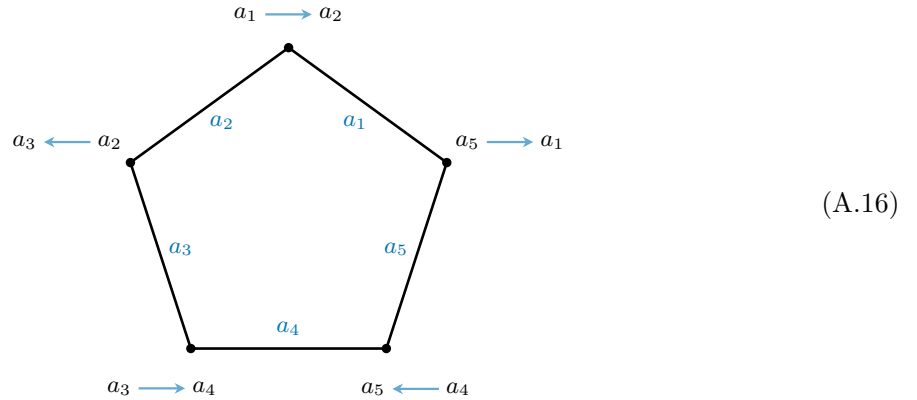
\begin{equation}
\raisebox{-70pt}{
\begin{tikzpicture}[
  scale=2.2,
  every node/.style={font=\footnotesize},
  vtx/.style={circle,fill=black,inner sep=1.25pt},
  pent/.style={line width=1.pt},
  qarrow/.style={
    draw=tubeBlue!60,
    -{Stealth[length=4pt,width=4pt]},
    line width=0.85pt
  }
]
\def\r{1.0}
\def\roff{0.4}
\def\foff{0.16}

\coordinate (A) at (90:\r);
\coordinate (B) at (162:\r);
\coordinate (C) at (234:\r);
\coordinate (D) at (306:\r);
\coordinate (E) at (18:\r);

\draw[pent] (A) -- (B) -- (C) -- (D) -- (E) -- cycle;

\node[vtx] at (A) {};
\node[vtx] at (B) {};
\node[vtx] at (C) {};
\node[vtx] at (D) {};
\node[vtx] at (E) {};

\newcommand{\seedright}[3]{%
  \begin{scope}[shift={#1}]
    \node[anchor=east,inner sep=1pt] (L) at (-0.14,0) {$#2$};
    \node[anchor=west,inner sep=1pt] (R) at ( 0.14,0) {$#3$};
    \draw[qarrow] (L.east) -- (R.west);
  \end{scope}
}
\newcommand{\seedleft}[3]{%
  \begin{scope}[shift={#1}]
    \node[anchor=east,inner sep=1pt] (L) at (-0.14,0) {$#2$};
    \node[anchor=west,inner sep=1pt] (R) at ( 0.14,0) {$#3$};
    \draw[qarrow] (R.west) -- (L.east);
  \end{scope}
}

\seedright{($(A)+(90:.5*\roff)$)}{a_1\,}{\,a_2}
\seedleft{($(B)+(162:\roff)$)}{a_3\,}{\,a_2}
\seedright{($(C)+(234:.74*\roff)$)}{a_3\,}{\,a_4}
\seedleft{($(D)+(306:.75*\roff)$)}{a_5\,}{\,a_4}
\seedright{($(E)+(18:\roff)$)}{a_5\,}{\,a_1}

\node at ($($(A)!0.5!(B)$)+(306:.8*\foff)$) {$\color{tubeBlue}{a_2}$};
\node at ($($(B)!0.5!(C)$)+(18:.8*\foff)$)  {$\color{tubeBlue}{a_3}$};
\node at ($($(C)!0.5!(D)$)+(90:.75*\foff)$)  {$\color{tubeBlue}{a_4}$};
\node at ($($(D)!0.5!(E)$)+(162:.8*\foff)$) {$\color{tubeBlue}{a_5}$};
\node at ($($(E)!0.5!(A)$)+(234:.8*\foff)$) {$\color{tubeBlue}{a_1}$};
\end{tikzpicture}
}
\label{eq:clusterassoc1}
\end{equation}
This {\it cluster polytope} encodes the mutation relations between seeds and the compatibility relations between cluster variables. 
Here cluster seeds that are connected by an edge can be obtained from each other by a single mutation. 
It is therefore natural to label the edges by the cluster variables that are left unchanged by mutation, and the facets by cluster variables themselves. Variables that belong to facets that intersect appear in some cluster together (they are said to be {\it compatible}) while variables corresponding to facets that do not meet never appear in a cluster together (they are {\it incompatible}).
In the $A_2$ example,
the five clusters corresponding to the vertices (which each have two variables) share each of their variables with a single other cluster, to which they are connected by the faces of the pentagon. Relatedly, each cluster variable has two variables that it appears in some cluster with and has two variables that it never appears with.

\vskip4pt
The $A_2$ cluster algebra also has a nice connection to triangulations of a pentagon. If we assign each cluster variable to a chord of a pentagon as
\be\label{eq: A2 variable chord assignments}
\begin{aligned}
    a_{1}&\equiv  a_{13} \,,   & \hspace{1.7cm} a_{2}& \equiv a_{14} \,,&\hspace{1.7cm} a_{3}&\equiv a_{24}\,,\\[1pt]
    a_{4}&\equiv a_{25}\,, & a_{5}&\equiv a_{35}\,,&&
\end{aligned}
\ee
where $a_{ij}$ corresponds to the diagonal between the $i^{\rm th}$ and $j^{\rm th}$ vertices of the pentagon,
then each cluster can be mapped to a triangulation of the pentagon. From this perspective, mutation corresponds to flipping a single diagonal. For example, the two possible clusters involving $a_1$ can be visualized as
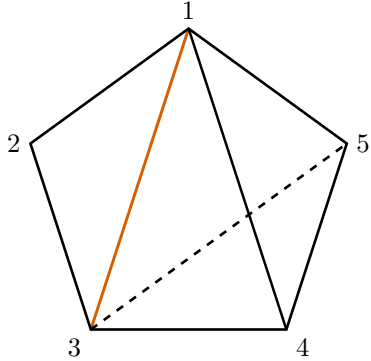
\begin{equation*}
\begin{tikzpicture}[
  scale=2.2,
  every node/.style={font=\small}
]
\coordinate (1) at (90:1);
\coordinate (2) at (162:1);
\coordinate (3) at (234:1);
\coordinate (4) at (306:1);
\coordinate (5) at (18:1);
\draw[line width=1.1pt,tubeRed] (1) -- (3);
\draw[line width=.95pt] (1) -- (4);
\draw[dashed,line width=1pt] (3) -- (5);
\node[above]      at (1) {$1$};
\node[left]       at (2) {$2$};
\node[below left] at (3) {$3$};
\node[below right] at (4) {$4$};
\node[right]      at (5) {$5$};
\draw[line width=1.pt] (1) -- (2) -- (3) -- (4) -- (5) -- cycle;
\end{tikzpicture}
\end{equation*}
Here the cluster $\{a_{13},a_{14}\}$ corresponds to the displayed triangulation, while the mutation to the cluster $\{a_{13},a_{35}\}$ corresponds to swapping the solid black diagonal to the dashed one, which also does not intersect the red diagonal. This provides a natural geometric interpretation of the compatibility of cluster variables: each diagonal has exactly two other diagonals that do not intersect it (with which it is compatible) and two that do (with which it is incompatible).

\vskip4pt
\noindent
\textit{Generalization: $A_N$}~\\
The $A_2$ cluster algebra admits an obvious generalization, which is to consider an $N$-site quiver of the form
\be
\raisebox{-5pt}{
\begin{tikzpicture}[
  quiver/.style={
    draw=tubeBlue!60,
    -{Stealth[length=5pt,width=5pt]},
    line width=0.8pt
  },
]
  \node (4) at (6,2) {$a_N$};
  \node (3) at (4,2) {$\cdots$};
  \node (2) at (2,2) {$a_2$};
  \node (1) at (0,2) {$a_1$};
\draw[quiver] (1) -- (2);
\draw[quiver] (2) -- (3);
\draw[quiver] (3) -- (4);
\end{tikzpicture} 
}\,,
\label{eq:Anquiver}
\ee
which corresponds to an initial seed with $N$ cluster variables. Note that this is a directed $A$-type Dynkin diagram.
We then consider all possible mutations of this seed. (Note that mutation will act on this quiver in a more interesting way than~\eqref{eq:A2seed1} because it will not just reverse directions of arrows, but can also involve the introduction of new arrows.)
Remarkably, the sequence of mutations will eventually truncate to produce $C_{N+1} \equiv (2N+2)!/((N+1)!(N+2)!)$ inequivalent seeds.
Altogether the $A_N$ cluster algebra has $N(N+3)/2$ cluster variables, which are arranged into $C_{N+1}$ clusters of size $N$.
Each of these clusters can be assigned to the vertices of an $N$-dimensional associahedron, which serves as the cluster polytope for this cluster algebra. The codimension-one facets of this polyhedron can be labeled by the cluster variables appearing in the algebra. Facets of higher codimension correspond to collections of cluster variables shared between seeds.

\vskip4pt
The $A_N$ cluster algebra is also naturally related to the triangulations of an $(N+3)$-gon.\footnote{Note that it is an accident of $N=2$ that the cluster polytope and this polygon happen to both be a pentagon.} We describe this more fully in Appendix~\ref{app: Structures of An and Bn}.

\vskip4pt
\noindent
\textit{$B_N$ Cluster Algebra}~\\
The other family of finite-type cluster algebras relevant to our investigation is the $B_N$ family. These algebras can be generated by an initial $N\times N$ exchange matrix of the form~\cite{FominZelevinsky2003ClusterII}
\begin{equation}
B_{ij}^{B_N}=
\renewcommand{\arraystretch}{0.8}
\setlength{\arraycolsep}{3pt}
\begin{pmatrix}
0 & 1 & 0& \cdots& 0\\
-1 & 0 & 1 & & \vdots\\
0& -1 & \ddots & \ddots & \\
\vdots& & \ddots & 0 & 1 \\
0 & \cdots& & -2 & 0
\end{pmatrix}\,,
\end{equation}
which differs from the initial exchange matrix of $A_N$ only by the entry $B_{N\, N-1}=-2$ (rather than $-1$). Since this exchange matrix is not antisymmetric, it cannot be represented directly by a quiver. Nevertheless, we can still perform mutations according to~\eqref{eq: quiver mutation rule} and~\eqref{eq: cluster var mutation rule}. The sequence of mutations will eventually terminate, producing $(2N)!/{(N!)}^2$ unique seeds populated by $N^2+N$ unique cluster variables.
Each seed can be assigned to the vertices of the $N$-dimensional cyclohedron, which is the cluster polytope of $B_N$ cluster algebras.\footnote{The $N$-dimensional cyclohedron is the graph associahedron of the cycle graph with $N+1$ vertices.} Relatedly, it is possible to associate the cluster variables and seeds of $B_N$ with centrally symmetric pairs of chords and triangulations of a $2(N+1)$-gon respectively, as we describe in Appendix~\ref{app: Structures of An and Bn}.

\vskip4pt
As an example, the $B_2$ cluster algebra can be generated by the initial seed
\begin{equation}
    B_{ij}^{B_2} =\left(
    \begin{array}{cc}
    0 & 1 \\
    -2 & 0 \\
    \end{array}
    \right) \,, \quad {\rm with}\quad A=\left(a_1,a_2\right)\,.
\end{equation}
Performing alternating mutations on the indices $1$ and $2$ generates 6 total cluster variables, the initial $(a_1,a_2)$ and
\begin{align}
  &a_3=\frac{a_2^2+1}{a_1}
  &&a_4=\frac{a_1+1}{a_2} \\
  &a_5=\frac{a_2^2+a_1+1}{a_1 a_2} 
  &&a_6=\frac{\left(a_1+1\right){}^2+a_2^2}{a_1 a_2^2} \nonumber \,.
\end{align}
These variables appear in 6 unique clusters, which can be used to label the $2$-dimensional cyclohedron, which is a hexagon:
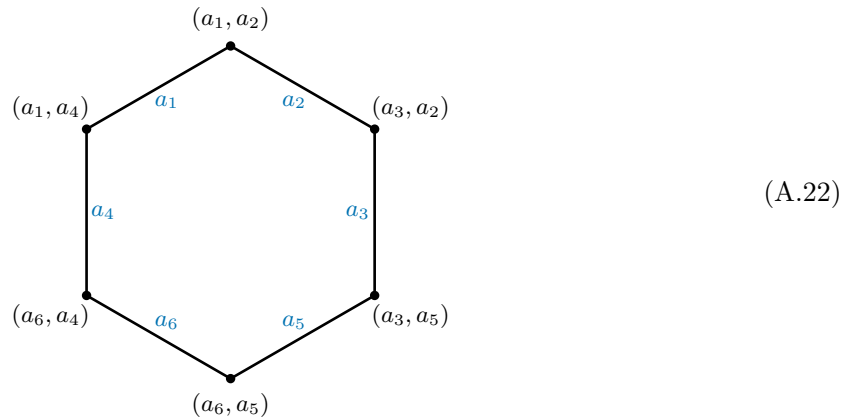
\begin{equation}
\raisebox{-85pt}{
\begin{tikzpicture}[
  scale=2.2,
  every node/.style={font=\footnotesize},
  vtx/.style={circle,fill=black,inner sep=1.25pt},
  hex/.style={line width=1.pt}
]
\def\r{1.0}
\def\roff{0.20}    
\def\foff{0.10}    
\def\foffs{0.10}

\coordinate (A) at ( 90:\r);   
\coordinate (B) at ( 30:\r);   
\coordinate (C) at (-30:\r);   
\coordinate (D) at (-90:\r);   
\coordinate (E) at (210:\r);   
\coordinate (F) at (150:\r);  

\draw[hex] (A) -- (B) -- (C) -- (D) -- (E) -- (F) -- cycle;

\foreach \p in {A,B,C,D,E,F}{\node[vtx] at (\p) {};}

\node at ($(A)+( 90:0.8*\roff)$)   {$(a_1,a_2)$};
\node at ($(B)+( 30:1.25*\roff)$)  {$(a_3,a_2)$};
\node at ($(C)+(-30:1.25*\roff)$)  {$(a_3,a_5)$};
\node at ($(D)+(-90:0.8*\roff)$)   {$(a_6,a_5)$};
\node at ($(E)+(210:1.25*\roff)$)  {$(a_6,a_4)$};
\node at ($(F)+(150:1.25*\roff)$)  {$(a_1,a_4)$};

\node at ($($(A)!0.5!(B)$)+(240:\foffs)$) {$\color{tubeBlue}{a_2}$};
\node at ($($(B)!0.5!(C)$)+(180:\foff)$)  {$\color{tubeBlue}{a_3}$};
\node at ($($(C)!0.5!(D)$)+(120:\foffs)$) {$\color{tubeBlue}{a_5}$};
\node at ($($(D)!0.5!(E)$)+( 60:\foffs)$) {$\color{tubeBlue}{a_6}$};
\node at ($($(E)!0.5!(F)$)+(  0:\foff)$)  {$\color{tubeBlue}{a_4}$};
\node at ($($(F)!0.5!(A)$)+(300:\foffs)$) {$\color{tubeBlue}{a_1}$};
\end{tikzpicture}
}
\label{eq:clusterpolytopeB2}
\end{equation}
Here the vertices are labeled by clusters and the edges represent mutations, labeled by the cluster variable left intact under the mutation.

\paragraph{Frozen Variables}~\\
An important generalization is to add {\it frozen} variables to a cluster algebra. These are variables that decorate the cluster algebra, but themselves do not mutate.
At the quiver level, this can be accomplished by considering an {\it ice quiver}, which is a quiver that has both mutable and non-mutable nodes. The variables assigned to non-mutable nodes are called frozen and appear in every cluster but cannot be mutated. They still enter into the mutation rules for the mutable (unfrozen) variables, and so add coefficients to these cluster variables.
Interestingly, this is the only modification of the cluster algebra---the introduction of frozen variables does not change the number of clusters and their adjacency relations.  The frozen variables therefore serve as a sort of deformation of the cluster algebra, introducing constants that can be chosen to take convenient values in some cases. On the locus where all the frozen variables are set to 1, the cluster algebra is identical to having never added any frozen variables.

\vskip4pt
Concretely,  given an initial cluster $A=\{a_1,\ldots,a_N,z_{1},\ldots,z_M\}$, with $N$ mutable $a$ variables and $M$ frozen $z$ variables, we associate to the cluster an \textit{extended exchange matrix} $\tilde{B}_{Ij}$, which is an $(M+N) \times N$ matrix with integer entries. The top $N\times N$ sub-matrix is skew-symmetrizable. The bottom $M$ rows may contain any integers, with $\tilde{B}_{N+i \, j}$ encoding the number of arrows from $z_i$ to $a_j$ in an ice quiver diagram. 
These bottom $M$ rows are called the \textit{coefficients} of the cluster algebra.  Using the extended exchange matrix, one can apply the same mutation rules to generate the cluster algebra involving frozen variables, which is the generalization of~\eqref{eq: cluster var mutation rule}
\begin{equation}
\label{eq: exchange relation w frozen variables}
        \mu_{k'}(a_k)= \begin{cases}
        \frac{1}{a_k} \left(p_+\prod_{B_{ik}>0} \, a_i^{B_{ik}}+p_-\prod_{B_{ik}<0} \,a_i^{-B_{ik}} \right) & {\rm if}~~ k=k' \\
        a_k & {\rm if}~~k \neq k'
        \end{cases}\,.
\end{equation} 
Here the coefficients appearing depend only on the frozen variables, and are defined by
\be
\label{eq:ppmdef}
\begin{aligned}
  p_+&=\prod_{i=1}^{M} z_i^{{\rm max}[B_{N+i\, k},0]}\\
        p_-&=\prod_{i=1}^{M} \, z_i^{{\rm max}[-B_{N+i\, k},0]}\,,
\end{aligned}
\ee

\vskip4pt
Aside from providing a deformation of cluster algebras, frozen variables provide useful tools to characterize the properties of cluster algebras.
There are many ways to introduce frozen variables, but there are two particularly simple schemes, which we now describe.

\paragraph{Principal Coefficients}~\\
A cluster algebra with {\it principal coefficients} is generated by a $2N \times N$ extended exchange matrix comprised of an $N \times N$ exchange matrix on its top $N$ rows stacked on an $N \times N$ identity matrix on the bottom $N$ rows~\cite{FominZelevinsky2007ClusterIV,Arkani-Hamed:2020tuz}. Represented as a quiver, this corresponds to adding $N$ frozen nodes, with each frozen node having one arrow pointing to one of the $N$ initial mutable nodes.

\vskip4pt
\noindent
{\it $g$-vectors and $F$-polynomials}~\\
The principal coefficients can be used to assign a unique $\mathbb{Z}^N$ vector to each mutable cluster variable, which is called a  \textit{$g$-vector}~\cite{FominZelevinsky2007ClusterIV,Arkani-Hamed:2020tuz}.
This provides a $\mathbb{Z}^N$ grading to a cluster algebra.\footnote{Mutable cluster variables are homogeneous with respect to this grading~\cite{FominZelevinsky2007ClusterIV}.}
We start by assigning the following vectors to the initial cluster variables and frozen variables, which define the formal degree of the variables
\begin{equation}
   \boldsymbol{g}_{a_i} =  \deg(a_i)\equiv \mathbf{e}_i,\qquad\qquad\quad \deg(z_i)\equiv-B_{ji}^0 \, \mathbf{e}_j \,.
\end{equation} 
Here $a_i$ and $z_i$ are the initial mutable cluster variables and frozen variables, respectively, with $i\in\{1,\cdots, N\}$. The vectors $\mathbf{e}_i$ are the $i^{\rm th}$ basis vectors in~$\mathbb{Z}^N$ and $B_{ij}^0$ is the $N\times N$ initial, unextended exchange matrix. We can then define the $g$-vector of the other cluster variables obtained by mutation to be $\boldsymbol{g}_{a_{k}}\equiv \deg(a_{k})$.
Mechanically, we can calculate $g$-vectors by assigning a set $\{\boldsymbol{g}_1,\boldsymbol{g}_2,\cdots,\boldsymbol{g}_N\}$ 
to each seed, which is the set of the $N$ $g$-vectors corresponding to the $N$ mutable variables in the cluster. Insisting on the homogeneity of the grading implies the following mutation rule for the $g$-vector $\boldsymbol{g}_k$~\cite{FominZelevinsky2007ClusterIV}: 
\begin{equation}
    \label{eq: g vector mutation rule}
    \mu_k(\boldsymbol{g}_k)=-\boldsymbol{g}_k+\sum_{i=1}^{N} {\rm max}\left[B_{ik},0\right] \boldsymbol{g}_i - \sum_{i,j=1}^{N}{\rm max}\left[B_{N+j\, k},0\right] B_{ij}^0 \, \mathbf{e}_i \,.
\end{equation}
In~\eqref{eq: g vector mutation rule}, $B_{ij}$ is the extended exchange matrix of the given seed before mutation, $\boldsymbol{g}_i$ is the $i^{\rm th}$ vector in the set $\{\boldsymbol{g}_1,\boldsymbol{g}_2,\cdots,\boldsymbol{g}_N\}$ for that seed, and $B_{ij}^0$ is the initial, unextended exchange matrix. Using these rules, one can obtain $g$-vectors for all cluster variables. (Note that the $g$-vector associated to a given cluster variable depends on the initial choice of seed used to define the grading.)

\vskip4pt
An interesting feature of $g$-vectors is that cluster variables can be written in the following factorized form in terms of the set of cluster variables of the original seed~\cite{FominZelevinsky2007ClusterIV}
\be
a_k = {\boldsymbol a}^{{\boldsymbol g}_{a_k}} F_{a_k}(\hat {\boldsymbol z})\,,
\label{eq:factorizationformula}
\ee
where ${\boldsymbol g}_{a_k}$ is the $g$-vector corresponding to the variable $a_k$. The notation ${\boldsymbol a}^{{\boldsymbol g}_{a_k}}$ means that we raise the $i^{\rm th}$ variable in the initial seed ${\boldsymbol a} = (a_1,\cdots, a_N)$ to the power that is the $i^{\rm th}$ component of the $g$-vector ${\boldsymbol g}_{a_k}$. The polynomial $F$, called an {\it $F$-polynomial}, is constructed from the rescaled frozen variables
\be
\hat z_k = z_k \prod_{r=1}^N \,a_r^{B^0_{rk}}\,,
\ee
which have been multiplied by powers of the original seed mutable variables so as to have degree zero. In this way,~\eqref{eq:factorizationformula} manifestly has the correct degree. The $F$-polynomial has the nice feature that $F(0) = 1$ for all $a_k$~\cite{FominZelevinsky2007ClusterIV}. This provides a simple way to extract the $g$-vectors: we mutate with principal coefficients and then set the frozen variables to zero. The exponents of the monomials in the original cluster variables are then the components of the $g$-vectors. The information in $g$-vectors and the $F$-polynomial (which can be obtained by setting all the mutable cluster variables to $1$) is then enough information to reconstruct the full cluster variable with coefficients.

\vskip4pt
\noindent
\textit{Example: $A_2$}~\\
It is useful to illustrate these constructions in a simple example, so let us reconsider the $A_2$ cluster algebra, but allow for additional frozen variables. The $A_2$ cluster algebra with principal coefficients can be generated from the initial (ice) quiver
\be
\raisebox{-5pt}{
\begin{tikzpicture}[
  quiver/.style={
    draw=tubeBlue!60,
    -{Stealth[length=5pt,width=5pt]},
    line width=0.8pt
  },
]
  \node (4) at (6,2) {$\color{darkblue}{z_2}$};
  \node (3) at (4,2) {$a_2$};
  \node (2) at (2,2) {$a_1$};
  \node (1) at (0,2) {$\color{darkblue}{z_1}$};
\draw[quiver] (1) -- (2);
\draw[quiver] (2) -- (3);
\draw[quiver] (4) -- (3);
\end{tikzpicture} 
}\,,
\ee
which corresponds to the extended exchange matrix
\begin{equation}
    \tilde B =\left(
    \begin{array}{cc}
    0 & 1 \\
    -1 & 0 \\
    1 &0 \\
    0 &1
    \end{array}
    \right) \,.
\end{equation}

Starting with the initial seed $(a_1,a_2)$, we 
 define the $g$-vectors of the initial mutable variables to be
\be
 \boldsymbol{g}_{a_1} =  
 \left(
    \begin{array}{c}
    1\\
    0
    \end{array}
    \right)
    \quad\quad\quad
    \boldsymbol{g}_{a_2} =  
 \left(
    \begin{array}{c}
    0\\
    1
    \end{array}
    \right)\,.
    \label{eq:a2gvec1}
\ee
By mutating, we obtain the other three cluster variables
\begin{align}
a_3 &= a_1^{-1}a_2 \left(1+\frac{z_1}{a_2}\right) \\
a_4 &= a_1^{-1}\left(1+\frac{z_1}{a_2}+\frac{z_1z_2 a_1}{a_2}\right)\\
a_5 &= a_2^{-1}\left(1+z_2a_1\right)\,,
\end{align}
which we have put in the form~\eqref{eq:factorizationformula}. Specializing to $z_1=z_2=0$, we can then read off the $g$-vectors
\be
 \boldsymbol{g}_{a_3} =  
 \left(
    \begin{array}{c}
    -1\\
    1
    \end{array}
    \right)
    \quad\quad\quad
    \boldsymbol{g}_{a_4} =  
 \left(
    \begin{array}{c}
    -1\\
    0
    \end{array}
    \right)
        \quad\quad\quad
    \boldsymbol{g}_{a_5} =  
 \left(
    \begin{array}{c}
    0\\
    -1
    \end{array}
    \right)\,.
        \label{eq:a2gvec2}
\ee
It is illuminating to plot these vectors
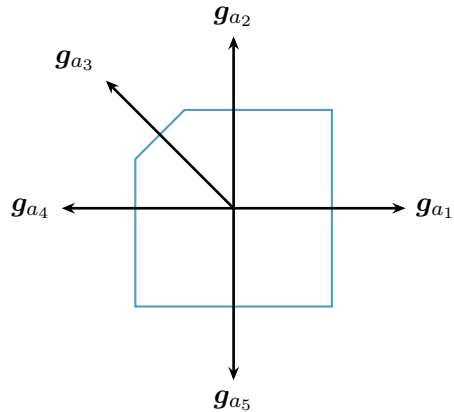
\begin{equation*}
\raisebox{-25pt}{
\begin{tikzpicture}[
  scale=1.3,
  >={Stealth[length=5pt,width=4pt]}
]
\coordinate (O) at (0,0);
\coordinate (P1) at (1,1);
\coordinate (P2) at (-0.5,1);
\coordinate (P3) at (-1,0.5);
\coordinate (P4) at (-1,-1);
\coordinate (P5) at (1,-1);
\draw[line width=.8pt,color=tubeBlue!70] (P1) -- (P2) -- (P3) -- (P4) -- (P5) -- cycle;
\draw[->, line width=1pt] (O) -- (1.75,0) node[anchor=west] {$ \boldsymbol{g}_{a_1}$};
\draw[->, line width=1pt] (O) -- (0,1.75) node[anchor=south] {$ \boldsymbol{g}_{a_2}$};
\draw[->, line width=1pt] (O) -- (-1.3,1.3) node[anchor=south east] {$ \boldsymbol{g}_{a_3}$};
\draw[->, line width=1pt] (O) -- (-1.75,0) node[anchor=east] {$ \boldsymbol{g}_{a_4}$};
\draw[->, line width=1pt] (O) -- (0,-1.75) node[anchor=north] {$ \boldsymbol{g}_{a_5}$};
\end{tikzpicture}
}
\end{equation*}
There are two features of this plot worth noting. First, the $g$-vectors that correspond to variables that appear together in clusters appear consecutively and bound cones. (These cones also tile the full space, which happens for finite-type cluster algebras.) 
In addition, these vectors are the normal vectors to the facets of a pentagon (which we have overlaid for illustration), which is precisely the cluster polytope.

\vskip4pt
While  $g$-vectors are intrinsically interesting objects with many useful properties, for our purposes they are primarily important to define so-called universal coefficients.

\paragraph{Universal Coefficients}~\\
We will actually be mostly interested in a slightly different (but related) assignment of frozen variables---so-called {\it universal coefficients}. 
A cluster algebra with universal coefficients is generated by an $(N+r) \times N$ extended exchange matrix $\tilde{B}_{Ij}$ where $r=|\mathcal{A}|$ is the total number of mutable cluster variables.
The top $N$ rows are again an $N \times N$ exchange matrix $B_{ij}$ and the bottom $r$ rows are taken to be the $r$ $g$-vectors of all of the cluster variables associated to the transpose exchange matrix $B^T$~\cite{FominZelevinsky2007ClusterIV,2012arXiv1209.3987R,Arkani-Hamed:2020tuz}.
This assigns a frozen variable $z_k$ to each $g$-vector $\boldsymbol{g}^{(T)}_{a_k}$.
These frozen variables are called universal because any other assignment of coefficients can be obtained from these by specializing the frozen variables to particular values.

\vskip4pt
\noindent
\textit{$u$-variables}~\\
Recall the form of generic exchange relations with frozen variables~\eqref{eq: exchange relation w frozen variables}. A \textit{primitive} exchange relation is of the form 
\begin{equation}
    \mu_k(a_k) \, a_k=  p_{\pm}\, S(a)+ p_{\mp}\,,
\end{equation}
for a given seed,
where $S(a)=\prod_{i=1}^{N} \, a_i^{{\rm max}[B_{ik},0]}$ or $\prod_{i=1}^{N} \, a_i^{{\rm max}[-B_{ik},0]}$ is a monomial in the mutable variables, and $p_\pm$ are defined as in~\eqref{eq:ppmdef}.
The novel feature of these exchange relations is that one of the two terms on the right-hand side contains only frozen variables ($p_\mp$).

\vskip4pt
Primitive exchange relations can be identified by enumerating all possible exchange relations, but this becomes fairly tedious for large algebras. Fortunately, 
for all finite-type cluster algebras there exists an automorphism on the space of mutable variables $\tau$, such that the exchange relation between $a_k$ and $\tau(a_k)=\mu_k(a_k)$ is primitive~\cite{Arkani-Hamed:2020tuz,Bazier-Matte:2018rat}. 
With our indexing of $A_N$ and $B_N$, this automorphism corresponds to the relabeling of indices $\tau(a_{i\,j})=a_{i-1\,j-1}$.
 For cluster algebras with universal coefficients, the primitive exchange relations then take the form~\cite{2008arXiv0804.3303Y}
\begin{equation}
    \label{eq: universal primitive exchange relation}
    \tau(a_k)\, a_k=z_k\, S(a) + \prod_{\ell \neq k} z_{\ell}^{(\ell || k)}\,,
\end{equation}
where $z_k$ is the frozen variable corresponding to the vector $g_{a_k}^{(T)}$, and the product is over all other frozen variables with the exponent the compatibility degree of $a_{\ell}$ and $a_{k}$. We can then define a variable $u_k$ for each cluster variable via its primitive exchange relation:
\begin{equation}
    u_k= \frac{z_k\, S(a)}{ \tau(a_k)\, a_k}\,.
\end{equation}
It is not a priori obvious that the $u$s defined this way will satisfy interesting equations, but notice that if we set all the mutable cluster variables in~\eqref{eq: universal primitive exchange relation} to $1$, these primitive exchange relations become 
\begin{equation}
    \label{eq: u equations in z universal}
    1=z_k + \prod_{\ell \neq k} z_{\ell}^{(\ell || k)}\,,
\end{equation}
which are simply $u$-equations for the variables $z_k$. There is a sense in which this is a particular ``gauge" for the $u$-variables, so they actually continue to obey these equations for generic $a$~\cite{Arkani-Hamed:2020tuz}.

\vskip4pt
\noindent
\textit{Example: $A_2$ universal coefficients and $u$-variables}~\\
It is again useful to illustrate things using $A_2$. We first need the $g$-vectors for the transpose matrix $B^T$, which has extended exchange matrix
\be
    \tilde B^T =\left(
    \begin{array}{cc}
    0 & -1 \\
    1 & 0 \\
    1&0\\
    0&1
    \end{array}
    \right) \,.
\ee
Following the same steps as in the previous section, we find the $g$-vectors:
\be
 \boldsymbol{g}^{(T)}_{a_1} =  
 \left(
    \begin{array}{c}
    1\\
    0
    \end{array}
    \right)
    \quad\quad
     \boldsymbol{g}^{(T)}_{a_2} =  
 \left(
    \begin{array}{c}
    0\\
    1
    \end{array}
    \right)
    \quad\quad
 \boldsymbol{g}^{(T)}_{a_3} =  
 \left(
    \begin{array}{c}
    -1\\
    0
    \end{array}
    \right)
    \quad\quad
    \boldsymbol{g}^{(T)}_{a_4} =  
 \left(
    \begin{array}{c}
    0\\
    -1
    \end{array}
    \right)
        \quad\quad
    \boldsymbol{g}^{(T)}_{a_5} =  
 \left(
    \begin{array}{c}
    1\\
    -1
    \end{array}
    \right)\,.
        \label{eq:a2gvec22}
\ee
Appending (the transpose of) each of these $g$-vectors~\eqref{eq:a2gvec22} to the $A_2$ exchange matrix, we obtain the extended exchange matrix
\begin{equation}
    \tilde B =\left(
    \begin{array}{cc}
    0 & 1 \\
    -1 & 0 \\
    1 &0 \\
    0 &1\\
    -1 & 0\\
     0& -1\\
    1& -1
    \end{array}
    \right) \,,
\end{equation}
which corresponds to the quiver
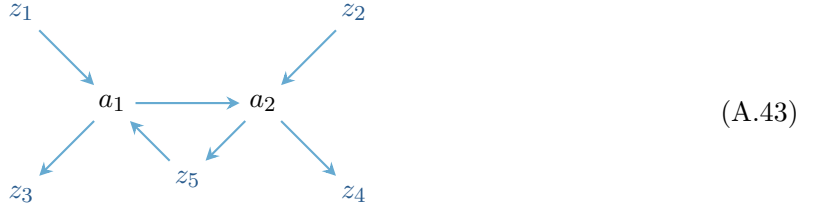
\be
\raisebox{-34pt}{
\begin{tikzpicture}[
  quiver/.style={
    draw=tubeBlue!60,
    -{Stealth[length=5pt,width=5pt]},
    line width=0.8pt
  },
]
  \node (a1) at (0,0) {$a_1$};
  \node (a2) at (2,0) {$a_2$};
  \node (z1) at (-1.2,1.2) {$\color{darkblue}{z_1}$};
  \node (z2) at (3.2,1.2) {$\color{darkblue}{z_2}$};
  \node (z5) at (1,-1) {$\color{darkblue}{z_5}$};
  \node (z3) at (-1.2,-1.2) {$\color{darkblue}{z_3}$};
  \node (z4) at (3.2,-1.2) {$\color{darkblue}{z_4}$};
  \draw[quiver] (a1) -- (a2);
  \draw[quiver] (z1) -- (a1);
  \draw[quiver] (z2) -- (a2);
  \draw[quiver] (a1) -- (z3);
  \draw[quiver] (a2) -- (z4);
  \draw[quiver] (z5) -- (a1);
  \draw[quiver] (a2) -- (z5);
\end{tikzpicture}
}
\ee
where $a_1,a_2$ are mutable variables and $z_1,\cdots, z_5$ are frozen variables related to the $g$-vectors. Performing the same mutations as above generates the exchange relations 
\begin{align}
    \label{eq: A2 universal exchange relations}
    a_3\, a_1&=a_2 z_3+z_1 z_5 \\
    a_4\, a_1&=a_5 z_1+z_3 z_4 \nonumber\\
    a_4\, a_2&=a_3 z_4+z_1 z_2 \nonumber\\
    a_5\, a_2&=a_1 z_2+z_4 z_5 \nonumber \\
    a_3\, a_5&=a_4 z_5+z_2 z_3 \nonumber\,.
\end{align}
From this, we see that
all exchange relations in $A_2$ are primitive. Using the index labeling in~\eqref{eq: A2 variable chord assignments}, we can identify the action of the automorphism $\tau$ on the cluster variables,  
\begin{align}
    \tau(a)&=\left\{a_{1\,3}\to a_{2\,5}\to a_{1\,4}\to a_{3\,5}\to a_{2\,4}\to a_{1\,3}\right\}\\
    &=\left\{a_1\to a_4\to a_2\to a_5\to a_3\to a_1\right\}\nonumber
\end{align}
and then define the $u$-variables 
\begin{align}
    u_{1}&=\frac{a_5 z_1}{a_1 a_4},\quad &u_{2}=\frac{a_1 z_2}{a_2 a_5}\\
    u_{3}&=\frac{a_2 z_3}{a_1 a_3},\quad &u_{4}=\frac{a_3 z_4}{a_2 a_4} \nonumber\\
    u_{5}&=\frac{a_4 z_5}{a_3 a_5}& \nonumber\,.
\end{align}
One can confirm that these variables satisfy the following $u$-equations:
\begin{align}
    1-u_{1}&=u_3 \, u_4,\quad &1-u_{2}=u_4\,  u_5\\
    1-u_{3}&=u_1 \, u_5,\quad &1-u_{4}= u_1\,  u_2 \nonumber\\
    1-u_{5}&=u_2 \, u_3\,.& \nonumber
\end{align}

\subsection{Some Properties}

Now that we know how to construct cluster algebras, we would like to describe some of their elementary properties.

\paragraph{Laurent Phenomenon}~\\
A non-obvious feature of mutation is that the
cluster variables generated by the rule~\eqref{eq: cluster var mutation rule} can always be written as a Laurent polynomial in the initial cluster variables~\cite{FominZelevinsky2002ClusterI}, which further has positive integer coefficients~\cite{lee2014positivityclusteralgebras}.
This interesting fact is known as the {\it Laurent Phenomenon}, and is fairly surprising because repeated mutation involves division by terms of increasing complexity, so it is not obvious that the denominators will not also grow in complexity.
This is already a sign that there is more structure to the mutation procedure than one might first guess.

\paragraph{Finite-Type Cluster Algebras}~\\
Notice that the definition of a cluster algebra involves iterating mutations in all possible ways. Given a randomly chosen initial quiver or exchange matrix, we should not expect this process to ever truncate, and so the generic expectation is that we would find an infinite number of seeds and an infinite number of cluster variables.
In fact, there are two different ways that we can have an infinite number of seeds. In some cases, a finite number of quivers will appear, but an infinite number of distinct cluster variables, and therefore clusters. In other cases, both the number of quivers and the number of variables can be infinite. 

\vskip4pt
In rare cases, the number of seeds is finite.
In this case, the cluster variables generate
so-called \textit{finite-type} cluster algebras. Remarkably, all such finite-type cluster algebras have been classified~\cite{FominZelevinsky2003ClusterII}. Essentially, a cluster algebra is of finite type if one of its seeds has a quiver that is an orientation of a Dynkin diagram.\footnote{More properly for algebras with only skew-symmetrizable $B_{ij}$, one defines the Cartan counterpart of $B$ as
\begin{equation}
    A_{ij}=\begin{cases}
        2& \text{if}\,\, i=j\\
        -\lvert B_{ij} \rvert & \text{if}\,\, i\neq j\,.
    \end{cases}
\end{equation}
Then, a cluster algebra is of finite type if the exchange matrix of one of its seeds has a Cartan counterpart that is a Cartan matrix of finite type (meaning it is positive-definite).
}
The finite type cluster algebras that will appear naturally in connection to cosmology are of $A_N$ and $B_N$ type.

\paragraph{Compatibility and the Exchange Graph}~\\
In the cosmological context, a critical role is played by the concept of compatibility between different cluster variables. Recall that two cluster variables are compatible if they appear together in some cluster.
It will be useful to refine the notion of compatibility by defining the \textit{compatibility degree} $(i || j)$ between two cluster variables, $a_i$ and $a_j$, which is $0$ if $a_i$ is compatible with $a_j$, and $>0$ otherwise. In a sense the compatibility degree measures how incompatible two cluster variables are, and the precise patterns of compatibility degrees depend on the particular cluster algebra. (In particular we will see that they have different properties for $A_N$ and $B_N$ type cluster algebras.)

\vskip4pt
As we saw for the $A_N$ examples, the relations between clusters and the compatibilities of cluster variables can be geometrically encoded in terms of what is called the exchange graph of the cluster algebra. Clusters that differ by a single cluster variable are adjacent, with mutation on that variable relating them. The exchange graph captures the mutation relations between different seeds. For the algebras of finite type that we consider, the vertices of this graph are the vertices of an $N$-dimensional polytope (the {\it cluster polytope}) with $k$ facets where $N$ is the rank of the algebra and $k$ is the number of unique cluster variables. Compatible cluster variables correspond to facets that meet on some lower-dimensional locus. 

\vskip4pt
The compatibility between cluster variables is also represented in the $g$-vectors of the cluster algebra. For a finite-type algebra, the $g$-vector fan tiles all of ${\mathbb R}^N$, with cones defined by the $g$-vectors of variables in a cluster. In addition to this, the $g$-vectors themselves are the normal vectors to the cluster polytope.

\subsection[$A_N$ and $B_N$ in Context]{\texorpdfstring{$\bm{A_N}$}{A_N} and \texorpdfstring{$\bm{B_N}$}{B_N} in Context}
\label{app: Structures of An and Bn}

Perhaps the most fascinating feature of cluster algebras is their ubiquity. They appear in many seemingly unrelated areas of mathematics and physics, providing unexpected connections. Indeed, their appearance in cosmology is another example of this phenomenon. Here we wish to describe some mathematical appearances of the finite-type cluster algebras $A_N$ and $B_N$ that are most closely related to their application to cosmological correlators.

\paragraph{Polygon Triangulations}~\\
A nice way to express cluster variables and their compatibilities is to realize them as chords of a polygon (in the non-exceptional cases). Different seeds/clusters then correspond to different triangulations~\cite{Arkani-Hamed:2020tuz}. Variables that are compatible correspond to diagonals that do not cross, while variables that correspond to crossing diagonals are incompatible. Both $A_N$ and $B_N$ cluster algebras can be realized simply in these terms, which provides intuition for their properties.\footnote{In fact, all finite-type cluster algebras can be realized in this way, with some additional structure~\cite{Arkani-Hamed:2020tuz}.}

\vskip4pt
\noindent
\textit{$A_N$ Cluster algebra:}~\\
The cluster variables of $A_N$ cluster algebras are naturally labeled by the diagonals of an $(N+3)$-gon, 
which are the fully internal lines that connect vertices of the $(N+3)$-gon. 
It is convenient to label these diagonals as $(i\,j)$ and the variables as $a_{i\,j}$, where $i$ and $j$ are the two vertices connected by the diagonal (which must not be adjacent).
An $(N+3)$-gon has $N(N+3)/2$ diagonals, which is the total number of cluster variables of $A_N$. 
 Aside from being a convenient way to label the cluster variables, the compatibilities of diagonals are shared by cluster variables.
The compatibility degree between variables $a_{i\,j}$ and $a_{k\,l}$ is $1$ if the chords cross and $0$ if they do not. (Meaning non-crossing diagonals are compatible, while crossing ones are not.)

 \vskip4pt
In this language, clusters correspond to maximal collections of compatible diagonals, which are triangulations of the $(N+3)$-gon. 
These are counted by the Catalan numbers $C_{N+1} \equiv (2N+2)!/[(N+1)!(N+2)!]$.
Mutation serves to flip a single diagonal, so that two clusters are adjacent in the exchange graph if they differ by a single diagonal.
As an example, one of the clusters of $A_4$ corresponds to the following triangulation of the heptagon
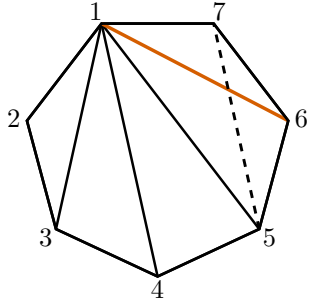
\begin{equation*}
\begin{tikzpicture}[
  scale=.65,
  boundary/.style={
    black,
    line width=1.pt,
  },
  tri/.style={
    black,
    line width=.95pt
  }
]
\coordinate (O) at (0,0);
\foreach \lab/\ang in {
  1/115,
  7/65,
  6/10,
  5/-40,
  4/-90,
  3/-140,
  2/170
}{
  \coordinate (v\lab) at (\ang:2.7);
  \node[font=\small] at ($(O)!1.1!(v\lab)$) {\lab};
}
\draw[boundary] (v1) -- (v7) -- (v6) -- (v5) -- (v4) -- (v3) -- (v2) -- cycle;
\draw[tri] (v1) -- (v3);
\draw[tri] (v1) -- (v4);
\draw[tri] (v1) -- (v5);
\draw[line width=1.3pt,color=tubeRed] (v1) -- (v6);
\draw[dashed, line width=1.1pt] (v5) -- (v7);
\draw[boundary] (v1) -- (v7) -- (v6) -- (v5) -- (v4) -- (v3) -- (v2) -- cycle;
\end{tikzpicture}
\end{equation*}
Mutating the variable $a_{16}$ corresponds to swapping the solid red diagonal for the dashed diagonal that crosses it, producing a new triangulation.

\vskip4pt
\noindent
\textit{$B_N$ Cluster algebra:}~\\
The compatibility relations and clusters of the $B_{N-1}$ cluster algebra can be simply captured by chords of a $2N$-gon. In contrast to $A_N$, where every chord of a polygon corresponds to a cluster variable, we now associate cluster variables to configurations of chords that are invariant under a $180^\circ$ rotation of the polygon. 
Specifically, cluster variables correspond to pairs of chords that map to each other under a $180^\circ$ rotation, or to a single diameter chord, which maps to itself under such a rotation. (Such configurations are called centrally symmetric.) For example, in the case of $B_2$, cluster variables are related to chords of the hexagon like the following
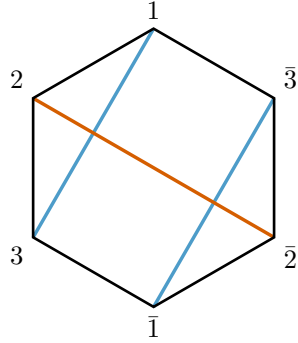
\begin{equation}
\raisebox{-75pt}{
\begin{tikzpicture}[
  scale=1.15,
  every node/.style={font=\small},
  vtx/.style={circle,fill=black,inner sep=1.2pt},
  poly/.style={line width=1.pt},
  diag/.style={line width=1.3pt, draw=tubeBlue!70}
]
\coordinate (1)  at (90:1.6);
\coordinate (2)  at (150:1.6);
\coordinate (3)  at (-150:1.6);
\coordinate (b1) at (-90:1.6);
\coordinate (b2) at (-30:1.6);
\coordinate (b3) at (30:1.6);
\foreach \p in {1,2,3,b1,b2,b3}
  \node[] at (\p) {};
\node[above]       at (1)  {$1$};
\node[above left]  at (2)  {$2$};
\node[below left]  at (3)  {$3$};
\node[below]       at (b1) {$\bar1$};
\node[below right] at (b2) {$\bar2$};
\node[above right] at (b3) {$\bar3$};
\draw[diag] (1) -- (3);
\draw[diag] (b1) -- (b3);
\draw[line width=1.3pt,color=tubeRed] (2) -- (b2);
\draw[poly] (1)--(2)--(3)--(b1)--(b2)--(b3)--cycle;
\end{tikzpicture}
}
\label{eq:BNexs}
\end{equation}
Here the pair of blue chords corresponds to a single cluster variable, as does the single (red) diameter. In order to conveniently label these chords, we define $\bar j \equiv j+N$, so that the vertices $j$ and $\bar j$ appear antipodal on the $2N$-gon.
We can then label cluster variables as $a_{[i,j]}$ where $[i,j]$ denotes the pair of chords $\{(i,j),(\bar i, \bar j)\}$ (note $\bar{\bar j} = j$). For a diameter, these two chords coincide so $a_{[i,j]}$ only corresponds to a single diameter. 

\vskip4pt
A new feature of $B_N$ is that---as a consequence of variables corresponding to two chords---there are now different degrees of incompatibility. The compatibility degree $(a_{[i,j]} \lvert \rvert a_{[k,l]})$ is given by the number of crossings between {\it either one} of the chords corresponding to $a_{[i,j]}$ with {\it all} the chords representing $a_{[k,l]}$~\cite{Arkani-Hamed:2020tuz}. Note that this definition is not symmetric. For example $(a_{[2,\bar{2}]}||a_{[1,3]})=2$, but $(a_{[1,3]}||a_{[2,\bar{2}]})=1$, as can be seen from~\eqref{eq:BNexs}.
As for $A_N$, clusters of $B_{N-1}$ correspond to complete triangulations of the $2N$-gon, subject to the restriction that they are symmetric under a $180^\circ$ rotation. Examples of valid triangulations of this kind of the hexagon and octagon are
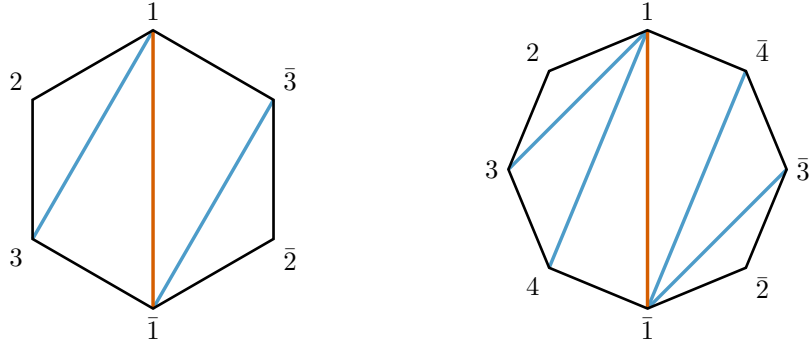
\begin{equation*}
\begin{tikzpicture}[
  scale=1.15,
  every node/.style={font=\small},
  vtx/.style={circle,fill=black,inner sep=1.2pt},
  poly/.style={line width=1.pt},
  diag/.style={line width=1.3pt, draw=tubeBlue!70}
]
\coordinate (1)  at (90:1.6);
\coordinate (2)  at (150:1.6);
\coordinate (3)  at (-150:1.6);
\coordinate (b1) at (-90:1.6);
\coordinate (b2) at (-30:1.6);
\coordinate (b3) at (30:1.6);
\foreach \p in {1,2,3,b1,b2,b3}
  \node[] at (\p) {};
\node[above]       at (1)  {$1$};
\node[above left]  at (2)  {$2$};
\node[below left]  at (3)  {$3$};
\node[below]       at (b1) {$\bar1$};
\node[below right] at (b2) {$\bar2$};
\node[above right] at (b3) {$\bar3$};
\draw[diag] (1) -- (3);
\draw[diag] (b1) -- (b3);
\draw[line width=1.3pt,color=tubeRed] (1) -- (b1);
\draw[poly] (1)--(2)--(3)--(b1)--(b2)--(b3)--cycle;
\end{tikzpicture}
\hspace{2.25cm}
\begin{tikzpicture}[
  scale=1.15,
  every node/.style={font=\small},
  vtx/.style={circle,fill=black,inner sep=1.2pt},
  poly/.style={line width=1.pt},
  diag/.style={line width=1.3pt, draw=tubeBlue!70}
]
\coordinate (1)  at (90:1.6);
\coordinate (2)  at (135:1.6);
\coordinate (3)  at (180:1.6);
\coordinate (4)  at (225:1.6);
\coordinate (b1) at (270:1.6);
\coordinate (b2) at (315:1.6);
\coordinate (b3) at (0:1.6);
\coordinate (b4) at (45:1.6);
\foreach \p in {1,2,3,4,b1,b2,b3,b4}
  \node[] at (\p) {};
\node[above]       at (1)  {$1$};
\node[above left]  at (2)  {$2$};
\node[left]        at (3)  {$3$};
\node[below left]  at (4)  {$4$};
\node[below]       at (b1) {$\bar1$};
\node[below right] at (b2) {$\bar2$};
\node[right]       at (b3) {$\bar3$};
\node[above right] at (b4) {$\bar4$};
\draw[diag] (1) -- (3);
\draw[diag] (b1) -- (b3);
\draw[diag] (1) -- (4);
\draw[diag] (b1) -- (b4);
\draw[line width=1.3pt,color=tubeRed] (1) -- (b1);
\draw[poly] (1)--(2)--(3)--(4)--(b1)--(b2)--(b3)--(b4)--cycle;
\end{tikzpicture}
\end{equation*}

From the perspective of polygon triangulations, it is natural to also assign variables to the facets of the polygon, which are frozen variables that are compatible with all the cluster variables. This will also have the benefit of making some formulas more uniform, but they can of course all be set to $1$ if desired.

\paragraph{Graph Tubings}~\\
The combinatorics of graphs plays an important role in the perturbative computation and representation of cosmological observables.
There is also a close connection between the properties of graphs and cluster algebras, which serves as an interface between the two subjects in the main text.

\vskip4pt
To a graph, ${\cal G}$, one can naturally assign a partial ordering to the space of connected (proper) subgraphs, $T_{\cal G}$, which are called {\it tubes}~\cite{carr2005coxetercomplexesgraphassociahedra,devadoss2006realizationgraphassociahedra}.
Tubes can be labeled by the vertices that they include, and so are typically visualized by encircling these vertices. For example, the following is a tube of a chain graph
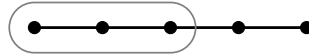
\begin{equation*}
\chaingraph{5}[][v1/v3]
\end{equation*}
We will often be interested in collections of tubes that are compatible with each other. Two tubes are said to intersect if they have a nonzero intersection as sets, and neither is a subset of the other. Two tubes are said to be adjacent if their intersection is null and their union is a connected subgraph of ${\cal G}$. Two tubes are then {\it compatible} if they neither intersect nor are adjacent. Pictorially, tubes are compatible if their circlings do not intersect, and are not directly next to each other.

\vskip4pt
A collection of compatible tubes is called a tubing, and a \textit{maximal tubing} is a maximal set of compatible tubes in $T_{\cal G}$. For a graph with $N$ vertices, such a maximal tubing consists of $N-1$ tubes. Tubings of a graph are a partially ordered set.
The relations between tubings can be visualized using a polytope called the \textit{graph associahedron}~\cite{carr2005coxetercomplexesgraphassociahedra,devadoss2006realizationgraphassociahedra}. As an example, the graph associahedron of the three-site chain graph is the pentagon:
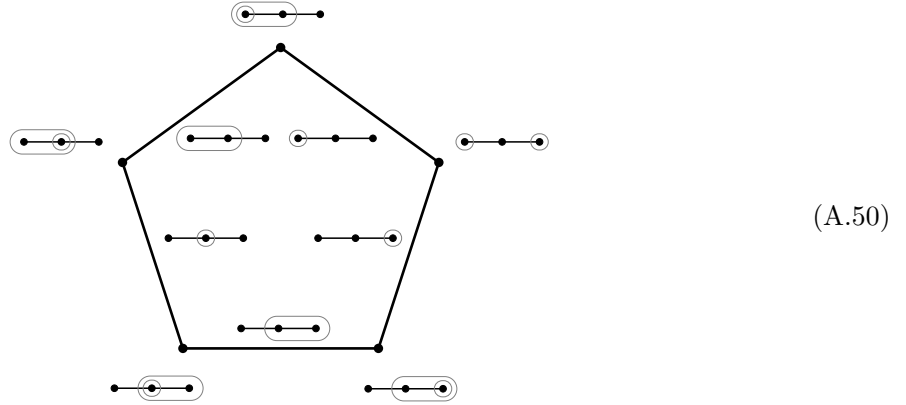
\begin{equation}
\raisebox{-70pt}{
\begin{tikzpicture}[
  scale=2.2,
  every node/.style={font=\footnotesize},
  vtx/.style={circle,fill=black,inner sep=1.25pt},
  pent/.style={line width=1.pt},
  qarrow/.style={
    draw=darkblue!60,
    -{Stealth[length=4pt,width=4pt]},
    line width=0.85pt
  }
]
\def\r{1.0}
\def\roff{0.4}
\def\foff{0.16} 
\coordinate (A) at (90:\r);
\coordinate (B) at (162:\r);
\coordinate (C) at (234:\r);
\coordinate (D) at (306:\r);
\coordinate (E) at (18:\r);
\draw[pent] (A) -- (B) -- (C) -- (D) -- (E) -- cycle;
\node[vtx] at (A) {};
\node[vtx] at (B) {};
\node[vtx] at (C) {};
\node[vtx] at (D) {};
\node[vtx] at (E) {};
\newcommand{\seedright}[3]{%
  \begin{scope}[shift={#1}]
    \node[anchor=east,inner sep=1pt] (L) at (-0.14,0) {$#2$};
    \node[anchor=west,inner sep=1pt] (R) at ( 0.14,0) {$#3$};
    \draw[qarrow] (L.east) -- (R.west);
  \end{scope}
}
\newcommand{\seedleft}[3]{%
  \begin{scope}[shift={#1}]
    \node[anchor=east,inner sep=1pt] (L) at (-0.14,0) {$#2$};
    \node[anchor=west,inner sep=1pt] (R) at ( 0.14,0) {$#3$};
    \draw[qarrow] (R.west) -- (L.east);
  \end{scope}
}
\node at ($(E)+(18:\roff)$) {\scalebox{.55}{{\chaingraph{3}[v1,v3][]}}};
\node at ($(A)+(90:.5*\roff)$) {\scalebox{.55}{{\chaingraph{3}[v1][v1/v2]}}};
\node at ($(B)+(162:\roff)$) {\scalebox{.55}{{\chaingraph{3}[v2][v1/v2]}}};
\node at ($(C)+(234:.74*\roff)$)  {\scalebox{.55}{{\chaingraph{3}[v2][v2/v3]}}};
\node at ($(D)+(306:.75*\roff)$){\scalebox{.55}{{\chaingraph{3}[v3][v2/v3]}}};

\node at ($($(E)!0.5!(A)$)+(234:1.55*\foff)$)  {\scalebox{.55}{{\chaingraph{3}[v1][]}}};
\node at ($($(A)!0.5!(B)$)+(306:1.55*\foff)$) {\scalebox{.55}{{\chaingraph{3}[][v1/v2]}}};
\node at ($($(B)!0.5!(C)$)+(18:2.1*\foff)$) {\scalebox{.55}{{\chaingraph{3}[v2][]}}};
\node at ($($(C)!0.5!(D)$)+(90:.75*\foff)$)  {\scalebox{.55}{{\chaingraph{3}[][v2/v3]}}};
\node at ($($(D)!0.5!(E)$)+(162:2.1*\foff)$) {\scalebox{.55}{{\chaingraph{3}[v3][]}}};
\end{tikzpicture}
}
\label{eq:clusterassoc2}
\end{equation}
Its vertices are labeled by the maximal tubings of the graph (which contain two tubes in this case), and its facets correspond to the individual admissible tubes of the graph. Traversing an edge connecting two maximal tubings corresponds to swapping a single tube to produce a new maximal tubing.

\vskip4pt
\noindent
{\it Tubes and Cluster Variables}~\\
There is a clear parallel between maximal graph tubings and clusters. In some cases this goes beyond a parallel to a precise isomorphism between tubings and cluster variables. In order for such an isomorphism to exist, it is necessary for the graph associahedron to coincide with the cluster polytope of a given cluster algebra. This does {\it not} happen in general, but does for some cases of interest.
The graph associahedron for $(N+1)$-site chain graphs is the $N$-dimensional associahedron~\cite{assoc}, which is the cluster polytope for $A_N$. Similarly, the graph associahedron for $(N+1)$-site cycle graphs is the $N$-dimensional cyclohedron~\cite{cyclo}, which is the cluster polytope for the cluster algebras $B_{N}$ and $C_N$.\footnote{The algebras $B_N$ and $C_N$ have the same exchange graph and structure of clusters, but the variables themselves are different Laurent polynomials in terms of the initial cluster variables for $N>2$.} Therefore in these cases we can find a map between the tubes of a graph and cluster variables~\cite{Arkani-Hamed:2020tuz,He:2020onr}. The cases of interest in the main text are $A_N$ and $B_N$, so we describe these mappings explicitly.

\vskip4pt
\noindent
{\it $A_N$ variables:}~We first describe the mapping between tubes and cluster variables for $A_N$, which corresponds to the tubes of an $(N+1)$-site chain graph. In order to simplify the map, it is convenient to label the possible tubes of the graph in a manner that is similar to the labeling of cluster variables. Recall that $A_N$ cluster variables can be simply labeled by the diagonals of an $(N+3)$-gon as $a_{ij}$. It is therefore useful to introduce edge-centric labels where we label the spaces between the sites of the graph from $1$ to $N+2$. In this scheme, we also add labels to the left of the first vertex and to the right of the last vertex. For example, the $3$-site chain is labeled as
\begin{equation*}
{\tikzset{every picture/.style={scale=1}}
\chaingraph{3}
  [][]
  [v1/x1/{$1$}/{-13pt}/{-8pt},
  v1/v2/{$2$}/{0pt}/{-8pt},
   v2/v3/{$3$}/{0pt}/{-8pt},
   v3/x3/{$4$}/{0pt}/{-8pt}
   ]
   }
\end{equation*}
Each tube can then be labeled by the leftmost and rightmost indices that enclose its vertices. The full set of these tubes can then be written as
\begin{equation}
    \label{eq: Chain tubes}
T_{\text{$(N+1)$-site chain}}= \Big\{
        T_{i\,j}  ~~\Big\vert~~
        1\leq i<j\leq N+2,\,~ (i,j)\neq (1,N+2)\Big\}\,.
\end{equation}
Note that the full graph is enclosed by the would-be tube, 
$T_{1\,N+2}$, which formally is not an allowed tube of the graph, because it is not a proper subgraph.

\vskip4pt
The benefit of the indexing~\eqref{eq: Chain tubes} is that the map between tubes and cluster variables takes the simple form:
\begin{equation}
    \label{eq: Tube to cluster variable map An}
    T_{ij} \,\cong\, a_{i\,j+1}\,.
\end{equation}
It is illuminating to enumerate the correspondence between tubings and cluster variables for the explicit example of the $3$-site chain, where the correspondence is
\begin{align}
a_{13} &\cong T_{12} = \,\raisebox{2pt}{\chaingraph{3}[v1][]}\\
a_{24} &\cong T_{23} = \,\raisebox{2pt}{\chaingraph{3}[v2][]}\\
a_{35} &\cong T_{34} = \,\raisebox{2pt}{\chaingraph{3}[v3][]}\\
a_{14} &\cong T_{13} = \,\raisebox{2pt}{\chaingraph{3}[v1/v2][]}\\
a_{25} &\cong T_{24} = \,\raisebox{2pt}{\chaingraph{3}[v2/v3][]}
\end{align}
It is straightforward to check that~\eqref{eq:clusterassoc2} and~\eqref{eq:clusterassoc1} encode the same information using this mapping along with the identification~\eqref{eq: A2 variable chord assignments}.

\vskip4pt
In a similar way we can map tubes of an arbitrary $(N+1)$-site chain graph to the cluster variables of $A_N$, preserving the notion of compatibility.

\vskip4pt
\noindent
{\it $B_N$ variables:}~It is also possible to map the tubes of an $N$-site loop to cluster variables, in this case corresponding to those of the  $B_{N-1}$ cluster algebra.

\vskip4pt
In this case, the labeling scheme is slightly more involved. In the same way that $B_{N-1}$ cluster variables are naturally associated to pairs of chords of a $2N$-gon, we assign a pair of labels to each edge $(i,\bar{i})$, clockwise, where $\bar{i}=i+N$. For example, the 4-site loop is labeled as 
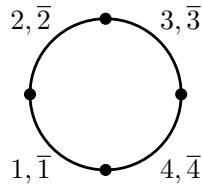
\begin{equation*}
    \loopdraw{4}
\end{equation*}
The set of tubes is then 
\begin{equation}
    T_{\text{$N$-site loop}}
   =
   \Big\{
    T_{[i,j]} ~~\Big\vert~~ [i,j]=\{(i,j),(\bar{i},\bar{j})\}  ~{\rm with}~1\leq i <j \leq N
\Big\}\,,
\end{equation} 
where indices can be either barred or unbarred.
Traversing the indices clockwise, $T_{[i,j]}$ is the tube enclosing sites between $i$ and $j$ or equivalently $\bar{i}$ and $\bar{j}$. These tubes have the same compatibility as $B_{N-1}$ cluster variables through the mapping
\be
    \label{eq: Tube to cluster variable map Bn}
    T_{[i,j]} \,\cong\, a_{[i,j+1]}\,.
\ee
In the explicit example of the 4-site loop and $B_3$ cluster variables, this correspondence is
\begin{align}
a_{[1,3]} &\cong T_{[1,2]} = \vcenter{\hbox{\loopdrawtube{1}{1}{4}}}, \quad &a_{[2,4]} &\cong T_{[2,3]} = \vcenter{\hbox{\loopdrawtube{2}{2}{4}}}\\
a_{[1,\bar{3}]} &\cong T_{[3,4]} = \vcenter{\hbox{\loopdrawtube{3}{3}{4}}},\quad &a_{[2,\bar{4}]} &\cong T_{[1,\bar{4}]} = \vcenter{\hbox{\loopdrawtube{4}{4}{4}}} \nonumber\\
a_{[1,4]} &\cong T_{[1,3]} = \vcenter{\hbox{\loopdrawtube{1}{2}{4}}}, \quad &a_{[1,\bar{2}]} &\cong T_{[2,4]} = \vcenter{\hbox{\loopdrawtube{2}{3}{4}}} \nonumber\\
a_{[2,\bar{3}]} &\cong T_{[1,\bar{3}]} = \vcenter{\hbox{\loopdrawtube{3}{4}{4}}},\quad &a_{[3,\bar{4}]} &\cong T_{[2,\bar{4}]} = \vcenter{\hbox{\loopdrawtube{4}{5}{4}}}\nonumber\\
a_{[1,\bar{1}]} &\cong T_{[1,4]} = \vcenter{\hbox{\loopdrawtube{1}{3}{4}}}, \quad &a_{[2,\bar{2}]} &\cong T_{[1,\bar{2}]} = \vcenter{\hbox{\loopdrawtube{2}{4}{4}}} \nonumber\\
a_{[3,\bar{3}]} &\cong T_{[2,\bar{3}]} = \vcenter{\hbox{\loopdrawtube{3}{5}{4}}},\quad &a_{[4,\bar{4}]} &\cong T_{[3,\bar{4}]} = \vcenter{\hbox{\loopdrawtube{4}{6}{4}}} \nonumber\,.
\end{align}

\newpage
\section{Polylogarithms and Symbols}
\label{app: Symbology}

The de Sitter wavefunction coefficients we study in the main text are integrals of rational functions composed of linear factors~\eqref{eq:dsWFint}. This makes it manifest that they (and the corresponding correlators) will evaluate to (generalized) polylogarithmic functions~\cite{10.2969/jmsj/02720248,Arkani-Hamed:2017fdk,Arkani-Hamed:2017ahv}.
These functions can be
written as (linear combinations of) iterated integrals of the schematic form
\be
    \label{eq: polylog iterated integral definition}
    f_{(n)} 
    = \int \rd \log R_{1} \circ \cdots \circ \rd \log R_{n} \,,
\ee
where the $R_I$ are rational factors. It is convenient to grade iterated integrals like this by the number of integrations, which we call the {\it transcendentality} (or sometimes weight) of the function.

\vskip4pt
Most of the information that we need for physical purposes (like the discontinuities and differentials of the functions) is actually determined by the $R_I$ and so it is useful to have a way to keep track of this information. These properties are efficiently encoded in the 
\textit{symbol}~\cite{goncharov2001multiple,goncharov2009simple,Goncharov:2010jf}
\begin{equation}
    \label{eq: symbol definition}
    \mathcal{S}\big(f_{(n)}\big) \equiv R_{1}\otimes R_{2}\otimes \cdots \otimes R_{n}\,,
\end{equation}
where $\otimes$ is a tensor product with the same multilinear properties as products of logarithms. The rational functions $R_I$ are referred to as {\it letters} and the product of letters appearing in a single term as in~\eqref{eq: symbol definition} is called a {\it word} (with the set of all letters appearing in a symbol called its alphabet).
The compression of information provided by the symbol is naturally lossy.\footnote{The symbol is part of a larger algebraic structure involving a coaction on polylogarithmic/hypergeometric functions (see e.g.,~\cite{brown2017notesmotivicperiods}).
Aspects of this coaction in the cosmology context have been studied recently in~\cite{McLeod:2026jpz,McLeod:2026kpo}.} Terms of lower transcendentality or transcendental numbers like $\pi$ or $\log 2$ are projected out by the symbol map.
This information can be recovered by ``integrating" the symbol, either by directly computing~\eqref{eq: polylog iterated integral definition} or by imposing suitable boundary conditions on an ansatz.

\vskip4pt
In this appendix, we review some of the mathematical features of the symbol that we utilize in the main text. More complete discussions can be found in~\cite{Duhr:2011zq,Duhr:2014woa}.

\paragraph{Basic Properties}~\\
The symbol of the general polylogarithmic functions appearing in the main text can be written in the schematic form
\begin{equation}\label{eq: Generic symbol polylog app}
    \mathcal{S}\big(F_{(n)}\big)=\sum_{I_1,\ldots, I_n} C_{I_1\cdots I_n} \, R_{I_1}\otimes \cdots \otimes R_{I_n}\,,
\end{equation}
with $C_{I_1\cdots I_n}$ constant coefficients. Let us describe some of its elementary properties.

\begin{itemize}
\item {\bf Multilinearity:}
The product $\otimes$ behaves like an ordinary product if we think of the $R_i$ as arguments of logarithms:
\be
\begin{aligned}
    \label{eq: symbol sum rule}
    a_1 \otimes a_2 - a_1 \otimes b_2 &= a_1 \otimes \frac{a_2}{b_2}\\
    a_1 \otimes a_2 + a_1 \otimes b_2 &= a_1 \otimes \left(a_2\,  b_2 \right)\,.
\end{aligned}
\ee

If we multiply two functions, their symbols multiply via the {\it shuffle product}, so that
\begin{equation}
    \mathcal{S}\big(F_{(n)}\times G_{(m)}\big)= \mathcal{S}\big(F_{(n)}\big) \shuffle\, \mathcal{S}\big(G_{(m)}\big) \,,
\end{equation}
is the symbol of the product of two functions $F_{(n)}$ and $G_{(m)}$ of transcendentality $n$ and $m$, respectively. The shuffle product sums over all possible ways of interleaving the entries of the two original symbols, while preserving the relative ordering of the entries of each constituent symbol. (The name is meant to evoke the shuffling of cards by interleaving two halves of the deck.)
Concretely, the shuffle product is defined recursively by
\be
\left(a\otimes u\right)\, \shuffle\,\left(b\otimes v \right)= a\otimes \Big( u \shuffle\left(b\otimes v\right)\Big)
+ b\otimes \Big(\left(a\otimes u\right)\shuffle v\Big)\,,
\ee
along with $a\shuffle {\mathds 1}={\mathds 1} \shuffle a= a$, with ${\mathds 1}$ the empty word.

\item {\bf Discontinuities:} The symbol efficiently encodes the discontinuities of the corresponding function. A function with a symbol of the form~\eqref{eq: Generic symbol polylog app} has branch points where $R_{I_1} $ has zeroes or poles. The symbol of the discontinuity across this branch cut is obtained by deleting the leftmost entry $R_{I_1}$:
\begin{equation}
    \mathcal{S}\Big((2\pi i)^{-1}{\rm Disc}_{R_{I_1}}\big[F_{(n)}(z)\big]\Big)=  \sum_{I_2,\cdots, I_n} C_{I_1\cdots I_n} \, R_{I_2}(z)\otimes \cdots \otimes R_{I_n}(z)\,,
\end{equation} 
which corresponds to a function with transcendentality $n-1$.

\item {\bf Differential:} The symbol also encodes the differential of the function $F_{(n)}$. In this case we move from right to left, so that the differential is given by
\be
\rd F_{(n)} = \sum_{I_1,\ldots, I_n} C_{I_1\cdots I_n} \, R_{I_1}\otimes \cdots \otimes R_{I_{n-1}}~\rd\log R_{I_n}\,,
\label{eq:differentialsymbol}
\ee
which again corresponds to a function of transcendentality $n-1$.

\item {\bf Integrability:}
Not every symbol of the form~\eqref{eq: Generic symbol polylog app} can be integrated to an actual function. This is because the symbol encodes differentials as in~\eqref{eq:differentialsymbol}. As such, the fact that $\rd^2= 0$ implies that not all of the entries are independent. In fact, the 
 coefficients $C_{I_1\cdots I_n}$ must obey the relation~\cite{brown2006multiplezetavaluesperiods,goncharov2009simple,Duhr:2011zq}
\begin{equation}
    \label{eq: Symbol integrability condition}
    \sum_{I_1,\ldots, I_n} C_{I_1,\ldots, I_n} \, R_{I_1}\otimes \cdots \otimes R_{I_{p-1}} \otimes R_{I_{p+2}}\otimes\cdots \otimes R_{I_n} \, \rd \log R_{I_p} \wedge  \rd \log R_{I_{p+1}}=0\,,
\end{equation}
for all $1\leq p\leq n-1$. This condition strongly constrains the space of legal symbol functions given an alphabet of energy singularities.

\end{itemize}

\paragraph{Wavefunction Symbol from its Integrand}~\\
It is not immediately obvious that the wavefunction coefficients that we study are (generalized) polylogarithmic functions, but this follows from the fact that they can be written as integrals of the flat-space wavefunction~\eqref{eq:dsWFint}, which itself is the canonical form of a projective polytope~\cite{Arkani-Hamed:2017fdk,Arkani-Hamed:2017ahv}.
This additional geometric structure also makes it possible to understand the sequences of discontinuities, and therefore the symbol, of the wavefunction by studying the residues of the flat-space wavefunction integrand~\cite{Arkani-Hamed:2017fdk}.

\vskip4pt
The canonical form of a polytope is the unique differential form with logarithmic singularities on its boundaries, which has a recursive structure so that the residue on any boundary is the canonical form of that lower-dimensional geometry~\cite{Arkani-Hamed:2017tmz}. (This implies that its maximal residues are $\pm 1$).
Consider a function
\begin{equation}
  F_{(n)}(X_v,Y_e)=\int_{0}^{\infty} \rd x_1 \ldots \rd x_n \, \Omega(x_v+X_v,Y_e)\, ,
\end{equation}
where $\Omega(X_v, Y_e)$ is a canonical function (the scalar prefactor of the canonical form). By definition, $\Omega$ and all its residues are rational functions with only simple poles.
We can then consider a sequence of $n$ poles $\mathcal{R}=\{R_1(X_v,Y_e),\ldots, R_n(X_v,Y_e)\}$ of $\Omega$ whose combined sequential residue is~$\pm 1$:
\begin{equation}\label{eq: residue sequence}
  \operatorname{Res}_{R_n} \cdots \, \operatorname{Res}_{R_1} \Omega(X_v, Y_e)=\pm 1\equiv \sign\,\mathcal{R} \,.
\end{equation}
Given all such non-vanishing sequences of residues, $\{\mathcal{R}\}$, we can construct the symbol by summing over them~\cite{Arkani-Hamed:2017fdk}
\begin{equation}\label{eq: symbol from residue sequences}
  \mathcal{S}\big(F_{(n)}\big)= \sum_{\{\mathcal{R}\} } \sign\,\mathcal{R}\,~  R_{1}\otimes R_2\otimes \cdots \otimes R_n\,.
\end{equation}
This formula can be understood through the above discussion of discontinuities. The branch points of $F_{(n)}$ arise from poles in the integrand, which define the set of candidate letters for $R_1$. Given some $R_1$, the letters $R_2$ which follow $R_1$ are given by the branch points of $\operatorname{Disc}_{R_1}[F_{(n)}]$, or the poles of $\operatorname{Res}_{R_1}\Omega$, and so on. The sequences in~\eqref{eq: residue sequence} therefore yield constituent symbol words of $F_{(n)}$, and fixing their relative signs fixes the symbol up to an overall sign. 

\vskip4pt
\noindent
{\it Two-site chain:}
As an example, let us derive the symbol of the two-site chain with this method. The wavefunction coefficient is given by the integral
\be
\begin{aligned}
  F_{(2)}(X_1,X_2,Y)&=\int_{0}^{\infty} \rd x_1\rd x_2 \, \Omega_2(x_1+X_1,x_2+X_2,Y) \\
  \Omega_2(X_1,X_2,Y)&=\frac{2 Y}{(X_1+X_2)(X_1+Y)(X_2+Y)} \,.
\end{aligned}
\ee
The canonical function
$\Omega_2$ has three poles with residues
\begin{align}
  \operatorname{Res}_{X_1+Y}\Omega_2 &= \frac{2Y}{(X_2+Y)(X_2-Y)}\\
  \operatorname{Res}_{X_2+Y}\Omega_2 &= \frac{2Y}{(X_1+Y)(X_1-Y)}\\
  \operatorname{Res}_{X_1+X_2}\Omega_2 &= \frac{-2Y}{(X_2+Y)(X_2-Y)}=\frac{-2Y}{(X_1+Y)(X_1-Y)}\,.
\end{align}
Note that for the pole at $X_1+X_2=0$ we have parameterized this locus both in terms of $X_1$ and $X_2$. In general, when a pole contains multiple $X_v$ variables, we need to consider all possible parameterizations in order to see the possible subsequent residues. Taking further residues, we find the following 8 non-vanishing sequences:
\be
\begin{aligned}
  \operatorname{Res}_{X_2+Y}\operatorname{Res}_{X_1+Y}\Omega_2 &= -1
  &\hspace{1.7cm}
   \operatorname{Res}_{X_2-Y}\operatorname{Res}_{X_1+Y}\Omega_2 &= +1 \\
  \operatorname{Res}_{X_1+Y}\operatorname{Res}_{X_2+Y}\Omega_2 &= -1
  &\operatorname{Res}_{X_1-Y}\operatorname{Res}_{X_2+Y}\Omega_2 &= +1 \\
  \operatorname{Res}_{X_2+Y}\operatorname{Res}_{X_1+X_2}\Omega_2 &= +1
  &\operatorname{Res}_{X_2-Y}\operatorname{Res}_{X_1+X_2}\Omega_2 &= -1 \\
  \operatorname{Res}_{X_1+Y}\operatorname{Res}_{X_1+X_2}\Omega_2 &= +1
  &\operatorname{Res}_{X_1-Y}\operatorname{Res}_{X_1+X_2}\Omega_2 &= -1
\end{aligned}
\ee
Using these sequences of residues in~\eqref{eq: symbol from residue sequences}, we recover~\eqref{eq:2sitesymbol}.

\newpage
\section{Interpreting Cluster Compatibility}\label{app: Steinmann}

While the notion of cluster compatibility we utilize in the main text imposes strong constraints on the symbol, the 
nontrivial map between energy singularities and cluster variables makes it difficult to give these constraints a precise physical interpretation. 
Here we wish to contextualize these constraints by comparing them to the notions of compatibility provided by 
Steinmann relations~\cite{Benincasa:2020aoj,Benincasa:2021qcb} and (a simplified version of) kinematic flow~\cite{Arkani-Hamed:2023kig,Arkani-Hamed:2023bsv}. 
(Subsets of these conditions have recently been used to bootstrap the symbol of the wavefunction~\cite{Paranjape:2026htn,Capuano:2026pgq,Ferro:2026oph}.) 
We find cluster compatibility is a stronger constraint, suggesting that it encodes causality and locality, along with additional physical information.

\paragraph{Steinmann Relations}~\\
 The Steinmann relations are constraints on sequential discontinuities of some observable due to causality and locality in the bulk theory. In their original flat-space formulation, this is the requirement that double discontinuities of an amplitude in overlapping channels vanish~\cite{Steinmann1960a,Steinmann1960b,Caron-Huot:2016owq}. 
Since these relations derive from causality and locality, it is natural to expect that they have a manifestation in other observables.
In the context of cosmology, the Steinmann relations impose a similar constraint on the double discontinuities of the wavefunction coefficient~\cite{Benincasa:2020aoj,Benincasa:2021qcb}. For a graph $\cal G$  and subgraphs $\mathfrak{g}_1$ and $\mathfrak{g}_2$,
\begin{equation}
    \label{eq: Steinmann relations}
    \operatorname{Disc}_{E_{\mathfrak{g}_1}}\operatorname{Disc}_{E_{\mathfrak{g}_2}} \psi_\mathcal{G}=0,  \quad \text{when } \mathfrak{g}_1 \text{ and } \mathfrak{g}_2 \text{ are incompatible.}
\end{equation}
In this context, $\mathfrak{g}_1$ and $\mathfrak{g}_2$ are incompatible if their intersection is non-empty and neither is a proper subset of the other. Graphically, the tubes that define the subgraphs literally intersect if the subgraphs are incompatible.

\vskip4pt
As is described in Appendix~\ref{app: Symbology}, the sequential discontinuities of the wavefunction are given by neighboring symbol entries, so Eq.~\eqref{eq: Steinmann relations} implies that no two neighboring symbol letters can correspond to partially overlapping tubes. In~\cite{Benincasa:2021qcb} it was shown that actually stronger extended Steinmann relations hold which bar any two letters in a symbol word from partially overlapping, even if they are not in adjacent slots.

\paragraph{Kinematic Flow}~\\
Kinematic flow~\cite{Arkani-Hamed:2023kig,Arkani-Hamed:2023bsv} is a set of graphical rules that determine the differential equations satisfied by wavefunction coefficients in power-law FLRW spacetimes for theories with polynomial interactions. 
While kinematic flow specifies the full differential of the wavefunction (and hence the symbol directly), we can impose a weaker particular consequence, which is the  constraint that energies in a symbol word, ordered left to right, must correspond to tubes which either shrink or are disjoint
\begin{equation}
    \label{eq: KF rules}
    {\cal S} \supset E_{\mathfrak{g}_1}\otimes \cdots \otimes E_{\mathfrak{g}_2}\otimes \cdots \, \implies \, \mathfrak{g}_1 \text{ and }\mathfrak{g}_2 \, \text{ are disjoint, or }\, \mathfrak{g}_2 \subset \mathfrak{g}_1 \,.
\end{equation}
For example, given the tubes $\textcolor{tubeRed}{\mathfrak{g}_1}$ and $\textcolor{tubeBlue}{\mathfrak{g}_2}$:
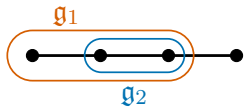
\begin{equation*}
{
\tikzset{
  chain small blob/.append style={color=tubeBlue},
  chain large blob/.append style={color=tubeRed}
}
\chaingraph{4}[v2/v3][v1/v3][v1/v2/{${\color{tubeRed}\mathfrak{g}_1}$}/{0pt}/{15pt},
    v2/v3/{${\color{tubeBlue}\mathfrak{g}_2}$}/{0pt}/{-15pt}]
}
\end{equation*}
The symbol word $E_{\textcolor{tubeRed}{\mathfrak{g}_1}}\otimes E_{\textcolor{tubeBlue}{\mathfrak{g}_2}}$ is compatible with kinematic flow while $E_{\textcolor{tubeBlue}{\mathfrak{g}_2}}\otimes E_{\textcolor{tubeRed}{\mathfrak{g}_1}}$ is not. Note that~\eqref{eq: KF rules} is a strictly stronger constraint than the Steinmann relations.

\paragraph{Constraint Landscape}~\\
We can now repeat the exercise of Section~\ref{sec:clusterbootstrap} and impose these conditions rather than cluster compatibility, and compare the size of the resulting parameter space. We impose the same mathematical and physical conditions described in Section~\ref{sec:clusterbootstrap}, and then consider imposing individually additional constraints from Steinmann relations, kinematic flow, or cluster compatibility (which is what was done in Section~\ref{sec:clusterbootstrap}). The results are tabulated below:
\begin{table}[h]
    \begin{center}
    \begin{tabular}{|l|llllll|}
        \hline
        \multirow{2}{3em}{~~~~~Graph}& \multirow{2}{3em}{Initial Coeffs.}& \multirow{2}{7em}{Math and Phys. Conditions}& \multirow{2}{5em}{+Steinmann}& \multirow{2}{3.5em}{+K. F.}& \multirow{2}{4em}{+Cluster}& \multirow{2}{6em}{+No Repeat Cluster Vars.}\\ 
        &&&&&&\\
        \hline
        2-Site Chain&36 &\hspace{.5cm}2&\hspace{.5cm}2&\hspace{.4cm}1&\hspace{.7cm}1&\hspace{.8cm}1\\
        3-Site Chain&3375&\hspace{.5cm}63&\hspace{.5cm}48&\hspace{.4cm}13&\hspace{.7cm}1&\hspace{.8cm}1\\
        4-Site Chain&614656&\hspace{.5cm}961&\hspace{.5cm}308&\hspace{.4cm}14&\hspace{.7cm}2&\hspace{.8cm}1\\
        5-Site Chain&184528125& \hspace{.5cm}29548& \hspace{.5cm}3954&\hspace{.4cm}53 & \hspace{.7cm}3&\hspace{.8cm}1\\
        2-Site Loop&169&\hspace{.5cm}13&\hspace{.5cm}12&\hspace{.4cm}5&\hspace{.7cm}1&\hspace{.8cm}1\\
        3-Site Loop&29791&\hspace{.5cm}348&\hspace{.5cm}151&\hspace{.4cm}22&\hspace{.7cm}1&\hspace{.8cm}1\\
        \hline
    \end{tabular}
\end{center}
\vspace{-15pt}
\end{table}

\noindent
In order to isolate the wavefunction uniquely, in addition to pure cluster compatibility, we can impose that no cluster variables repeat if they first appear in the second entry or later. This distinguishes the wavefunction from products of lower-point functions.

\vskip4pt
The most notable inference from this exercise is that cluster compatibility is substantially more stringent than either Steinmann relations or the simplified version of kinematic flow that we impose separately. It is not surprising that cluster compatibility is stronger than Steinmann relations, since $E_{ij} \propto a_{ij}$, we cannot have Steinmann-incompatible symbol entries without them also being cluster incompatible. It is less clear how kinematic flow is encoded in cluster compatibility, especially since cluster compatibility does not obviously have a directionality like kinematic flow does. It would be very interesting to further understand their relationship.

\vskip4pt
Another perspective is provided by~\cite{Paranjape:2026htn,Capuano:2026pgq}. In~\cite{Paranjape:2026htn} it was shown that the wavefunction can be bootstrapped by imposing the same mathematical and physical conditions we do (except not forbidding repeated energies), along with Steinmann relations, exchange symmetries, and vanishing when internal energies $Y_I$ are taken to zero.
In~\cite{Capuano:2026pgq} it was shown  that one can bootstrap the symbol of the chain graph by imposing the conditions in Section~\ref{sec:clusterbootstrap}, kinematic flow, and requiring words with a first letter corresponding to an outermost graph site take the form of the recursion relation of~\cite{Hillman:2019wgh}.
It is similarly interesting, though not obvious to us, how these conditions are encoded in our notion of cluster compatibility. The benefit of these relations versus cluster compatibility is that they generalize to all graphs. For example, we have checked that the symbol of the four-site star wavefunction coefficient is the unique expression satisfying integrability, our first letter conditions, kinematic flow, and the above physical constraints (along with vanishing at $Y\to 0$). (This agrees with the findings in~\cite{Capuano:2026pgq}.)

\newpage
\section{Correlators}
\label{app: Correlator vs WF}

In the main text we have primarily focused on the cluster compatibility of wavefunction coefficients. It is natural to expect that the structures that we have explored have analogues for correlation functions, because correlators are governed by the same physical principles as the wavefunction. Of course, it is also natural to expect that there will be differences, since correlators involve a reshuffling of the information in the wavefunction and have a strictly smaller singularity alphabet~\cite{Chowdhury:2023arc,Benincasa:2024ptf,Glew:2025arc,Arkani-Hamed:2025mce,Pimentel:2026kqc,Chowdhury:2026dwm}. We find that in-in correlators corresponding to chain graphs share the cluster structure of the wavefunction, while those for loop graphs do not. (In addition, even for chains there are interesting differences in structure depending on whether the number of sites is even or odd~\cite{Chowdhury:2026dwm}.)
Here we explain these features.

\paragraph{From Wavefunction to Correlator}~\\
In-in correlators can be obtained from the wavefunction
by squaring and integrating, in accordance with the Born rule:
\be
\langle \varphi_{\vec k_1}\cdots \varphi_{\vec k_N}\rangle = \frac{\displaystyle\int{\cal D}\varphi\, \varphi_{\vec k_1}\cdots \varphi_{\vec k_N}\,\left\lvert\Psi[\varphi]\right\rvert^2}{\displaystyle\int{\cal D}\varphi\, \left\lvert\Psi[\varphi]\right\rvert^2}\,.
\ee
In perturbation theory, we can expand the wavefunction~\eqref{eq:WFdef2} and perform gaussian integrals over $\varphi$ in order to express this in-in correlator in terms of the wavefunction coefficients~\eqref{eq:WFcoeff}. The result of this procedure has a simple graphical interpretation. The correlator corresponding to a Feynman graph ${\cal G}$ can be written in terms of the (real parts of) wavefunction coefficients related to all possible ways of partitioning the graph into pieces by deleting internal lines~\cite{Benincasa:2024leu,Figueiredo:2025daa}. Concretely we have (for an $N$-point $L$-loop graph)
\begin{equation}\label{eq: correlator in terms of cut WF coeffs}
  \langle \varphi_{\vec k_1}\cdots \varphi_{\vec k_N}\rangle'=\frac{\lambda^{n_{\cal G}} H^{2-2L} \eta_*^N}{\prod_{a=1}^{N} 2 k_a}\left(\prod_{e\in \cal G}\frac{1}{2 Y_e}\right)\left(2{\rm Re}[(-i)^{n_{\cal G}}F_{\cal G}]+\sum_{\rm cuts}\prod_{\mathfrak{g}\in \{\mathcal{G}_{\rm cut}\}}2{\rm Re}[(-i)^{n_{\mathfrak g}}F_{\mathfrak{g}}]\right)\,,
\end{equation}
where the sum runs over all possible ``cuts" of the diagram (ways of deleting internal lines), and we have removed the momentum-conserving delta function. Here ${\mathfrak g}$ denotes the connected subgraphs produced by these deletions, and $n_{\cal G}$ and $n_{\mathfrak g}$ denote the number of vertices of the full graph and subgraph, respectively. If a graph is partitioned into multiple pieces by these cuts, we take the product of $2{\rm Re}[(-i)^{n_{\mathfrak g}}F_{\mathfrak{g}}]$ for each connected component. We have written these formulas in terms of the functions $F$ studied in the main text instead of $\psi$; the translation between the two involves factors of $2Y$ and $(-i)^n$ where $n$ is the number of vertices in the (sub)graph.

\vskip4pt
We can illustrate this with some simple examples. First consider the two-site correlator, which can be written as a sum of two terms
\be
\langle\varphi^4\rangle~ = \hspace{-8pt}
\raisebox{9pt}{
{
\tikzset{
  chain small blob/.append style={
    draw=tubeGreen,
    color=tubeGreen
  }
}
\chaingraph{2}[v1/v2][][v1/v2/{${\color{tubeGreen} {\cal G}_2}$}/{0pt}/{14pt}]
}
}
+
\!\!\!
\raisebox{9pt}{
{
\tikzset{
  chain dashed edges={1},
  chain small blob/.append style={draw=tubeBlue,color=tubeBlue}
}
\chaingraph{2}[v1,v2][][v1/v1/{${\color{tubeBlue} {\mathfrak g_{1,L}}}$}/{0pt}/{14pt},v2/v2/{${\color{tubeBlue} {\mathfrak g_{1,R}}}$}/{0pt}/{14pt}]
}
}
\ee
This implies that the correlator can be written in terms of the two-site wavefunction and a product of two one-site wavefunctions
as
\begin{equation}\label{eq: 2 site chain corr}
    \langle \varphi_{1}\varphi_{2}\varphi_{3}\varphi_{4}\rangle'=\frac{\lambda^2 H^2 \eta_*^4}{\prod_{a=1}^{4} 2 k_a} \frac{1}{2 Y}\Big(-2\Re[F_{\textcolor{tubeGreen}{\cal G}}]+4\Re[iF_{\textcolor{tubeBlue}{\mathfrak{g}_{1,L}}}]\Re[iF_{\textcolor{tubeBlue}{\mathfrak{g}_{1,R}}}]\Big)\,,
\end{equation}
where we should recall that the one-site wavefunction is given by
\be
    -iF_{\textcolor{tubeBlue}{\mathfrak{g}_1}}[X]= \lim_{\eta_* \rightarrow 0}i\int_{-\infty_\epsilon}^{\eta_*}\frac{\rd\eta }{\eta}\E^{i X \eta}=  i\log(-i X \eta_*)\,,
\ee
so that $2\Re[(-i)F_{\textcolor{tubeBlue}{\mathfrak{g}_1}}[X]]=-\pi$. 
We therefore see in this case the correlator and wavefunction only differ by a multiple of $\pi^2$, in agreement with~\cite{Arkani-Hamed:2015bza,Hillman:2019wgh}.
The three-site case is conceptually similar and can be written diagrammatically as
\be
\langle\varphi^5\rangle~ = \hspace{-8pt}
\raisebox{9pt}{
{
\tikzset{
  chain small blob/.append style={
    draw=tubeRed,
    color=tubeRed
  }
}
\chaingraph{3}[v1/v3][][v1/v3/{${\color{tubeRed} {\cal G}_3}$}/{0pt}/{14pt}]
}
}
+
\!\!\!
\raisebox{10pt}{
{
\tikzset{
  chain dashed edges={2},
  every picture/.append style={
    execute at end picture={
      \node[
        chain small blob,
        draw=tubeGreen,
        color=tubeGreen,
        fit=(Lv1)(Rv2)
      ] {};

      \node[
        chain small blob,
        draw=tubeBlue,
        color=tubeBlue,
        fit=(Lv3)(Rv3)
      ] {};

      \node at ($(v1)!0.5!(v2)+(0,14pt)$)
        {${\color{tubeGreen}{\mathfrak g}_{2,L}}$};

      \node at ($(v3)+(0,14pt)$)
        {${\color{tubeBlue}{\mathfrak g}_{1,R}}$};
    }
  }
}
\chaingraph{3}
  [v1/v2,v3/v3]
  []
  []
}
}
+
\!\!\!\!\!
\raisebox{10pt}{
{
\tikzset{
  chain dashed edges={1},
  every picture/.append style={
    execute at end picture={
      \node[
        chain small blob,
        draw=tubeBlue,
        color=tubeBlue,
        fit=(Lv1)(Rv1)
      ] {};

      \node[
        chain small blob,
        draw=tubeGreen,
        color=tubeGreen,
        fit=(Lv2)(Rv3)
      ] {};

      \node at ($(v1)+(0,14pt)$)
        {\color{tubeBlue}${\mathfrak g}_{1,L}$};

      \node at ($(v2)!0.5!(v3)+(0,14pt)$)
        {\color{tubeGreen}${\mathfrak g}_{2,R}$};
    }
  }
}
\chaingraph{3}[][][]
}
}
+
\!\!\!\!\!
\raisebox{10pt}{
{
\tikzset{
  chain dashed edges={1,2},
  every picture/.append style={
    execute at end picture={
      \node[
        chain small blob,
        draw=tubeBlue,
        color=tubeBlue,
        fit=(Lv1)(Rv1)
      ] {};

      \node[
        chain small blob,
        draw=tubeBlue,
        color=tubeBlue,
        fit=(Lv2)(Rv2)
      ] {};

      \node[
        chain small blob,
        draw=tubeBlue,
        color=tubeBlue,
        fit=(Lv3)(Rv3)
      ] {};

      \node at ($(v1)+(0,14pt)$)
        {\color{tubeBlue}${\mathfrak g}_{1,L}$};

      \node at ($(v2)+(0,14pt)$)
        {\color{tubeBlue}${\mathfrak g}_{1,M}$};

      \node at ($(v3)+(0,14pt)$)
        {\color{tubeBlue}${\mathfrak g}_{1,R}$};
    }
  }
}
\chaingraph{3}[][][]
}
}
\,,
\ee
which can be translated into the equation
\be
\label{eq: 3 site chain corr}
\begin{aligned}
    \langle \varphi_1 \varphi_2\varphi_3\varphi_4\varphi_5\rangle'=\frac{\lambda^3 H^2 \eta_*^5}{\prod_{a=1}^{5} 2 k_a} \frac{1}{4 Y_{12} Y_{23}} \bigg(& 2\Re[iF_{\textcolor{tubeRed}{ {\cal G}_3}}]+4\Re[F_{\textcolor{tubeGreen}{\mathfrak{g}_{2,L}}}]\Re[iF_{\textcolor{tubeBlue}{\mathfrak{g}_{1,R}}}]+4\Re[iF_{\textcolor{tubeBlue}{\mathfrak{g}_{1,L}}}]\Re[F_{\textcolor{tubeGreen}{\mathfrak{g}_{2,R}}}]\\
    &-8\Re[iF_{\textcolor{tubeBlue}{\mathfrak{g}_{1,L}}}]\Re[iF_{\textcolor{tubeBlue}{\mathfrak{g}_{1,M}}}]\Re[iF_{\textcolor{tubeBlue}{\mathfrak{g}_{1,R}}}]\bigg) \,.
\end{aligned}
\ee
Recall from the representation~\eqref{eq:integralrepF} that the $F$ functions are real functions for real kinematics. (The exception is $F_{\mathfrak g_1}$, which has an imaginary part due to its infrared divergence.) This implies that $\Re[iF_{\textcolor{tubeRed}{ {\cal G}_3}}] = 0$ for real kinematics. Since $2\Re[iF_{\textcolor{tubeBlue}{\mathfrak{g}_1}}]=\pi$, we see that all the contributions to the three-site correlator are factors of $\pi$ times functions of lower transcendentality.

\vskip4pt
At four sites, a conceptually new thing happens. As before, we can diagrammatically catalog all of the contributions to the correlator:
\begin{align}
\nonumber
\langle\varphi^6\rangle
~ &= \hspace{-8pt}
\raisebox{9pt}{
{
\tikzset{
  chain small blob/.append style={
    draw=tubeOrange,
    color=tubeOrange
  }
}
\chaingraph{4}
  [v1/v4]
  []
  [v1/v4/{${\color{tubeOrange}{\cal G}_4}$}/{0pt}/{14pt}]
}
}
+
\!\!\!
\raisebox{10pt}{
{
\tikzset{
  chain dashed edges={3},
  every picture/.append style={
    execute at end picture={
      \node[
        chain small blob,
        draw=tubeRed,
        color=tubeRed,
        fit=(Lv1)(Rv3)
      ] {};
      \node[
        chain small blob,
        draw=tubeBlue,
        color=tubeBlue,
        fit=(Lv4)(Rv4)
      ] {};
      \node at ($(v1)!0.5!(v3)+(0,14pt)$)
        {\color{tubeRed}${\mathfrak g}_{3,L}$};
      \node at ($(v4)+(0,14pt)$)
        {\color{tubeBlue}${\mathfrak g}_{1,R}$};
    }
  }
}
\chaingraph{4}[][][]
}
}
+
\!\!\!
\raisebox{10pt}{
{
\tikzset{
  chain dashed edges={2},
  every picture/.append style={
    execute at end picture={
      \node[
        chain small blob,
        draw=tubeGreen,
        color=tubeGreen,
        fit=(Lv1)(Rv2)
      ] {};
      \node[
        chain small blob,
        draw=tubeGreen,
        color=tubeGreen,
        fit=(Lv3)(Rv4)
      ] {};
      \node at ($(v1)!0.5!(v2)+(0,14pt)$)
        {\color{tubeGreen}${\mathfrak g}_{2,L}$};
      \node at ($(v3)!0.5!(v4)+(0,14pt)$)
        {\color{tubeGreen}${\mathfrak g}_{2,R}$};
    }
  }
}
\chaingraph{4}[][][]
}
}
\\[4pt]
&~~~~
+
\!\!\!
\raisebox{10pt}{
{
\tikzset{
  chain dashed edges={1},
  every picture/.append style={
    execute at end picture={
      \node[
        chain small blob,
        draw=tubeBlue,
        color=tubeBlue,
        fit=(Lv1)(Rv1)
      ] {};
      \node[
        chain small blob,
        draw=tubeRed,
        color=tubeRed,
        fit=(Lv2)(Rv4)
      ] {};
      \node at ($(v1)+(0,14pt)$)
        {\color{tubeBlue}${\mathfrak g}_{1,L}$};
      \node at ($(v2)!0.5!(v4)+(0,14pt)$)
        {\color{tubeRed}${\mathfrak g}_{3,R}$};
    }
  }
}
\chaingraph{4}[][][]
}
}
+
\!\!\!
\raisebox{10pt}{
{
\tikzset{
  chain dashed edges={2,3},
  every picture/.append style={
    execute at end picture={
      \node[
        chain small blob,
        draw=tubeGreen,
        color=tubeGreen,
        fit=(Lv1)(Rv2)
      ] {};
      \node[
        chain small blob,
        draw=tubeBlue,
        color=tubeBlue,
        fit=(Lv3)(Rv3)
      ] {};
      \node[
        chain small blob,
        draw=tubeBlue,
        color=tubeBlue,
        fit=(Lv4)(Rv4)
      ] {};
      \node at ($(v1)!0.5!(v2)+(0,14pt)$)
        {\color{tubeGreen}${\mathfrak g}_{2,L}$};
      \node at ($(v3)+(0,14pt)$)
        {\color{tubeBlue}${\mathfrak g}_{1,M}$};
      \node at ($(v4)+(0,14pt)$)
        {\color{tubeBlue}${\mathfrak g}_{1,R}$};
    }
  }
}
\chaingraph{4}[][][]
}
}
+
\!\!\!
\raisebox{10pt}{
{
\tikzset{
  chain dashed edges={1,3},
  every picture/.append style={
    execute at end picture={
      \node[
        chain small blob,
        draw=tubeBlue,
        color=tubeBlue,
        fit=(Lv1)(Rv1)
      ] {};
      \node[
        chain small blob,
        draw=tubeGreen,
        color=tubeGreen,
        fit=(Lv2)(Rv3)
      ] {};
      \node[
        chain small blob,
        draw=tubeBlue,
        color=tubeBlue,
        fit=(Lv4)(Rv4)
      ] {};
      \node at ($(v1)+(0,14pt)$)
        {\color{tubeBlue}${\mathfrak g}_{1,L}$};
      \node at ($(v2)!0.5!(v3)+(0,14pt)$)
        {\color{tubeGreen}${\mathfrak g}_{2,M}$};
      \node at ($(v4)+(0,14pt)$)
        {\color{tubeBlue}${\mathfrak g}_{1,R}$};
    }
  }
}
\chaingraph{4}[][][]
}
}
\\[4pt]
\nonumber
&~~~~
+
\!\!\!
\raisebox{10pt}{
{
\tikzset{
  chain dashed edges={1,2},
  every picture/.append style={
    execute at end picture={
      \node[
        chain small blob,
        draw=tubeBlue,
        color=tubeBlue,
        fit=(Lv1)(Rv1)
      ] {};
      \node[
        chain small blob,
        draw=tubeBlue,
        color=tubeBlue,
        fit=(Lv2)(Rv2)
      ] {};
      \node[
        chain small blob,
        draw=tubeGreen,
        color=tubeGreen,
        fit=(Lv3)(Rv4)
      ] {};
      \node at ($(v1)+(0,14pt)$)
        {\color{tubeBlue}${\mathfrak g}_{1,L}$};
      \node at ($(v2)+(0,14pt)$)
        {\color{tubeBlue}${\mathfrak g}_{1,M}$};
      \node at ($(v3)!0.5!(v4)+(0,14pt)$)
        {\color{tubeGreen}${\mathfrak g}_{2,R}$};
    }
  }
}
\chaingraph{4}[][][]
}
}
+
\!\!\!
\raisebox{10pt}{
{
\tikzset{
  chain dashed edges={1,2,3},
  every picture/.append style={
    execute at end picture={
      \node[
        chain small blob,
        draw=tubeBlue,
        color=tubeBlue,
        fit=(Lv1)(Rv1)
      ] {};
      \node[
        chain small blob,
        draw=tubeBlue,
        color=tubeBlue,
        fit=(Lv2)(Rv2)
      ] {};
      \node[
        chain small blob,
        draw=tubeBlue,
        color=tubeBlue,
        fit=(Lv3)(Rv3)
      ] {};
      \node[
        chain small blob,
        draw=tubeBlue,
        color=tubeBlue,
        fit=(Lv4)(Rv4)
      ] {};
      \node at ($(v1)+(0,14pt)$)
        {\color{tubeBlue}${\mathfrak g}_{1,L}$};
      \node at ($(v2)+(0,14pt)$)
        {\color{tubeBlue}${\mathfrak g}_{1,M_1}$};
      \node at ($(v3)+(0,14pt)$)
       {\color{tubeBlue}${\mathfrak g}_{1,M_2}$};
      \node at ($(v4)+(0,14pt)$)
        {\color{tubeBlue}${\mathfrak g}_{1,R}$};
    }
  }
}
\chaingraph{4}[][][]
}
}
\!\!\!\!
\,.
\end{align}
This corresponds to the following expression for the correlator in terms of wavefunction coefficients
\begin{align}
\label{eq: 4 site chain corr}
\langle \varphi^6\rangle'
=\,&
\frac{\lambda^4 H^2 \eta_*^6}
{\prod_{a=1}^{6}2k_a}\,
\frac{1}{8Y_{12}Y_{23}Y_{34}}
\Bigg(
2\Re[
F_{\textcolor{tubeOrange}{{\cal G}_4}}
]
-4\Re[
iF_{\textcolor{tubeRed}{\mathfrak g_{3,L}}}
]
\Re[
iF_{\textcolor{tubeBlue}{\mathfrak g_{1,R}}}
]
+4\Re[
F_{\textcolor{tubeGreen}{\mathfrak g_{2,L}}}
]
\Re[
F_{\textcolor{tubeGreen}{\mathfrak g_{2,R}}}
]
\nonumber\\[4pt]
\nonumber
&\!\!\!
-4\Re[
iF_{\textcolor{tubeBlue}{\mathfrak g_{1,L}}}
]
\Re[
iF_{\textcolor{tubeRed}{\mathfrak g_{3,R}}}
]
-8\Re[
F_{\textcolor{tubeGreen}{\mathfrak g_{2,L}}}
]
\Re[
iF_{\textcolor{tubeBlue}{\mathfrak g_{1,M}}}
]
\Re[
iF_{\textcolor{tubeBlue}{\mathfrak g_{1,R}}}
]
-8\Re[
iF_{\textcolor{tubeBlue}{\mathfrak g_{1,L}}}
]
\Re[
F_{\textcolor{tubeGreen}{\mathfrak g_{2,M}}}
]
\Re[
iF_{\textcolor{tubeBlue}{\mathfrak g_{1,R}}}
]
\\[4pt]
&\!\!\!
-8\Re[
iF_{\textcolor{tubeBlue}{\mathfrak g_{1,L}}}
]
\Re[
iF_{\textcolor{tubeBlue}{\mathfrak g_{1,M}}}
]
\Re[
F_{\textcolor{tubeGreen}{\mathfrak g_{2,R}}}
]
+16\Re[
iF_{\textcolor{tubeBlue}{\mathfrak g_{1,L}}}
]
\Re[
iF_{\textcolor{tubeBlue}{\mathfrak g_{1,M_1}}}
]
\Re[
iF_{\textcolor{tubeBlue}{\mathfrak g_{1,M_2}}}
]
\Re[
iF_{\textcolor{tubeBlue}{\mathfrak g_{1,R}}}
]
\Bigg)\,.
\end{align}
Here we notice that there is a contribution to the correlator of the form $\Re[
F_{\textcolor{tubeGreen}{\mathfrak g_{2}}}
]^2
$,
which is a product of two two-site wavefunctions. This disconnected piece has the same transcendentality (four) as $\Re[
F_{\textcolor{tubeOrange}{{\cal G}_4}}
]$. It turns out that both of these contributions are cluster compatible, and so there is a one-parameter family of objects consistent with cluster compatibility. The four-site chain wavefunction is the first time that this happens, because it requires us to be able to delete an internal line to produce two graphs with an even number of sites. (The real part of odd-site wavefunctions has lower transcendentality.)

\vskip4pt
For completeness, we give one example involving a loop. The two-site loop graph correlator receives contributions from the following diagrams
\begin{equation}
\begin{tikzpicture}[baseline=(current bounding box.center)]

  \def\R{0.75}
  \def\dr{0.25}

  \pgfmathsetmacro{\Rout}{\R+\dr}
  \pgfmathsetmacro{\Rin}{\R-\dr}
  \pgfmathsetmacro{\dotR}{1.1*\R}

  \def\colsep{3.0}
  \pgfmathsetmacro{\cA}{0}
  \pgfmathsetmacro{\cB}{\colsep}
  \pgfmathsetmacro{\cC}{2*\colsep}
  \pgfmathsetmacro{\cD}{3*\colsep}
  \pgfmathsetmacro{\pA}{0.5*\colsep}
  \pgfmathsetmacro{\pB}{1.5*\colsep}
  \pgfmathsetmacro{\pC}{2.5*\colsep}

  \begin{scope}[shift={(\cA,0)}, rotate=90]
    \draw[color=tubePurple, line width=0.9pt] (0,0) circle (\Rout);
    \draw[line width=1.05pt] (0,0) circle (\R);
    \fill (0, \R) circle (\dotR mm);
    \fill (0,-\R) circle (\dotR mm);
  \end{scope}
  \node[color=tubePurple] at (\cA, {\Rout+0.3})
    {$\mathcal{G}_2^{(1)}$};

  \node at (\pA,0) {$+$};

  \begin{scope}[shift={(\cB,0)}, rotate=90]

    \draw[color=tubeGreen, line width=0.9pt]
      ({\Rout*cos(90)}, {\Rout*sin(90)})
      arc[start angle=90,  end angle=-90,  radius=\Rout]
      arc[start angle=-90, end angle=-270, radius=\dr]
      arc[start angle=-90, end angle=90,   radius=\Rin]
      arc[start angle=270, end angle=90,   radius=\dr]
      -- cycle;

    \draw[line width=1.05pt]
      (0,\R) arc[start angle=90, end angle=-90, radius=\R];

    \draw[dashed, line width=1.05pt]
      (0,\R) arc[start angle=90, end angle=270, radius=\R];

    \fill (0, \R) circle (\dotR mm);
    \fill (0,-\R) circle (\dotR mm);
  \end{scope}
  \node[color=tubeGreen, above=3pt] at (\cB,\Rout)
    {$\mathfrak{g}_{2,U}$};

  \node at (\pB,0) {$+$};

  \begin{scope}[shift={(\cC,0)}, rotate=90]

    \draw[color=tubeGreen, line width=0.9pt]
      ({\Rout*cos(90)}, {\Rout*sin(90)})
      arc[start angle=90,  end angle=270, radius=\Rout]
      arc[start angle=-90, end angle=90,  radius=\dr]
      arc[start angle=270, end angle=90,  radius=\Rin]
      arc[start angle=-90, end angle=90,  radius=\dr]
      -- cycle;

    \draw[line width=1.05pt]
      (0,\R) arc[start angle=90, end angle=270, radius=\R];

    \draw[dashed, line width=1.05pt]
      (0,\R) arc[start angle=90, end angle=-90, radius=\R];

    \fill (0, \R) circle (\dotR mm);
    \fill (0,-\R) circle (\dotR mm);
  \end{scope}
  \node[color=tubeGreen, below=3pt] at (\cC,{-\Rout})
    {$\mathfrak{g}_{2,D}$};

  \node at (\pC,0) {$+$};

  \begin{scope}[shift={(\cD,0)}, rotate=90]

    \node[
      draw=tubeBlue,
      color=tubeBlue,
      line width=0.9pt,
      circle,
      inner sep=5pt
    ] at (0,\R) {};

    \node[
      draw=tubeBlue,
      color=tubeBlue,
      line width=0.9pt,
      circle,
      inner sep=5pt
    ] at (0,-\R) {};

    \draw[dashed, line width=1.05pt]
      (0,\R) arc[start angle=90, end angle=-90, radius=\R];

    \draw[dashed, line width=1.05pt]
      (0,\R) arc[start angle=90, end angle=270, radius=\R];

    \fill (0, \R) circle (\dotR mm);
    \fill (0,-\R) circle (\dotR mm);
  \end{scope}

  \node[color=tubeBlue, above=5pt]
    at ({\cD-\R-.25},{0.22})
    {$\mathfrak{g}_{1,L}$};

  \node[color=tubeBlue, above=5pt]
    at ({\cD+\R+.25},{0.22})
    {$\mathfrak{g}_{1,R}$};

\end{tikzpicture}\,,
\end{equation}
which corresponds to the sum of terms
\begin{equation}
\label{eq: 2 site loop corr}
\langle \varphi^2\rangle'{}^{(1)}
=
\frac{\lambda^2 \eta_*^2}
{4k_1k_2}
\frac{1}{4Y_{12}Y_{21}}
\Big(
-2\Re[F_{\textcolor{tubePurple}{{\cal G}_2^{(1)}}}]
-
2\Re[F_{\textcolor{tubeGreen}{\mathfrak g_{2,U}}}]
-
2\Re[F_{\textcolor{tubeGreen}{\mathfrak g_{2,D}}}]
+
4\Re[iF_{\textcolor{tubeBlue}{\mathfrak g_{1,L}}}]
 \Re[iF_{\textcolor{tubeBlue}{\mathfrak g_{1,R}}}]
\Big)\,.
\end{equation}
Here we see that the two-site wavefunctions have the same transcendentality as the two-site loop, similar to the four-site chain.
It is straightforward to continue to more complicated graphs and write the correlator in terms of wavefunction coefficients using~\eqref{eq: correlator in terms of cut WF coeffs}.

\paragraph{Symbols and Cluster Algebras}~\\
Since it is possible to write in-in correlators in terms of wavefunction coefficients, it is natural to examine whether the cluster structure we have investigated in the main text for wavefunctions has a counterpart in correlators. To facilitate this, we first note that the real part of the one-site wavefunction is $2\Re[iF_{\textcolor{tubeBlue}{\mathfrak{g}_1}}]= \pi$. This means that any terms in the correlator involving a factor of $\Re[iF_{\textcolor{tubeBlue}{\mathfrak{g}_1}}]$ will drop out of the symbol. Another important simplification comes from the fact that for physical kinematics the functions $F$ are real. This means that terms involving an odd number of interaction vertices in~\eqref{eq: correlator in terms of cut WF coeffs} will vanish when we take the real part (for real kinematics). As a result, the in-in correlator has lower transcendentality than the corresponding wavefunction in these cases~\cite{Chowdhury:2026dwm}. (Note that if we imagine complexifying kinematics---which is often interesting---the correlator again has the same transcendentality as the wavefunction.)

\vskip4pt
With these simplifications, the symbol of in-in correlators is then given by that of the wavefunction along with the shuffle product of terms involving dashed internal lines that split the graph into pieces that have no one-site graphs (projecting out graphs with odd numbers of vertices). For example, the cases we considered correspond to (for real kinematics) 
\be
\label{eq:corrsymfromWF}
\begin{aligned}
  \mathcal{S}\left(\langle \varphi^4\rangle\right)&\propto\, \mathcal{S}\left(F_{\textcolor{tubeGreen}{{\cal G}_2}}\right)\\
  \mathcal{S}\left(\langle \varphi^5\rangle\right)&= 0\\
  \mathcal{S}\left(\langle \varphi^6\rangle\right)&\propto\, \mathcal{S}\left(F_{\textcolor{tubeOrange}{{\cal G}_4}}\right)+2\, \mathcal{S}\left(F_{\textcolor{tubeGreen}{\mathfrak{g}_{2,L}}}\right)\shuffle\mathcal{S}\left(F_{\textcolor{tubeGreen}{\mathfrak{g}_{2,R}}}\right) \\
  \mathcal{S}\big(\langle \varphi^2\rangle^{(1)}\big)&\propto\,\mathcal{S}\big(F_{\textcolor{tubePurple}{{\cal G}_2^{(1)}}}\big)+\mathcal{S}\left(F_{\textcolor{tubeGreen}{\mathfrak{g}_{2,U}}}\right)+\mathcal{S}\left(F_{\textcolor{tubeGreen}{\mathfrak{g}_{2,D}}}\right)\,.
\end{aligned}
\ee
Here $0$ should be interpreted to mean that the leading transcendental part of the symbol vanishes.
We therefore see that starting at four sites, the chain correlator has contributions from products of lower-point functions, while such contributions appear starting with the two-site loop.

\vskip4pt
\noindent
{\it Cluster Compatibility:}~\\
We see from~\eqref{eq:corrsymfromWF} that generically the symbol of a correlator receives contributions from subgraphs of the original Feynman graph. As we saw in the main text, the symbols corresponding to each of these subgraphs (including the full graph) are individually cluster compatible with respect to some $A_N$ or $B_N$ cluster algebra. This does not, however, guarantee that these various algebras will be compatible with each other when we combine the symbols corresponding to different subgraphs. The basic failure mode is that the cluster algebra corresponding to the tubings of the subgraph could fail to be a subalgebra of the cluster algebra of the full graph.
It turns out that for chains, both the wavefunction and correlator are cluster compatible, while for loops only the wavefunction is compatible.

\vskip4pt
Imagine we have an $m$-site chain subgraph labeled with indices $\{i_1,\ldots,i_{2m}\}$,
\begin{equation*}
\;\cdots\;
{\scalebox{.9}{%
\markedchaingraph{2}
  [][]
  [v1/x1/{$i_1$}/{-21pt}/{-8pt},
   v1/x1/{$i_2$}/{0pt}/{-8pt},
   x1/v2/{$i_3$}/{0pt}/{-8pt}
   ][1.8]%
}}
\;\cdots\;
{\scalebox{.9}{%
\markedchaingraph{2}
  [][]
  [
   x1/v2/{$\cdots$ ~~~~$i_{2m-1}$}/{-13pt}/{-8pt},
   v2/x2/{$i_{2m}$}/{20pt}/{-8pt}
   ][1.8]%
}
}
\;\cdots\;
\end{equation*}
which is
embedded in some larger $n$-site chain or loop graph. The $m$-site chain is cluster compatible with respect to $A_{2m-2}$, while the larger graph has a cluster algebra structure that is either $A_{2n-2}$ or $B_{2n-1}$. We want to see if these two algebraic structures are compatible with each other. 
To do this, we can write the energies of the subgraph in terms of cluster variables of the full graph by writing
\be
E_{i_{k} i_{l}} = \frac{a_{i_{k} i_{l}}}{\tilde a_{i_k}\tilde a_{i_l}}\,,
\ee
where $\tilde{a}_i=a_{i\,2n+1}$ for $A_{2n-2}$ and $\tilde{a}_i= a_{[i,\, i+2n]}$ for $B_{2n-1}$ (we have scaled $E_{[i,\bar i]}$ to 1). Notice that compared to the original $i_1,\cdots, i_{2m}$ indices, this definition involves a single additional index $(2n+1)$ for the $A_N$ case and $2m$ new indices in the $B_N$ case.
In the $A_N$ case, each cluster variable corresponding to the subgraph can be associated to a chord in a $(2m+1)$-gon, with vertices labeled by $\{i_1,\ldots,i_{2m}\}\cup \{2n+1\}$, producing an $A_{2m-2}$ subalgebra of $A_{2n-2}$ (correspondingly, the vertices correspond to a sub-polygon). This means that chain correlators have the same cluster compatibility that the wavefunction does.
With the $B_{2n-1}$ prescription, the cluster variable indices lie in $\{i_1,\ldots,i_{2m}\}\cup \{i_{1}+2n,\ldots,i_{2m}+2n\}$, which does not form an $A_{2m-2}$ subalgebra in $B_{2n-1}$. Correspondingly, the product of these lower-site chains is {\it not} cluster compatible in the sense of the loop graph. The result is that the correlator in these cases does not have the $B_N$ cluster compatibility that the wavefunction does.

\vskip4pt
This gives a natural interpretation of the families of solutions to the constraints imposed in Section~\ref{sec:clusterbootstrap}. 
For $n$-site chain graphs starting at four sites, we expect both the wavefunction and products of lower-site chains to be weight-$n$ transcendental functions with an $A_{2n-2}$ cluster algebra structure. To isolate the wavefunction from this family, we can impose the condition that no cluster variables in a symbol word can repeat amongst the second through last symbol letters. (It would be nice to find a similarly simple characterization of the correlator in cluster language.) For loop graphs, we only expect a single cluster compatible function, the wavefunction coefficient of the loop, which is what we find.

\newpage
\phantomsection
\addcontentsline{toc}{section}{References}
\bibliographystyle{utphys}
{\linespread{.98}
	\bibliography{clusterbib}

\providecommand{\href}[2]{#2}\begingroup\raggedright\begin{thebibliography}{100}

\bibitem{Baumann:2022jpr}
D.~Baumann, D.~Green, A.~Joyce, E.~Pajer, G.~L. Pimentel, C.~Sleight, and
  M.~Taronna, ``{Snowmass White Paper: The Cosmological Bootstrap},''
  \href{http://dx.doi.org/10.21468/SciPostPhysCommRep.1}{{\em SciPost Phys.
  Comm. Rep.} {\bfseries 2024} (2024) 1},
  \href{http://arxiv.org/abs/2203.08121}{{\ttfamily arXiv:2203.08121
  [hep-th]}}.

\bibitem{Benincasa:2022gtd}
P.~Benincasa, ``{Amplitudes meet Cosmology: A (Scalar) Primer},''
  \href{http://arxiv.org/abs/2203.15330}{{\ttfamily arXiv:2203.15330
  [hep-th]}}.

\bibitem{Lee:2024sks}
M.~H.~G. Lee, E.~Pajer, M.~Giroux, H.~S. Hannesdottir, S.~Mizera, and
  C.~Pasiecznik, ``{Records from the S-Matrix Marathon: A Timeless History of
  Time},''
\newblock 9, 2024.
\newblock \href{http://arxiv.org/abs/2410.00227}{{\ttfamily arXiv:2410.00227
  [hep-th]}}.

\bibitem{corrlectures}
D.~Baumann and A.~Joyce, {\em
  {\href{https://github.com/ddbaumann/cosmo-correlators}{Lectures on
  Cosmological Correlations}}}.

\bibitem{Arkani-Hamed:2017fdk}
N.~Arkani-Hamed, P.~Benincasa, and A.~Postnikov, ``{Cosmological Polytopes and
  the Wavefunction of the Universe},''
  \href{http://arxiv.org/abs/1709.02813}{{\ttfamily arXiv:1709.02813
  [hep-th]}}.

\bibitem{Hillman:2019wgh}
A.~Hillman, ``{Symbol Recursion for the dS Wave Function},''
  \href{http://arxiv.org/abs/1912.09450}{{\ttfamily arXiv:1912.09450
  [hep-th]}}.

\bibitem{Arkani-Hamed:2023kig}
N.~Arkani-Hamed, D.~Baumann, A.~Hillman, A.~Joyce, H.~Lee, and G.~L. Pimentel,
  ``{Differential equations for cosmological correlators},''
  \href{http://dx.doi.org/10.1007/JHEP09(2025)009}{{\em JHEP} {\bfseries 09}
  (2025) 009}, \href{http://arxiv.org/abs/2312.05303}{{\ttfamily
  arXiv:2312.05303 [hep-th]}}.

\bibitem{Goncharov:2010jf}
A.~Goncharov, M.~Spradlin, C.~Vergu, and A.~Volovich, ``{Classical
  Polylogarithms for Amplitudes and Wilson Loops},''
  \href{http://dx.doi.org/10.1103/PhysRevLett.105.151605}{{\em Phys. Rev.
  Lett.} {\bfseries 105} (2010) 151605},
  \href{http://arxiv.org/abs/1006.5703}{{\ttfamily arXiv:1006.5703 [hep-th]}}.

\bibitem{Steinmann1960a}
O.~Steinmann, ``Über den zusammenhang zwischen den wightmanfunktionen und den
  retardierten kommutatoren,'' {\em Helvetica Physica Acta} {\bfseries 33}
  (1960) 257--275.

\bibitem{Steinmann1960b}
O.~Steinmann, ``Wightman-funktionen und retardierten kommutatoren ii,'' {\em
  Helvetica Physica Acta} {\bfseries 33} (1960) 347--362.

\bibitem{Caron-Huot:2016owq}
S.~Caron-Huot, L.~J. Dixon, A.~McLeod, and M.~von Hippel, ``{Bootstrapping a
  Five-Loop Amplitude Using Steinmann Relations},''
  \href{http://dx.doi.org/10.1103/PhysRevLett.117.241601}{{\em Phys. Rev.
  Lett.} {\bfseries 117} no.~24, (2016) 241601},
  \href{http://arxiv.org/abs/1609.00669}{{\ttfamily arXiv:1609.00669
  [hep-th]}}.

\bibitem{Caron-Huot:2019bsq}
S.~Caron-Huot, L.~J. Dixon, F.~Dulat, M.~Von~Hippel, A.~J. McLeod, and
  G.~Papathanasiou, ``{The Cosmic Galois Group and Extended Steinmann Relations
  for Planar $\mathcal{N} = 4$ SYM Amplitudes},''
  \href{http://dx.doi.org/10.1007/JHEP09(2019)061}{{\em JHEP} {\bfseries 09}
  (2019) 061}, \href{http://arxiv.org/abs/1906.07116}{{\ttfamily
  arXiv:1906.07116 [hep-th]}}.

\bibitem{Caron-Huot:2019vjl}
S.~Caron-Huot, L.~J. Dixon, F.~Dulat, M.~von Hippel, A.~J. McLeod, and
  G.~Papathanasiou, ``{Six-Gluon amplitudes in planar $ \mathcal{N} $ = 4
  super-Yang-Mills theory at six and seven loops},''
  \href{http://dx.doi.org/10.1007/JHEP08(2019)016}{{\em JHEP} {\bfseries 08}
  (2019) 016}, \href{http://arxiv.org/abs/1903.10890}{{\ttfamily
  arXiv:1903.10890 [hep-th]}}.

\bibitem{FominZelevinsky2002ClusterI}
S.~Fomin and A.~Zelevinsky, ``Cluster algebras {I}: Foundations,''
  \href{http://dx.doi.org/10.1090/S0894-0347-01-00385-X}{{\em Journal of the
  American Mathematical Society} {\bfseries 15} no.~2, (2002) 497--529},
  \href{http://arxiv.org/abs/math/0104151}{{\ttfamily arXiv:math/0104151
  [math.RT]}}.

\bibitem{FominZelevinsky2003ClusterII}
S.~Fomin and A.~Zelevinsky, ``Cluster algebras {II}: Finite type
  classification,'' \href{http://dx.doi.org/10.1007/s00222-003-0302-y}{{\em
  Inventiones Mathematicae} {\bfseries 154} no.~1, (2003) 63--121},
  \href{http://arxiv.org/abs/math/0208229}{{\ttfamily arXiv:math/0208229
  [math.RA]}}.

\bibitem{BerensteinFominZelevinsky2005ClusterIII}
A.~Berenstein, S.~Fomin, and A.~Zelevinsky, ``Cluster algebras {III}: Upper
  bounds and double {B}ruhat cells,''
  \href{http://dx.doi.org/10.1215/S0012-7094-04-12611-9}{{\em Duke Mathematical
  Journal} {\bfseries 126} no.~1, (2005) 1--52},
  \href{http://arxiv.org/abs/math/0305434}{{\ttfamily arXiv:math/0305434
  [math.RT]}}.

\bibitem{FominZelevinsky2007ClusterIV}
S.~Fomin and A.~Zelevinsky, ``Cluster algebras {IV}: Coefficients,''
  \href{http://dx.doi.org/10.1112/S0010437X06002521}{{\em Compositio
  Mathematica} {\bfseries 143} no.~1, (2007) 112--164},
  \href{http://arxiv.org/abs/math/0602259}{{\ttfamily arXiv:math/0602259
  [math.RA]}}.

\bibitem{Golden:2013xva}
J.~Golden, A.~B. Goncharov, M.~Spradlin, C.~Vergu, and A.~Volovich, ``{Motivic
  Amplitudes and Cluster Coordinates},''
  \href{http://dx.doi.org/10.1007/JHEP01(2014)091}{{\em JHEP} {\bfseries 01}
  (2014) 091}, \href{http://arxiv.org/abs/1305.1617}{{\ttfamily arXiv:1305.1617
  [hep-th]}}.

\bibitem{Drummond:2019cxm}
J.~Drummond, J.~Foster, {\"O}.~G{\"u}rdogan, and C.~Kalousios, ``{Algebraic
  singularities of scattering amplitudes from tropical geometry},''
  \href{http://dx.doi.org/10.1007/JHEP04(2021)002}{{\em JHEP} {\bfseries 04}
  (2021) 002}, \href{http://arxiv.org/abs/1912.08217}{{\ttfamily
  arXiv:1912.08217 [hep-th]}}.

\bibitem{Arkani-Hamed:2019rds}
N.~Arkani-Hamed, T.~Lam, and M.~Spradlin, ``{Non-perturbative geometries for
  planar $ \mathcal{N} $ = 4 SYM amplitudes},''
  \href{http://dx.doi.org/10.1007/JHEP03(2021)065}{{\em JHEP} {\bfseries 03}
  (2021) 065}, \href{http://arxiv.org/abs/1912.08222}{{\ttfamily
  arXiv:1912.08222 [hep-th]}}.

\bibitem{Drummond:2017ssj}
J.~Drummond, J.~Foster, and {\"O}.~G{\"u}rdo{\u{g}}an, ``{Cluster Adjacency
  Properties of Scattering Amplitudes in $N=4$ Supersymmetric Yang-Mills
  Theory},'' \href{http://dx.doi.org/10.1103/PhysRevLett.120.161601}{{\em Phys.
  Rev. Lett.} {\bfseries 120} no.~16, (2018) 161601},
  \href{http://arxiv.org/abs/1710.10953}{{\ttfamily arXiv:1710.10953
  [hep-th]}}.

\bibitem{Drummond:2018dfd}
J.~Drummond, J.~Foster, and {\"O}.~G{\"u}rdo{\u{g}}an, ``{Cluster adjacency
  beyond MHV},'' \href{http://dx.doi.org/10.1007/JHEP03(2019)086}{{\em JHEP}
  {\bfseries 03} (2019) 086}, \href{http://arxiv.org/abs/1810.08149}{{\ttfamily
  arXiv:1810.08149 [hep-th]}}.

\bibitem{Caron-Huot:2018dsv}
S.~Caron-Huot, L.~J. Dixon, M.~von Hippel, A.~J. McLeod, and G.~Papathanasiou,
  ``{The Double Pentaladder Integral to All Orders},''
  \href{http://dx.doi.org/10.1007/JHEP07(2018)170}{{\em JHEP} {\bfseries 07}
  (2018) 170}, \href{http://arxiv.org/abs/1806.01361}{{\ttfamily
  arXiv:1806.01361 [hep-th]}}.

\bibitem{Dixon:2016nkn}
L.~J. Dixon, J.~Drummond, T.~Harrington, A.~J. McLeod, G.~Papathanasiou, and
  M.~Spradlin, ``{Heptagons from the Steinmann Cluster Bootstrap},''
  \href{http://dx.doi.org/10.1007/JHEP02(2017)137}{{\em JHEP} {\bfseries 02}
  (2017) 137}, \href{http://arxiv.org/abs/1612.08976}{{\ttfamily
  arXiv:1612.08976 [hep-th]}}.

\bibitem{Drummond:2014ffa}
J.~M. Drummond, G.~Papathanasiou, and M.~Spradlin, ``{A Symbol of Uniqueness:
  The Cluster Bootstrap for the 3-Loop MHV Heptagon},''
  \href{http://dx.doi.org/10.1007/JHEP03(2015)072}{{\em JHEP} {\bfseries 03}
  (2015) 072}, \href{http://arxiv.org/abs/1412.3763}{{\ttfamily arXiv:1412.3763
  [hep-th]}}.

\bibitem{Dixon:2020cnr}
L.~J. Dixon and Y.-T. Liu, ``{Lifting Heptagon Symbols to Functions},''
  \href{http://dx.doi.org/10.1007/JHEP10(2020)031}{{\em JHEP} {\bfseries 10}
  (2020) 031}, \href{http://arxiv.org/abs/2007.12966}{{\ttfamily
  arXiv:2007.12966 [hep-th]}}.

\bibitem{Chicherin:2020umh}
D.~Chicherin, J.~M. Henn, and G.~Papathanasiou, ``{Cluster algebras for Feynman
  integrals},'' \href{http://dx.doi.org/10.1103/PhysRevLett.126.091603}{{\em
  Phys. Rev. Lett.} {\bfseries 126} no.~9, (2021) 091603},
  \href{http://arxiv.org/abs/2012.12285}{{\ttfamily arXiv:2012.12285
  [hep-th]}}.

\bibitem{Caron-Huot:2020bkp}
S.~Caron-Huot, L.~J. Dixon, J.~M. Drummond, F.~Dulat, J.~Foster,
  {\"O}.~G{\"u}rdo{\u{g}}an, M.~von Hippel, A.~J. McLeod, and G.~Papathanasiou,
  ``{The Steinmann Cluster Bootstrap for $N$ = 4 Super Yang-Mills
  Amplitudes},'' \href{http://dx.doi.org/10.22323/1.376.0003}{{\em PoS}
  {\bfseries CORFU2019} (2020) 003},
  \href{http://arxiv.org/abs/2005.06735}{{\ttfamily arXiv:2005.06735
  [hep-th]}}.

\bibitem{Dixon:2022rse}
L.~J. Dixon, O.~Gurdogan, A.~J. McLeod, and M.~Wilhelm, ``{Bootstrapping a
  stress-tensor form factor through eight loops},''
  \href{http://dx.doi.org/10.1007/JHEP07(2022)153}{{\em JHEP} {\bfseries 07}
  (2022) 153}, \href{http://arxiv.org/abs/2204.11901}{{\ttfamily
  arXiv:2204.11901 [hep-th]}}.

\bibitem{Mazloumi:2025pmx}
P.~Mazloumi and X.~Xu, ``{Cluster algebras for cosmological correlators},''
  \href{http://dx.doi.org/10.1007/JHEP03(2026)256}{{\em JHEP} {\bfseries 03}
  (2026) 256}, \href{http://arxiv.org/abs/2512.14854}{{\ttfamily
  arXiv:2512.14854 [hep-th]}}.

\bibitem{Capuano:2025myy}
M.~Capuano, L.~Ferro, T.~Lukowski, and A.~Palazio, ``{Cosmology meets cluster
  algebra},'' \href{http://arxiv.org/abs/2512.14859}{{\ttfamily
  arXiv:2512.14859 [hep-th]}}.

\bibitem{Paranjape:2026htn}
S.~Paranjape, M.~Skowronek, M.~Spradlin, A.~Volovich, and H.-C. Weng,
  ``{Cluster Bootstrap for Cosmological Correlators},''
  \href{http://arxiv.org/abs/2603.08670}{{\ttfamily arXiv:2603.08670
  [hep-th]}}.

\bibitem{Capuano:2026pgq}
M.~Capuano, L.~Ferro, T.~Lukowski, A.~Palazio, and Y.-Q. Zhang, ``{Generalised
  Cluster Adjacency for Cosmology},''
  \href{http://arxiv.org/abs/2603.09965}{{\ttfamily arXiv:2603.09965
  [hep-th]}}.

\bibitem{Ferro:2026oph}
L.~Ferro, T.~Lukowski, L.~Ren, M.~Spradlin, A.~Volovich, H.-C. Weng, and Y.-Q.
  Zhang, ``{de Sitter Wavefunction from Quadrangular Polylogarithms: Chain
  Graphs},'' \href{http://arxiv.org/abs/2605.06542}{{\ttfamily arXiv:2605.06542
  [hep-th]}}.

\bibitem{Benincasa:2020aoj}
P.~Benincasa, A.~J. McLeod, and C.~Vergu, ``{Steinmann Relations and the
  Wavefunction of the Universe},''
  \href{http://dx.doi.org/10.1103/PhysRevD.102.125004}{{\em Phys. Rev. D}
  {\bfseries 102} (2020) 125004},
  \href{http://arxiv.org/abs/2009.03047}{{\ttfamily arXiv:2009.03047
  [hep-th]}}.

\bibitem{Benincasa:2021qcb}
P.~Benincasa and W.~J.~T. Bobadilla, ``{Physical representations for scattering
  amplitudes and the wavefunction of the universe},''
  \href{http://dx.doi.org/10.21468/SciPostPhys.12.6.192}{{\em SciPost Phys.}
  {\bfseries 12} no.~6, (2022) 192},
  \href{http://arxiv.org/abs/2112.09028}{{\ttfamily arXiv:2112.09028
  [hep-th]}}.

\bibitem{Hillman:2023vas}
A.~Hillman, {\em {On the Infrared and Ultraviolet Behavior of Scattering
  Amplitudes and Wavefunctions}}.
\newblock PhD thesis, Princeton U., 2023.

\bibitem{Koba:1969rw}
Z.~Koba and H.~B. Nielsen, ``{Reaction amplitude for n mesons: A Generalization
  of the Veneziano-Bardakci-Ruegg-Virasora model},''
  \href{http://dx.doi.org/10.1016/0550-3213(69)90331-9}{{\em Nucl. Phys. B}
  {\bfseries 10} (1969) 633--655}.

\bibitem{Koba:1969kh}
Z.~Koba and H.~B. Nielsen, ``{Manifestly crossing invariant parametrization of
  n meson amplitude},''
  \href{http://dx.doi.org/10.1016/0550-3213(69)90071-6}{{\em Nucl. Phys. B}
  {\bfseries 12} (1969) 517--536}.

\bibitem{brown2006multiplezetavaluesperiods}
F.~C.~S. Brown, ``Multiple zeta values and periods of moduli spaces
  $\mathfrak{M}_{0,n}$,'' 2006.
\newblock \url{https://arxiv.org/abs/math/0606419}.

\bibitem{Arkani-Hamed:2019mrd}
N.~Arkani-Hamed, S.~He, and T.~Lam, ``{Stringy canonical forms},''
  \href{http://dx.doi.org/10.1007/JHEP02(2021)069}{{\em JHEP} {\bfseries 02}
  (2021) 069}, \href{http://arxiv.org/abs/1912.08707}{{\ttfamily
  arXiv:1912.08707 [hep-th]}}.

\bibitem{Arkani-Hamed:2019vag}
N.~Arkani-Hamed, S.~He, G.~Salvatori, and H.~Thomas, ``{Causal diamonds,
  cluster polytopes and scattering amplitudes},''
  \href{http://dx.doi.org/10.1007/JHEP11(2022)049}{{\em JHEP} {\bfseries 11}
  (2022) 049}, \href{http://arxiv.org/abs/1912.12948}{{\ttfamily
  arXiv:1912.12948 [hep-th]}}.

\bibitem{Arkani-Hamed:2023lbd}
N.~Arkani-Hamed, H.~Frost, G.~Salvatori, P.-G. Plamondon, and H.~Thomas, ``{All
  loop scattering as a counting problem},''
  \href{http://dx.doi.org/10.1007/JHEP08(2025)194}{{\em JHEP} {\bfseries 08}
  (2025) 194}, \href{http://arxiv.org/abs/2309.15913}{{\ttfamily
  arXiv:2309.15913 [hep-th]}}.

\bibitem{Arkani-Hamed:2023swr}
N.~Arkani-Hamed, Q.~Cao, J.~Dong, C.~Figueiredo, and S.~He, ``{Hidden zeros for
  particle/string amplitudes and the unity of colored scalars, pions and
  gluons},'' \href{http://dx.doi.org/10.1007/JHEP10(2024)231}{{\em JHEP}
  {\bfseries 10} (2024) 231}, \href{http://arxiv.org/abs/2312.16282}{{\ttfamily
  arXiv:2312.16282 [hep-th]}}.

\bibitem{Arkani-Hamed:2019plo}
N.~Arkani-Hamed, S.~He, T.~Lam, and H.~Thomas, ``{Binary geometries,
  generalized particles and strings, and cluster algebras},''
  \href{http://dx.doi.org/10.1103/PhysRevD.107.066015}{{\em Phys. Rev. D}
  {\bfseries 107} no.~6, (2023) 066015},
  \href{http://arxiv.org/abs/1912.11764}{{\ttfamily arXiv:1912.11764
  [hep-th]}}.

\bibitem{Arkani-Hamed:2020tuz}
N.~Arkani-Hamed, S.~He, and T.~Lam, ``{Cluster Configuration Spaces of Finite
  Type},'' \href{http://dx.doi.org/10.3842/SIGMA.2021.092}{{\em SIGMA}
  {\bfseries 17} (2021) 092}, \href{http://arxiv.org/abs/2005.11419}{{\ttfamily
  arXiv:2005.11419 [math.AG]}}.

\bibitem{He:2020onr}
S.~He, Z.~Li, P.~Raman, and C.~Zhang, ``{Stringy canonical forms and binary
  geometries from associahedra, cyclohedra and generalized permutohedra},''
  \href{http://dx.doi.org/10.1007/JHEP10(2020)054}{{\em JHEP} {\bfseries 10}
  (2020) 054}, \href{http://arxiv.org/abs/2005.07395}{{\ttfamily
  arXiv:2005.07395 [hep-th]}}.

\bibitem{Arkani-Hamed:2025zuf}
N.~Arkani-Hamed, H.~Frost, P.-G. Plamondon, G.~Salvatori, and H.~Thomas,
  ``{Configuration Spaces of Finite Representation Type Algebras},''
  \href{http://arxiv.org/abs/2512.24870}{{\ttfamily arXiv:2512.24870
  [math.RT]}}.

\bibitem{He:2021esx}
S.~He, Z.~Li, and Q.~Yang, ``{Notes on cluster algebras and some all-loop
  Feynman integrals},'' \href{http://dx.doi.org/10.1007/JHEP06(2021)119}{{\em
  JHEP} {\bfseries 06} (2021) 119},
  \href{http://arxiv.org/abs/2103.02796}{{\ttfamily arXiv:2103.02796
  [hep-th]}}.

\bibitem{rudenko2022}
D.~Rudenko, ``On the goncharov depth conjecture and a formula for volumes of
  orthoschemes,'' 2022.
\newblock \url{https://arxiv.org/abs/2012.05599}.

\bibitem{matveiakin2022}
A.~Matveiakin and D.~Rudenko, ``Cluster polylogarithms i: Quadrangular
  polylogarithms,'' 2022.
\newblock \url{https://arxiv.org/abs/2208.01564}.

\bibitem{Arkani-Hamed:2015bza}
N.~Arkani-Hamed and J.~Maldacena, ``{Cosmological Collider Physics},''
  \href{http://arxiv.org/abs/1503.08043}{{\ttfamily arXiv:1503.08043
  [hep-th]}}.

\bibitem{Baumann:2019oyu}
D.~Baumann, C.~Duaso~Pueyo, A.~Joyce, H.~Lee, and G.~L. Pimentel, ``{The
  cosmological bootstrap: weight-shifting operators and scalar seeds},''
  \href{http://dx.doi.org/10.1007/JHEP12(2020)204}{{\em JHEP} {\bfseries 12}
  (2020) 204}, \href{http://arxiv.org/abs/1910.14051}{{\ttfamily
  arXiv:1910.14051 [hep-th]}}.

\bibitem{Baumann:2020dch}
D.~Baumann, C.~Duaso~Pueyo, A.~Joyce, H.~Lee, and G.~L. Pimentel, ``{The
  Cosmological Bootstrap: Spinning Correlators from Symmetries and
  Factorization},'' \href{http://dx.doi.org/10.21468/SciPostPhys.11.3.071}{{\em
  SciPost Phys.} {\bfseries 11} (2021) 071},
  \href{http://arxiv.org/abs/2005.04234}{{\ttfamily arXiv:2005.04234
  [hep-th]}}.

\bibitem{Arkani-Hamed:2024jbp}
N.~Arkani-Hamed, C.~Figueiredo, and F.~Vaz{\~a}o, ``{Cosmohedra},''
  \href{http://dx.doi.org/10.1007/JHEP11(2025)029}{{\em JHEP} {\bfseries 11}
  (2025) 029}, \href{http://arxiv.org/abs/2412.19881}{{\ttfamily
  arXiv:2412.19881 [hep-th]}}.

\bibitem{Arkani-Hamed:2017ahv}
N.~Arkani-Hamed and E.~Y. Yuan, ``{One-Loop Integrals from Spherical
  Projections of Planes and Quadrics},''
  \href{http://arxiv.org/abs/1712.09991}{{\ttfamily arXiv:1712.09991
  [hep-th]}}.

\bibitem{Chowdhury:2025ohm}
C.~Chowdhury, A.~Lipstein, J.~Marshall, J.~Mei, and I.~Sachs, ``{Cosmological
  dressing rules},'' \href{http://dx.doi.org/10.1007/JHEP03(2026)076}{{\em
  JHEP} {\bfseries 03} (2026) 076},
  \href{http://arxiv.org/abs/2503.10598}{{\ttfamily arXiv:2503.10598
  [hep-th]}}.

\bibitem{Chowdhury:2026dwm}
C.~Chowdhury, S.~He, Y.-X. Su, and D.~Yang, ``{On the simplicity of de Sitter
  correlators},'' \href{http://arxiv.org/abs/2604.26421}{{\ttfamily
  arXiv:2604.26421 [hep-th]}}.

\bibitem{goncharov2001multiple}
A.~Goncharov, ``{Multiple Polylogarithms and Mixed Tate Motives},''
  \href{http://arxiv.org/abs/0103059}{{\ttfamily arXiv:0103059 [math.AG]}}.

\bibitem{goncharov2009simple}
A.~Goncharov, ``{A Simple Construction of Grassmannian Polylogarithms},''
  \href{http://arxiv.org/abs/0908.2238}{{\ttfamily arXiv:0908.2238 [math.AG]}}.

\bibitem{Chen:2006nt}
X.~Chen, M.-x. Huang, S.~Kachru, and G.~Shiu, ``{Observational signatures and
  non-Gaussianities of general single field inflation},''
  \href{http://dx.doi.org/10.1088/1475-7516/2007/01/002}{{\em JCAP} {\bfseries
  01} (2007) 002}, \href{http://arxiv.org/abs/hep-th/0605045}{{\ttfamily
  arXiv:hep-th/0605045}}.

\bibitem{Holman:2007na}
R.~Holman and A.~J. Tolley, ``{Enhanced Non-Gaussianity from Excited Initial
  States},'' \href{http://dx.doi.org/10.1088/1475-7516/2008/05/001}{{\em JCAP}
  {\bfseries 05} (2008) 001}, \href{http://arxiv.org/abs/0710.1302}{{\ttfamily
  arXiv:0710.1302 [hep-th]}}.

\bibitem{Flauger:2013hra}
R.~Flauger, D.~Green, and R.~A. Porto, ``{On squeezed limits in single-field
  inflation. Part I},''
  \href{http://dx.doi.org/10.1088/1475-7516/2013/08/032}{{\em JCAP} {\bfseries
  08} (2013) 032}, \href{http://arxiv.org/abs/1303.1430}{{\ttfamily
  arXiv:1303.1430 [hep-th]}}.

\bibitem{Arkani-Hamed:2018kmz}
N.~Arkani-Hamed, D.~Baumann, H.~Lee, and G.~L. Pimentel, ``{The Cosmological
  Bootstrap: Inflationary Correlators from Symmetries and Singularities},''
  \href{http://dx.doi.org/10.1007/JHEP04(2020)105}{{\em JHEP} {\bfseries 04}
  (2020) 105}, \href{http://arxiv.org/abs/1811.00024}{{\ttfamily
  arXiv:1811.00024 [hep-th]}}.

\bibitem{Green:2020whw}
D.~Green and R.~A. Porto, ``{Signals of a Quantum Universe},''
  \href{http://dx.doi.org/10.1103/PhysRevLett.124.251302}{{\em Phys. Rev.
  Lett.} {\bfseries 124} no.~25, (2020) 251302},
  \href{http://arxiv.org/abs/2001.09149}{{\ttfamily arXiv:2001.09149
  [hep-th]}}.

\bibitem{carr2005coxetercomplexesgraphassociahedra}
M.~Carr and S.~L. Devadoss, ``Coxeter complexes and graph-associahedra,'' 2005.
\newblock \url{https://arxiv.org/abs/math/0407229}.

\bibitem{devadoss2006realizationgraphassociahedra}
S.~L. Devadoss, ``A realization of graph-associahedra,'' 2006.
\newblock \url{https://arxiv.org/abs/math/0612530}.

\bibitem{Baumann:2026atn}
D.~Baumann, A.~Joyce, H.~Lee, and K.~Salehi~Vaziri, ``{Differential Equations
  for Massive Correlators},'' \href{http://arxiv.org/abs/2604.08658}{{\ttfamily
  arXiv:2604.08658 [hep-th]}}.

\bibitem{Benincasa:2018ssx}
P.~Benincasa, ``{From the flat-space S-matrix to the Wavefunction of the
  Universe},'' \href{http://arxiv.org/abs/1811.02515}{{\ttfamily
  arXiv:1811.02515 [hep-th]}}.

\bibitem{Chowdhury:2023arc}
C.~Chowdhury, A.~Lipstein, J.~Mei, I.~Sachs, and P.~Vanhove, ``{The subtle
  simplicity of cosmological correlators},''
  \href{http://dx.doi.org/10.1007/JHEP03(2025)007}{{\em JHEP} {\bfseries 03}
  (2025) 007}, \href{http://arxiv.org/abs/2312.13803}{{\ttfamily
  arXiv:2312.13803 [hep-th]}}.

\bibitem{Glew:2025arc}
R.~Glew, ``{Correlators from Amplitubes},''
  \href{http://arxiv.org/abs/2507.07199}{{\ttfamily arXiv:2507.07199
  [hep-th]}}.

\bibitem{Arkani-Hamed:2025mce}
N.~Arkani-Hamed, R.~Glew, and F.~Vaz{\~a}o, ``{Correlators are simpler than
  wavefunctions},'' \href{http://arxiv.org/abs/2512.23795}{{\ttfamily
  arXiv:2512.23795 [hep-th]}}.

\bibitem{Benincasa:2024ptf}
P.~Benincasa, G.~Brunello, M.~K. Mandal, P.~Mastrolia, and F.~Vaz{\~a}o,
  ``{One-loop corrections to the Bunch-Davies wave function of the universe},''
  \href{http://dx.doi.org/10.1103/PhysRevD.111.085016}{{\em Phys. Rev. D}
  {\bfseries 111} no.~8, (2025) 085016},
  \href{http://arxiv.org/abs/2408.16386}{{\ttfamily arXiv:2408.16386
  [hep-th]}}.

\bibitem{Pimentel:2026kqc}
G.~L. Pimentel and T.~Westerdijk, ``{On Cosmological Correlators at One
  Loop},'' \href{http://arxiv.org/abs/2601.00952}{{\ttfamily arXiv:2601.00952
  [hep-th]}}.

\bibitem{Donath:2024utn}
Y.~Donath and E.~Pajer, ``{The in-out formalism for in-in correlators},''
  \href{http://dx.doi.org/10.1007/JHEP07(2024)064}{{\em JHEP} {\bfseries 07}
  (2024) 064}, \href{http://arxiv.org/abs/2402.05999}{{\ttfamily
  arXiv:2402.05999 [hep-th]}}.

\bibitem{lam2016laurentphenomenonalgebras}
T.~Lam and P.~Pylyavskyy, ``Laurent phenomenon algebras,'' 2016.
\newblock \url{https://arxiv.org/abs/1206.2611}.

\bibitem{McLeod:2026jpz}
A.~McLeod, A.~Pokraka, and L.~Ren, ``{A Graphical Coaction for FRW Wavefunction
  Coefficients},'' \href{http://arxiv.org/abs/2603.25703}{{\ttfamily
  arXiv:2603.25703 [hep-th]}}.

\bibitem{McLeod:2026kpo}
A.~J. McLeod, A.~Pokraka, and L.~Ren, ``{A Graphical Coaction for FRW Integrals
  from Partial/Relative Twisted (Co)homology},''
  \href{http://arxiv.org/abs/2606.13627}{{\ttfamily arXiv:2606.13627
  [hep-th]}}.

\bibitem{Gasparotto:2024bku}
F.~Gasparotto, P.~Mazloumi, and X.~Xu, ``{Differential equations for tree-level
  cosmological correlators with massive states},''
  \href{http://dx.doi.org/10.1007/JHEP09(2025)043}{{\em JHEP} {\bfseries 09}
  (2025) 043}, \href{http://arxiv.org/abs/2411.05632}{{\ttfamily
  arXiv:2411.05632 [hep-th]}}.

\bibitem{Arkani-Hamed:2023bsv}
N.~Arkani-Hamed, D.~Baumann, A.~Hillman, A.~Joyce, H.~Lee, and G.~L. Pimentel,
  ``{Kinematic Flow and the Emergence of Time},''
  \href{http://dx.doi.org/10.1103/dsjm-tckw}{{\em Phys. Rev. Lett.} {\bfseries
  135} no.~3, (2025) 031602}, \href{http://arxiv.org/abs/2312.05300}{{\ttfamily
  arXiv:2312.05300 [hep-th]}}.

\bibitem{Baumann:2025qjx}
D.~Baumann, H.~Goodhew, A.~Joyce, H.~Lee, G.~L. Pimentel, and T.~Westerdijk,
  ``{Geometry of kinematic flow},''
  \href{http://dx.doi.org/10.1007/JHEP05(2026)211}{{\em JHEP} {\bfseries 05}
  (2026) 211}, \href{http://arxiv.org/abs/2504.14890}{{\ttfamily
  arXiv:2504.14890 [hep-th]}}.

\bibitem{De:2023xue}
S.~De and A.~Pokraka, ``{Cosmology meets cohomology},''
  \href{http://dx.doi.org/10.1007/JHEP03(2024)156}{{\em JHEP} {\bfseries 03}
  (2024) 156}, \href{http://arxiv.org/abs/2308.03753}{{\ttfamily
  arXiv:2308.03753 [hep-th]}}.

\bibitem{De:2024zic}
S.~De and A.~Pokraka, ``{A physical basis for cosmological correlators from
  cuts},'' \href{http://dx.doi.org/10.1007/JHEP03(2025)040}{{\em JHEP}
  {\bfseries 03} (2025) 040}, \href{http://arxiv.org/abs/2411.09695}{{\ttfamily
  arXiv:2411.09695 [hep-th]}}.

\bibitem{Fan:2024iek}
B.~Fan and Z.-Z. Xianyu, ``{Cosmological amplitudes in power-law FRW
  universe},'' \href{http://dx.doi.org/10.1007/JHEP12(2024)042}{{\em JHEP}
  {\bfseries 12} (2024) 042}, \href{http://arxiv.org/abs/2403.07050}{{\ttfamily
  arXiv:2403.07050 [hep-th]}}.

\bibitem{He:2024olr}
S.~He, X.~Jiang, J.~Liu, Q.~Yang, and Y.-Q. Zhang, ``{Differential equations
  and recursive solutions for cosmological amplitudes},''
  \href{http://dx.doi.org/10.1007/JHEP01(2025)001}{{\em JHEP} {\bfseries 01}
  (2025) 001}, \href{http://arxiv.org/abs/2407.17715}{{\ttfamily
  arXiv:2407.17715 [hep-th]}}.

\bibitem{Glew:2025ypb}
R.~Glew and A.~Pokraka, ``{Kinematic flow from the flow of cuts},''
  \href{http://dx.doi.org/10.1007/JHEP06(2026)158}{{\em JHEP} {\bfseries 06}
  (2026) 158}, \href{http://arxiv.org/abs/2508.11568}{{\ttfamily
  arXiv:2508.11568 [hep-th]}}.

\bibitem{Capuano:2025ehm}
M.~Capuano, L.~Ferro, T.~Lukowski, and A.~Palazio, ``{Canonical Differential
  Equations for Cosmology from Positive Geometries},''
  \href{http://arxiv.org/abs/2505.14609}{{\ttfamily arXiv:2505.14609
  [hep-th]}}.

\bibitem{Fu:2026dqb}
Y.~Fu and J.~Liu, ``{Notes on Diagrammatic Coaction for Cosmological
  Wavefunction Coefficients: A Two-Site Prelude},''
  \href{http://arxiv.org/abs/2603.25698}{{\ttfamily arXiv:2603.25698
  [hep-th]}}.

\bibitem{williams2014cluster}
L.~Williams, ``Cluster algebras: an introduction,'' {\em Bulletin of the
  American Mathematical Society} {\bfseries 51} no.~1, (2014) 1--26.

\bibitem{Daniel_2015}
D.~Parker, ``Cluster algebra structures for scattering amplitudes in $n=4$
  super yang--mills,'' 2015.
\newblock
  \url{https://danielericparker.github.io/documents/parker_senior_thesis.pdf}.

\bibitem{le2017approachclusterstructuresmoduli}
I.~Le, ``An approach to cluster structures on moduli of local systems for
  general groups,'' 2017.
\newblock \url{https://arxiv.org/abs/1606.00961}.

\bibitem{2016arXiv160805735F}
S.~{Fomin}, L.~{Williams}, and A.~{Zelevinsky}, ``{Introduction to Cluster
  Algebras. Chapters 1-3},'' \href{http://arxiv.org/abs/1608.05735}{{\ttfamily
  arXiv:1608.05735 [math.CO]}}.

\bibitem{Papathanasiou:2022lan}
G.~Papathanasiou, ``{The SAGEX review on scattering amplitudes Chapter 5:
  Analytic bootstraps for scattering amplitudes and beyond},''
  \href{http://dx.doi.org/10.1088/1751-8121/ac7e8e}{{\em J. Phys. A} {\bfseries
  55} no.~44, (2022) 443006}, \href{http://arxiv.org/abs/2203.13016}{{\ttfamily
  arXiv:2203.13016 [hep-th]}}.

\bibitem{2012arXiv1209.3987R}
N.~{Reading}, ``{Universal geometric cluster algebras},''
  \href{http://arxiv.org/abs/1209.3987}{{\ttfamily arXiv:1209.3987 [math.RA]}}.

\bibitem{Bazier-Matte:2018rat}
V.~Bazier-Matte, N.~Chapelier-Laget, G.~Douville, K.~Mousavand, H.~Thomas, and
  E.~Y{\i}ld{\i}r{\i}m, ``{ABHY Associahedra and Newton polytopes of
  F-polynomials for cluster algebras of simply laced finite type},''
  \href{http://dx.doi.org/10.1112/jlms.12817}{{\em J. Lond. Math. Soc.}
  {\bfseries 109} no.~1, (2024) e12817},
  \href{http://arxiv.org/abs/1808.09986}{{\ttfamily arXiv:1808.09986
  [math.RT]}}.

\bibitem{2008arXiv0804.3303Y}
S.-W. {Yang} and A.~{Zelevinsky}, ``{Cluster algebras of finite type via
  Coxeter elements and principal minors},''
  \href{http://arxiv.org/abs/0804.3303}{{\ttfamily arXiv:0804.3303 [math.RA]}}.

\bibitem{lee2014positivityclusteralgebras}
K.~Lee and R.~Schiffler, ``Positivity for cluster algebras,'' 2014.
\newblock \url{https://arxiv.org/abs/1306.2415}.

\bibitem{assoc}
J.~D. Stasheff, ``Homotopy associativity of h-spaces. ii,'' {\em Transactions
  of the American Mathematical Society} {\bfseries 108} no.~2, (1963) 293--312.

\bibitem{cyclo}
R.~Bott and C.~Taubes, ``On the self‐linking of knots,''
  \href{http://dx.doi.org/10.1063/1.530750}{{\em Journal of Mathematical
  Physics} {\bfseries 35} no.~10, (10, 1994) 5247--5287}.

\bibitem{10.2969/jmsj/02720248}
K.~Aomoto, ``{On vanishing of cohomology attached to certain many valued
  meromorphic functions},'' \href{http://dx.doi.org/10.2969/jmsj/02720248}{{\em
  Journal of the Mathematical Society of Japan} {\bfseries 27} no.~2, (1975)
  248 -- 255}.

\bibitem{brown2017notesmotivicperiods}
F.~Brown, ``Notes on motivic periods,'' 2017.
\newblock \url{https://arxiv.org/abs/1512.06410}.

\bibitem{Duhr:2011zq}
C.~Duhr, H.~Gangl, and J.~R. Rhodes, ``{From polygons and symbols to
  polylogarithmic functions},''
  \href{http://dx.doi.org/10.1007/JHEP10(2012)075}{{\em JHEP} {\bfseries 10}
  (2012) 075}, \href{http://arxiv.org/abs/1110.0458}{{\ttfamily arXiv:1110.0458
  [math-ph]}}.

\bibitem{Duhr:2014woa}
C.~Duhr, \href{http://dx.doi.org/10.1142/9789814678766_0010}{``{Mathematical
  aspects of scattering amplitudes},''} in {\em {Theoretical Advanced Study
  Institute in Elementary Particle Physics}: {Journeys Through the Precision
  Frontier: Amplitudes for Colliders}}, pp.~419--476.
\newblock 2015.
\newblock \href{http://arxiv.org/abs/1411.7538}{{\ttfamily arXiv:1411.7538
  [hep-ph]}}.

\bibitem{Arkani-Hamed:2017tmz}
N.~Arkani-Hamed, Y.~Bai, and T.~Lam, ``{Positive Geometries and Canonical
  Forms},'' \href{http://dx.doi.org/10.1007/JHEP11(2017)039}{{\em JHEP}
  {\bfseries 11} (2017) 039}, \href{http://arxiv.org/abs/1703.04541}{{\ttfamily
  arXiv:1703.04541 [hep-th]}}.

\bibitem{Benincasa:2024leu}
P.~Benincasa and G.~Dian, ``{The geometry of cosmological correlators},''
  \href{http://dx.doi.org/10.21468/SciPostPhys.18.3.105}{{\em SciPost Phys.}
  {\bfseries 18} no.~3, (2025) 105},
  \href{http://arxiv.org/abs/2401.05207}{{\ttfamily arXiv:2401.05207
  [hep-th]}}.

\bibitem{Figueiredo:2025daa}
C.~Figueiredo and F.~Vaz{\~a}o, ``{Correlator polytopes},''
  \href{http://dx.doi.org/10.1103/bq27-6d26}{{\em Phys. Rev. D} {\bfseries 113}
  no.~2, (2026) 025005}, \href{http://arxiv.org/abs/2506.19907}{{\ttfamily
  arXiv:2506.19907 [hep-th]}}.

\end{thebibliography}\endgroup
}

\end{document}